\documentclass[12pt]{article}

\usepackage{cite}

\usepackage{graphicx}
\usepackage{subfigure}

\usepackage{amssymb}
\usepackage{amsmath}
\usepackage{amscd}
\usepackage{latexsym}

\long\def\symbolfootnote[#1]#2{\begingroup%
\def\thefootnote{\fnsymbol{footnote}}\footnote[#1]{#2}\endgroup} 

\newcommand{\be}{\begin{equation}}
\newcommand{\ee}{\end{equation}}

\newcommand{\al}{\alpha}
\newcommand{\ra}{\rightarrow}

\newcommand{\gm}{\gamma}

\begin{document}

\begin{center}
{\Large{\bf Modeling symbiosis by interactions through species carrying capacities} \\ [5mm]

V.I. Yukalov$^{1,2,}$\symbolfootnote[1]{{\bf Corresponding author}: \\
V.I. Yukalov \\ Bogolubov Laboratory of Theoretical Physics \\ 
Joint Institute for Nuclear Research, Dubna 141980, Russia \\ 
{\bf E-mail}: yukalov@theor.jinr.ru \\ {\bf Tel}: +7(496) 21 63947, 
{\bf Fax}: +7(496) 21 65084}, E.P. Yukalova$^{1,3}$, 
and D. Sornette$^{1,4}$} \\ [3mm]

{\it
$^1$Department of Management, Technology and Economics, \\
Swiss Federal Institute of Technology, Z\"urich CH-8032, Switzerland \\ [3mm]

$^2$Bogolubov Laboratory of Theoretical Physics, \\
Joint Institute for Nuclear Research, Dubna 141980, Russia \\

$^3$Laboratory of Information Technologies, \\
Joint Institute for Nuclear Research, Dubna 141980, Russia \\

$^4$Swiss Finance Institute, c/o University of Geneva, \\
40 blvd. Du Pont d'Arve, CH 1211 Geneva 4, Switzerland }

\end{center}

\vskip 1cm

\begin{abstract}
We introduce a mathematical model of symbiosis between different species 
by taking into account the influence of each species on the carrying 
capacities of the others. The modeled entities can pertain to biological 
and ecological societies or to social, economic and financial
societies. Our model includes three basic types: symbiosis with direct
mutual interactions, symbiosis with asymmetric interactions, and symbiosis
without direct interactions. In all cases, we provide a complete
classification of all admissible dynamical regimes. The proposed model of
symbiosis turned out to be very rich, as it exhibits four qualitatively
different regimes: convergence to stationary states, unbounded exponential
growth, finite-time singularity, and finite-time death or extinction of
species.
\end{abstract}

\vskip 1cm


\vskip 1cm


\newpage

\section{Introduction}

The term symbiosis describes close and usually long-term interactions
between different biological species. Symbiotic relationships are
well known in biological and ecological societies. Numerous examples
can be found in the books \cite{1,2,3,4,5}. Arguably, the symbiosis that
is the most important to us as humans is the one between the typical
human body and the multitudes of commensal organisms representing
members of five of the six kingdoms of life. Specifically, the microbiome,
with more than 2000 bacterial species and some $10^{14}$ microorganisms
(about ten times the number of cells of the human body), is a key partner
to the human immune system, and to the human metabolism as it participates
in the synthesis of essential vitamins and amino acids, as well as in
the degradation of otherwise indigestible plant material, and of certain
drugs and pollutants in the guts \cite{6}. Let us also mention the
associations between plant roots and fungi, that are a feature of many
terrestrial ecosystems, and without which the kingdom of plants would
not exist as we know it.

In biology, one distinguishes three main types of symbiotic relations:
mutualism, commensalism, and parasitism. Below, we briefly define these 
types of symbiosis, as they are described in the books \cite{1,2,3,4,5},
and give some examples illustrating these relations.

{\it Mutualism} is any relationship between individuals of different 
species where both individuals derive a benefit. Generally, only lifelong 
interactions involving close physical and biochemical contact can properly 
be considered symbiotic. A large percentage of herbivores have mutualistic 
gut fauna that help them digest plant matter, which is more difficult to 
digest than animal prey. Coral reefs are the result of mutualisms between 
coral organisms and various types of algae that live inside them. Most 
land plants and land ecosystems rely on mutualisms between the plants, which 
fix carbon from the air, and mycorrhizal fungi, which help in extracting 
minerals from the ground. An example of mutual symbiosis is the relationship 
between the ocellaris clownfish that dwell among the tentacles of Ritteri 
sea anemones. The territorial fish protects the anemone from anemone-eating 
fish, and in turn the stinging tentacles of the anemone protect the clownfish 
from its predators. A special mucus on the clownfish protects it from the 
stinging tentacles. See details and more examples in Refs. \cite{1,2,3,4,5}.

{\it Commensalism} is a class of relationship between two organisms where 
one organism benefits but the other is neutral (there is no harm or benefit). 
An example of commensalism: cattle egrets foraging in fields among cattle or 
other livestock. As cattle, horses and other livestock graze on the field, 
they cause movements that stir up various insects. As the insects are stirred 
up, the cattle egrets following the livestock catch and feed upon them. The 
egrets benefit from this relationship because the livestock have helped them 
find their meals, while the livestock are typically unaffected by it. Another 
example of commensalism is between tigers and golden jackals. In India, lone 
golden jackals expelled from their pack have been known to form commensal 
relationships with tigers. These solitary jackals will attach themselves to 
a particular tiger, trailing it at a safe distance in order to feed on the 
big cat's kills. Tigers have been known to tolerate these jackals though 
having no profit from them.  One more example of commensalism: birds following 
army ant raids on a forest floor. As the army ant colony travels on the forest 
floor, they stir up various flying insect species. As the insects flee from 
the army ants, the birds following the ants catch the fleeing insects. In this 
way, the army ants and the birds are in a commensal relationship because the 
birds benefit while the army ants are unaffected \cite{1,2,3,4,5}.

{\it Parasitism} is a type of symbiotic relationship between organisms of 
different species where one organism, the parasite, benefits at the expense 
of the other, the host. Traditionally parasite referred to organisms with 
lifestages that needed more than one host. These are now called macroparasites 
(typically protozoa and helminths). The word parasite now also refers to 
microparasites, which are typically smaller, such as viruses and bacteria, 
and can be directly transmitted between hosts of the same species. Parasites 
that live on the surface of the host are called ectoparasites (e.g. some mites). 
Those that live inside the host are called endoparasites (including all 
parasitic worms). A typical example of endoparasites are bacteria and 
viruses \cite{1,2,3,4,5}.

Generally speaking, symbiosis is not restricted to biological systems.
Many relations in social and economic societies can also be interpreted
as examples of symbiosis. For instance, the economic and intellectual
levels of human societies can be considered as symbiotic to each other.
The interconnections between basic and applied research are symbiotic
as well \cite{7,8,9}. Economic and financial relations between different
firms or between firms and banks can be treated as symbiotic. The
notion of symbiosis has a very wide applicability to various relations,
whether in biology, ecology, economy, or finance. Therefore, when we
shall refer to societies, we keep in mind different kinds of societies,
including biological, human, economic, and others, whose interactions
can be considered from the point of view of symbiotic relations.

Extending the notion of symbiosis to social and economic systems, one 
meets the same types of symbiosis in many different forms and structures. 
As in biology, there can occur relations between different more or less
independent parts or species, illustrated by the relations between
plant roots and fungi \cite{10}, or in economics, the relations
between firms and banks, or between economies and arts. Social systems
are also often characterized by the existence of complicated relations,
e.g., corresponding to the relations between a general system and a
particular subsystem. Examples are the relations between the economy
of a country and a particular branch of the economy, or between
culture and language. Biological counterparts are widespread, as
illustrated by the endosymbiotic origin of mitochondria and plastids
of eukaryotic cells, by some marine annelid worms lacking mouth, guts
and anus, and which rely on multiple extracellular bacterial endosymbionts
for their excretory system \cite{11}, and more generally between the
microbiome network and the mammal body \cite{12}.

Recently, the novel term {\it industrial symbiosis} has appeared to describe
a certain type of eco-industrial development within the larger framework 
of industrial ecology. Industrial ecology is a relatively new 
field that is based on the ideology of nature. It claims that industrial 
ecosystem may behave similar to the natural ecosystem. In the same way 
as in nature, where symbiosis refers to an association between at least 
two different species, industrial symbiosis is an association between 
two or more industrial facilities or companies. For example, the wastes 
or byproducts of one company can become the raw materials for another 
\cite{Chertow,Graedel,Pearce}. Tyre shred, plastic pellets or waste 
steam from a factory are examples of outputs that can be sold on to 
other businesses. Mutual collaboration between different firms in many 
cases can be characterized as mutualistic symbiosis \cite{Press}. 
In this way, the notion of symbiosis is now applicable to biological 
as well as to economical societies.  

Interactions between different co-existing species are usually modeled 
by equations of the Lotka-Volterra type \cite{13,14}. Such equations are 
adequate for predator-prey relations between different species, whether 
one studies the simple co-existence of two species \cite{13,14},
or more complicated cases corresponding to high-dimensional dynamical
systems, such as multiply connected food webs \cite{15,16}, or the
inter-relations between different cells and pathogens inside biological
organisms \cite{17,18}. However, the symbiotic relations are known to be
principally different from the predator-prey relations \cite{1,2,3,4,5}.
Hence, they require another mathematical representation.

To be precise, we keep in mind the dynamical representation of symbiotic
relations and the evolution of coexisting symbiotic species characterized
by their concentrations. There are particular models describing the
biological co-evolution of species by considering equations for some
distributions, whether over size, or over age, or over phenotypes
\cite{19,20,21,22}. But our aim is to obtain the equations for the
concentrations themselves, so that these equations could describe the
variety of possible symbiotic relations.

Monod \cite{23,24} considered the microbial growth in a chemostat,
which is a liquid phase of a chemical substrate nutrient. The Monod
equations have been widely used for describing fermentation processes
\cite{25,26}. These equations can also describe the situation when
two or more bacterial species compete for the same growth substrate
\cite{27,28,29,30}. When the populations of microorganisms, inhabiting
a common nonliving environment, compete for nutrients, this is, of
course, an example of coexistence and coevolution. This, however, is not
symbiosis in its direct sense, though it may share many features of the
latter. The usual end of the evolution, when two species compete for
the same food substrate, is that the one attaining the higher growth
rate, under the given conditions, competes more successfully and
ultimately displaces the slower-growing competitor \cite{28,29,30}.
It is well known \cite{1,2,3,4,5,31} that competition processes are
quite different from symbiosis. Therefore several bacterial species
competing for the same food substrate \cite{27,28,29,30}, strictly
speaking, do not illustrate symbiosis or, in the best case, represent
its very particular form.

The Holling \cite{H} second type functional relation between predators
and prey describes the situation when predators meet, not the amount of
prey proportional to their total number, but an effective number of
attacked prey that is proportional to the prey density $D$ over $1 + c D$,
with a parameter $c > 0$. This relation takes into account that predators,
in order to consume prey, need to search for it, chase, kill, eat, and
digest. This is why predators attack not all prey but a limited number
of them, which saturates to a constant when the prey density increases
\cite{K}. Such predator-prey relations, clearly, do not characterize
symbiosis.

The main aim of the present paper is to suggest a mathematical model 
of symbiosis characterizing the overall dynamics of this process for 
various systems. We concentrate in the present paper on the general 
mathematical properties of the considered model and on the classification 
of admissible types of symbiotic behavior in the frame of this model. 

We stress that our aim is not the reinvention of a novel qualitative
classification of symbiotic relations. Such a classification is well
developed and widely employed in the biological literature
\cite{1,2,3,4,5}. And we accept this classification, as most other
authors do. Our aim is to suggest a model of symbiosis that
would fit well the known classification and would provide the
possibility for studying {\it dynamical} as well as {\it mathematical}
consequences of different symbiosis types.

The basic point that allows us to suggest a model of symbiosis
is the idea that in symbiotic relations it is not the species that 
interact directly with each other, as in the Lotka-Volterra equation,
but that {\it symbiotic species act on the carrying capacities of 
each other}. 

The carrying capacity of a biological species in an environment is 
generally understood as the maximum population size of the species that 
the environment can sustain indefinitely, given the food, habitat, water 
and other necessities available in the environment. In population biology, 
carrying capacity is defined as the environment maximal load \cite{Hui}.
Historically, carrying capacity has been treated as a given fixed value 
\cite{Zimmerer,Sayre}. But later it has been understood that the carrying 
capacity of an environment may vary for different numbers of species and 
may change over time due to a variety of factors, including food 
availability, water supply, environmental conditions, living space, and,
the most important, population activity. 

Mutual coexistence and symbiosis of several species strongly influence
the carrying capacities of the species, with the changes being, to a first
approximation, proportional to the species numbers. For example, humans 
have increased the carrying capacity of the environment for a few other 
species, including those with which we live in a mutually beneficial 
symbiosis. Those companion species include more than about 20 billion 
domestic animals such as cows, horses, pigs, sheep, goats, dogs, cats, 
and chickens, as well as certain plants such as wheat, rice, barley, 
maize, tomato, and cabbage. Clearly, humans and their selected companions 
have benefited greatly through active management of mutual carrying 
capacities \cite{Begon}.

Interactions between two or more biological species are known to essentially 
influence the carrying capacity of each other, by either increasing it, 
when species derive a mutual benefit, or decreasing it, when their interactions
are antagonistic \cite{Boucher, Callaway, Stachowicz}. The same applies to 
economic and financial interactions between firms, which also form a kind of 
symbiosis, where the interacting firms develop the carrying capacity of each 
other also roughly proportionally to their sizes \cite{Press}. When species 
coexist, their carrying capacities are influenced by the species mutual 
interactions, either facilitating the capacity development or damaging it. 
Being functions of the species populations, such nonequilibrium carrying 
capacities can be naturally represented as polynomials over the population 
numbers \cite{Richerson}. 

Thus, it is now generally accepted that symbiotic organisms influence the 
carrying capacities of each other, hence the carrying capacities of 
symbiotic species are not fixed quantities, but should be considered as 
functions of population sizes \cite{Brown,Goff}. In the present paper, we 
suggest a mathematical formulation of this idea and investigate its 
consequences. 

The paper proceeds as follows. Section 2 presents the general
structure of the model. In Section 3, we specify the classification
of different types of symbiotic interactions, in the frame of the
suggested mathematical model. Sections 4, 5, and 6 focus on the
structure of the model of symbiosis in the presence of mutual,
respectively, asymmetric, and in the absence of direct interactions.
Section 7 concludes, explaining the principal difference of our
model from the Lotka-Volterra model, and proposes a way to avoid the
finite-time singularities arising under parasitic interactions.
Also, we described several concrete examples demonstrating that
the suggested mathematical models can be directly applied to
different types of symbiotic relations, whether in biological
symbiosis or in economic and industrial symbiosis. A discussion is 
given on the relation of the suggested approach to the processes  
describing the coexistence of microbial populations.

\section{General model of symbiosis}

Our basic suggestion, making our approach principally new, is to describe 
symbiotic relations by the mutual influence of the co-existing species 
on their respective carrying capacities. Indeed, in the presence of 
symbiotic relations, the species act on the livelihood of each other 
by creating resources that others exploit, by using resources created 
by others without feedbacks, or by destroying the livelihood in the 
case of parasitism. In mathematical language, the livelihood is nothing 
but the carrying capacity. This is the principal point of the approach 
we suggest, as compared to the predator-prey models, where the species 
directly predate or compete. Such direct interactions are appropriate for
describing the predator-prey relations, when predators eat prey.
But this is not what one usually calls symbiosis. Predation and
competition are commonly treated as different processes as compared
to symbiosis \cite{31}. The standard description of symbiosis
\cite{1,2,3,4,5} corresponds exactly to the mutual influence of
species on the carrying capacities of each other. For example, in the
tree-fungi symbiosis, neither of the species eats another, but they
do influence the carrying capacities of each other, producing the
chemical elements helping the growth of both species.

Let us consider several species, or system parts, quantified by their
population number $N_i$, with the index $i$ enumerating the species.
For instance, in biology, $N_i$ can be the number of individuals of
the $i$-th species or, in finance, this can be the amount of money
or, in economics, this could be the quantity of some goods. To study
the mutual influence of varying carrying capacities, we start
with the logistic-type equations
\be
\label{1}
\frac{dN_i}{dt} = \gm_i N_i \; - \;
\frac{C_i N_i^2}{K_i} \; ,
\ee
for the variables $N_i = N_i(t)$ as functions of time $t \geq 0$.
Here $\gamma_i$ is the birth rate of the biological species $i$
or the growth rate in economic systems. The coefficient $C_i$
characterizes the intensity of mutual competition between the agents
of the $i$-th species. We will thus consider only the case where
\be
\label{2}
\gm_i > 0 \; , \qquad C_i > 0 \; .
\ee
The principal difference from the standard logistic equation is that
the carrying capacity is here considered to be a function
\be
\label{3}
K_i = A_i + B_i S_i ( \{ N_1,N_2,\ldots \} )
\ee
of the quantities $N_i$, in order to account for the co-existing
symbiotic species. The first term $A_i$ is the carrying capacity
of the given surrounding livelihood. The second term characterizes
the carrying capacity produced by other species. We thus capture
the typical feature of symbiotic relations via the mutual influence
of each species on the carrying capacities of others. The
{\it symbiotic coefficient} $B_i$ defines the intensity of producing,
or destroying, the carrying capacity in the process of symbiotic
relations. We refer to $B_i$ as the {\it production coefficient},
when $B_i$ is positive, or as the {\it destruction coefficient},
when $B_i$ is negative. Thence, we shall consider that
\be
\label{4}
A_i > 0 \; , \qquad B_i \in (-\infty, \infty) \;  .
\ee
Since the nature of the mutual interactions is characterized by
the sign of the symbiotic coefficient $B_i$, the {\it symbiosis
function}
\be
\label{5}
S_i(\{ N_1,N_2,\ldots \} ) \geq 0 \; ,
\ee
describing the type of symbiotic relations, can be taken
non-negative. 

Accepting that the symbiotic functions depend on the species populations,
we exemplify this in what follows by the simplest form of such a 
dependence, assuming that the effective carrying capacity is a linear 
combination of the natural carrying capacity, provided by nature, and by 
the carrying capacity produced (or destroyed) by the mutual species 
activity. Generally, we assume that the symbiotic functions are analytical
functions, hence, they can be expanded in power series over the species 
populations. We shall consider the following particular cases of such 
expansions, corresponding to different types of mutual influence of two 
symbiotic species. When the carrying capacity of an $i$-species is
influenced by the mutual interactions with a $j$-species, the effective 
carrying capacity is represented as
$$
K_i = A_i + B_i N_i N_j \; .
$$
And if the carrying capacity of the $i$-species is influenced by the 
$j$-species without direct interactions, as it happens in the case of 
commensalism, then the effective carrying capacity is given by the form
$$
K_i = A_i + B_i N_j \; .
$$
The natural carrying capacity $A_i$ is supposed to be nonzero, which 
implies that the species could exist without their symbionts, though
the existence of the latter can drastically change the species behavior.
In general, there can happen situations when one symbiont is obligatory
for another, so that one of the species cannot survive without their
counterpart. Such a situation would correspond to zero carrying capacity.
But this rare case is not treated here.

The mathematical structure of Eq. (\ref{1}) with
(\ref{3}), describing symbiosis, is principally different from
equations of the Lotka-Volterra type and cannot be reduced to
the latter, as we shall show in the Discussion section. 

As we mentioned in the Introduction, in biology and ecology,
three main categories of symbiotic relations can be distinguished
(mutualism, parasitism, and commensalism), depending on whether
the influence of one species on another is positive, negative,
or neutral \cite{1,2,3,4,5}. This classification is straightforwardly
linked to our model. Below, we explain this for the case of two
species, when $i = 1, 2$, which will be treated in what follows.

{\it Mutualism} implies the relations in which both species extract some
benefit from their relationship. In our model, this is equivalent to
$$
   B_1 > 0\; , \qquad B_2 > 0   \qquad  (mutualism) \; .
$$

{\it Parasitism} means that one of the species can benefit, while the
other is harmed in the process. Generally, one can say that
parasitism is the relation in which at least one of the species is
harmed. This means that one of the following inequalities is valid:
either
$$
B_1 > 0 \; , \qquad  B_2 < 0 \; ,
$$
or
$$
B_1 < 0 \; , \qquad B_2 > 0  \; ,
$$
or
$$
B_1 < 0 \; , \qquad  B_2 < 0 \qquad  (parasitism) \; .
$$

{\it Commensalism} is the relation in which one of the species
benefits, while another is not affected. Hence, one of the
following inequalities is satisfied: either
$$
B_1 > 0 \; , \qquad  B_2 = 0  \; ,
$$
or
$$
B_1 = 0 \; , \qquad  B_2 > 0 \qquad   (commensalism) \; .
$$
In addition to this classification, it is possible to distinguish
different kinds of mutual interactions embodied in the form of the
symbiosis function $S_i(\{ N_i \} )$ obeying inequality (\ref{5}),
which will be specified later.

Before specifying this function $S_i(\{ N_i \} )$, it is convenient to
introduce dimensionless units. For this purpose, we may choose two
characteristic scales, $N_{eff}$ and $Z_{eff}$, to serve as the measuring
units for the species variables $N_1$ and $N_2$, respectively. We thus
define the dimensionless species characteristics
\be
\label{6}
x \equiv \frac{N_1}{N_{eff}} \; , \qquad z \equiv
\frac{N_2}{Z_{eff}}
\ee
and the dimensionless carrying capacities
\be
\label{7}
y_1 \equiv \frac{\gm_1 K_1}{C_1 N_{eff}} \; ,
\qquad  y_2 \equiv \frac{\gm_1 K_2}{C_2 Z_{eff}}\; .
\ee
Let us measure time in units of $1/ \gamma_1$ and introduce the ratio
\be
\label{8}
\al \equiv \frac{\gm_2}{\gm_1} \; .
\ee
If the life times of the co-existing species were essentially different,
then their mutual influence would be rather limited, inducing just local
in time perturbations. The most interesting case is when the species
co-exist in the long time scale comparable to their life times. This
implies that the most important case is when the ratio (\ref{8}) is
close to one, which we shall take into account in what follows.

Then, for two symbiotic species, Eqs. (\ref{1}) take the form
\be
\label{9}
\frac{dx}{dt} = x \; - \; \frac{x^2}{y_1} \; , \qquad
\frac{dz}{dt} = z \; - \; \frac{z^2}{y_2} \; .
\ee
These equations, controlling the time evolution of the functions
$x=x(t)$ and $z = z(t)$, must be complemented by the initial conditions
\be
\label{10}
x(0) = x_0 \; , \qquad z(0) = z_0 \;  .
\ee
By definition, the quantities $N_i$, measuring the amount
of the corresponding species, are positive. Hence, we are interested only
in the non-negative solutions of Eqs. (\ref{9}):
\be
\label{11}
x(t) \geq 0 \; , \qquad z(t) \geq 0 \;  .
\ee

We then need to specify the forms of the symbiosis functions
$S_i(\{ N_i \})$ obeying inequality (\ref{5}) and, respectively,
the forms of the carrying capacities (\ref{3}). The forms of functions
(\ref{5}) can vary, depending on the kind of symbiotic relations between
the considered species.

At the end of this section, formulating the general idea of our approach,   
it is worth stressing again the meaning of the chosen type of the equations. 
We start with Eqs. (\ref{1}) having the form of logistic equations, with 
all terms enjoying the known straightforward underpinning. The quantity 
$K_i$ here is the carrying capacity of the $i-$th species. There can be 
numerous interpretations of what in particular cases could be this 
capacity, depending on whether biological, or social, or financial systems 
are considered. For concreteness, let us talk about food. So, the carrying
capacity is the amount of food available for the species.

Our basic point is that the carrying capacity is not a fixed number,
characterizing the amount of given food, but it consists of two parts, 
as in Eq. (\ref{3}). The first part is the naturally provided amount of 
food, while the second part is the additional food produced during the 
interaction of symbiotic species. It is possible to give an infinite number 
of examples illustrating this. The most classical example is the symbiosis 
of trees with mushrooms, or, more generally, plants with fungi. In such
interactions, as is well known, the elements are produced that serve as 
additional food for both symbiotic species. That is, there appears 
additional food created by the collaboration of the species.

We consider the simple case where more agents in species lead to the 
increase of food production during their interactions. In a reduced 
formalism at the macro-level, the simplest and most robust way to account 
for this effect is to take the symbiotically produced amount of food to 
be proportional to the number of agents in each of the species. This is 
equivalent to writing the additionally produced food for the $1-$th 
species as $B_1 N_1 N_2$, where $N_i$ is the number of agents in the 
$i-$th species group. Similarly, for the second species the additional 
food is $B_2 N_2 N_1$. From this definition, it immediately follows that 
the symbiotic coefficient $B_i$ is the amount of food produced in the 
process of symbiotic relations by the collaboration of an agent from the 
first type of species with an agent from the second species type. 

The idea that the amount of food can be created in the process of symbiotic 
relations, or destroyed, when these relations are parasitic, is the pivotal 
new point we have advanced. This is exactly what happens in symbiotic relations. 
And this provides a novel mathematical approach to describing symbiotic 
relations. As we demonstrate in the following sections, our model allows for 
the description of all types of symbiosis.

\section{Classification of forms of symbiotic relations}

There can exist three main types of symbiotic relations.

\vskip 3mm
\subsection{Influencing livelihood through mutual interactions}

\vskip 2mm
A common case is when the symbiotic species influence the livelihood
of each other by means of mutual interactions. Known biological
examples are plant trees and mushrooms and many other plants and
fungi \cite{1,2,3,4,5}. In the social sciences, this could be the
relationship between the economic level of a country and the
intellectual level of society. Or this can be the relation between
two firms, producing different kinds of goods in close collaboration
with each other. Another example is the relation between basic and
applied research \cite{7,8,9}. The simplest form of the carrying
capacities, representing livelihoods that are influenced by mutual
interactions, can be written as
\be
\label{12}
K_1 = A_1 + B_1N_1 N_2 \; , \qquad
K_2 = A_2 + B_2N_2 N_1 \; .
\ee
The first terms $A_1$ and $A_2$ define the given carrying capacities,
which in dimensionless units are
\be
\label{13}
a_1 \equiv \frac{\gm_1 A_1}{C_1 N_{eff}} \; , \qquad
a_2 \equiv \frac{\gm_1 A_2}{C_2 Z_{eff}} \; .
\ee
The second terms characterize the produced carrying capacities,
corresponding to the mutual influence of symbiotic species on the
livelihood of each other. The dimensionless symbiotic
coefficients are
\be
\label{14}
b \equiv \frac{\gm_1 B_1Z_{eff}}{C_1} \; , \qquad
g \equiv \frac{\gm_1 B_2N_{eff}}{C_2} \; .
\ee
The dimensionless carrying capacities (\ref{7}) become
\be
\label{15}
y_1 = a_1 + b xz \; , \qquad y_2 = a_2 + g xz \; .
\ee
The given carrying capacities are taken positive,
\be
\label{16}
a_1 > 0 \; , \qquad a_2 > 0 \; ,
\ee
which means that each of the two species can survive in the absence
of the other one. The interaction coefficients $b$ and $g$ can be
of any sign, depending on whether the relations are beneficial or
parasitic.

The scaling units $N_{eff}$ and $Z_{eff}$ can be chosen arbitrarily.
It is convenient to choose them as
\be
\label{17}
N_{eff} = \frac{\gm_1 A_1}{C_1} \; , \qquad
Z_{eff} = \frac{\gm_1 A_2}{C_2} \; .
\ee
Then one has
\be
\label{18}
a_1 = a_2 = 1 \; ,
\ee
and the symbiotic coefficients are
\be
\label{19}
b \equiv \frac{\gm_1^2 B_1 A_2}{C_1C_2} \; , \qquad
g \equiv \frac{\gm_1^2 A_1 B_2}{C_1C_2} \; .
\ee
Therefore, the carrying capacities (\ref{15}) become
\be
\label{20}
y_1 = 1 + b xz \; , \qquad y_2 = 1 + g xz \; .
\ee

\vskip 3mm
\subsection{Influencing livelihood through asymmetric interactions}

\vskip 2mm
Another possibility is when the interactions of symbiotic species
are not symmetric, such that the carrying capacity of one of them
is influenced by mutual interactions, while the carrying capacity
of another is influenced solely by the other species, not involving
their mutual interactions. This type of symbiosis is common to
many biological as well as social systems, when one of them is a
subsystem of a larger one. For example, the relation between the
economic level of a country and the development level of a particular
branch of economics, say between the country gross domestic product
and the level of science; or the relation between culture and language;
or the relation between the size of social groups and the size of
brain \cite{32,33}. In that asymmetric case, the carrying capacities
are
\be
\label{21}
K_1 = A_1 + B_1 N_1 N_2 \; , \qquad
K_2 = A_2 + B_2 N_1 \; .
\ee
In dimensionless units, $a_1$ and $a_2$ have the same form as in
Eq. (\ref{13}), while the symbiotic coefficients are
\be
\label{22}
b \equiv \frac{\gm_1 B_1Z_{eff}}{C_1} \; , \qquad
g \equiv \frac{\gm_1 B_2N_{eff}}{Z_{eff}} \; .
\ee
The carrying capacities (\ref{7}) are given by the expressions
\be
\label{23}
y_1 = a_1 + b xz \; , \qquad y_2 = a_2 + g x \; .
\ee

Opting for the scaling units (\ref{17}) yields normalization
(\ref{18}), with the symbiotic coefficients
\be
\label{24}
b \equiv \frac{\gm_1^2 B_1 A_2}{C_1C_2} \; , \qquad
g \equiv \frac{\gm_1^2 A_1 B_2}{C_1A_2} \; .
\ee
Thus, the dimensionless carrying capacities (\ref{23}) become
\be
\label{25}
y_1 = 1 + b xz \; , \qquad y_2 = 1 + g x \; .
\ee

\vskip 3mm
\subsection{Influencing livelihood without direct interactions}

\vskip 2mm
This is, probably, the most common of biological and ecological
symbiosis. Numerous examples can be found in the books
\cite{1,2,3,4,5}. In this case, the livelihood of each species
is influenced by the presence of another species without involving
their direct interactions. The carrying capacities are correspondingly
given by
\be
\label{26}
K_1 = A_1 + B_1 N_2 \; , \qquad
K_2 = A_2 + B_2 N_1 \; .
\ee
The terms $A_1$ and $A_2$ are the a priori given carrying
capacities, in the absence of the other species. Their
dimensionless forms are the same as in Eq. (\ref{13}), but
the symbiotic coefficients are
\be
\label{27}
b \equiv \frac{\gm_1 B_1Z_{eff}}{C_1N_{eff}} \; , \qquad
g \equiv \frac{\gm_1 B_2N_{eff}}{C_2Z_{eff}} \; .
\ee
The dimensionless carrying capacities (\ref{7}) take the forms
\be
\label{28}
y_1 = a_1 + b z \; , \qquad y_2 = a_2 + g x \; .
\ee
Choosing again the same scaling units (\ref{17}) gives the same
normalization (\ref{18}), but with the symbiotic coefficients
\be
\label{29}
b \equiv \frac{\gm_1 B_1 A_2}{A_1C_2} \; , \qquad
g \equiv \frac{\gm_1 A_1 B_2}{C_1A_2} \; .
\ee
This results in the dimensionless carrying capacities
\be
\label{30}
y_1 = 1 + b z \; , \qquad y_2 = 1 + g x \; .
\ee

Note that here, as well as in the previous cases, the symbiotic
coefficients $b$ and $g$ can take different values, depending on
the strength of the mutual influence between the symbiotic species.
And these coefficients can be of different signs, describing either
beneficial or parasitic relations. In the following sections, we
give a detailed analysis of each type of symbiosis classified above.

\section{Symbiosis with mutual interactions}

We study the equations (\ref{9}), with the carrying capacities
(\ref{20}), leading to the coupled system of two ordinary
differential equations:
\be
\label{31}
\frac{dx}{dt} = x \; - \; \frac{x^2}{1 + b x z} \; , \qquad
\frac{dz}{dt} = z \; - \; \frac{z^2}{1 + g x z} \;\; .
\ee
This system of equations is symmetric with respect to the change
$x\ra z$ and $b\ra g$. Three kinds of qualitatively different
solutions are found for these equations, which we describe in turn.

\vskip 3mm
\subsection{Convergence to Stationary States}

\vskip 2mm
Formally, there are five fixed points for Eqs. (\ref{31}).
However, as expressed by Eq. (\ref{11}), we are looking for
non-negative solutions. For each of the non-negative solutions,
we accomplish the Lyapunov stability analysis and select the
stable solutions. We assume that the reader is sufficiently qualified
in the Lyapunov technique, so that we do not discuss the details of
the stability analysis and present only the final results of the
analysis for the stable stationary solutions that are non-negative.

Equations (\ref{31}) can possess a single stable fixed point given
by the expressions
\be
\label{32}
x^* = \frac{1-b+g-\sqrt{(1+b-g)^2-4b}}{2g} \; , \qquad
z^* = \frac{1+b-g-\sqrt{(1+b-g)^2-4b}}{2b} \;  .
\ee
This point is stable in the {\it stability region} shown in Fig. 1
characterized by one of the conditions, either
\be
\label{33}
b< 0 \; , \qquad - \infty < g < \infty \; ,
\ee
or
\be
\label{34}
0 \leq b < 1 \; , \qquad g \leq g_c \equiv
\left ( \sqrt{b}-1\right )^2 \leq 1 \; ,
\ee
or
\be
\label{35}
b \geq 1 \; , \qquad g \leq 0 \; .
\ee
On the boundaries of the stability region, we have
$$
x^* = 1 \; , \qquad z^* = \frac{1}{1-g} \qquad
( b=0, \; g < 1) \; ,
$$
$$
x^* = \frac{1}{1-b} \; , \qquad z^* = 1 \qquad
( b < 1, \; g=0) \; ,
$$
\be
\label{36}
x^* = \frac{1}{1-\sqrt{b}}\; , \qquad
z^* = \frac{1}{\sqrt{b}}
\qquad (0 < b < 1, \; g = g_c)  \; .
\ee

The solution to Eqs. (\ref{31}) tends to the stationary point
(\ref{32}), provided that the symbiosis parameters $b$ and $g$
are in the stability region defined above, and when the initial
conditions are in the attraction basin of this fixed point. For
positive $b$ and $g$, the attraction basin can be found only
numerically while, if at least one of these parameters is negative,
the attraction basin is characterized by one of the conditions,
when either
\be
\label{37}
x_0z_0 \; < \; \frac{1}{|b|}  \qquad
( b < 0 , \; g > 0 )\; ,
\ee
or when
\be
\label{38}
x_0z_0 \; < \; \frac{1}{|g|}  \qquad
( b > 0 , \; g < 0 )\; ,
\ee
or when
\be
\label{39}
x_0 z_0 < {\rm min} \left \{ \frac{1}{|b|} \; , \;
\frac{1}{|g|} \right \} \qquad ( b < 0 , \; g < 0) \; .
\ee

The approach to the stationary solution (\ref{32}) can be either
monotonic or non-monotonic, from above or from below, depending
on the parameters $b$ and $g$ and on the initial conditions $x_0$
and $z_0$. In Figs. 2, 3, and 4, the typical behavior of the
solutions is shown for different parameters and initial conditions:
when $x_0>x^*$ and $z_0>z^*$ (Fig. 2), when $x_0> x^*$ but $z_0<z^*$
(Fig. 3), and when $x_0<x^*$ with $z_0<z^*$ (Fig. 4). The case
when $x_0<x^*$ but $z_0>z^*$ is similar to the case when $x_0>x^*$
but $z_0< z^*$, with changing $x$ by $z$ and $b$ by $g$.

It is instructive to compare the behavior of the symbiotic
solutions, described by the coupled Eqs. (\ref{31}), with that
of the solutions of the decoupled equations ($b=g=0$)
\be
\frac{dx}{dt} = x - x^2 \; , \qquad
\frac{dz}{dt} = z - z^2  \qquad (b = g = 0) \; .
\label{thyhbwrq}
\ee
The solutions for these non-symbiotic species tend to the stable
fixed point $x^*=z^*=1$. The comparison, explicitly showing the
role of symbiosis, is demonstrated in Figs. 5 and 6. These two
figures illustrate the general property that beneficial (respectively,
parasitic) symbiosis leads to the increase (respectively, decrease)
of the stationary solutions. Another important message is that
stationary states can exist even in the presence of parasites.

\vskip 3mm
\subsection{Unbounded Exponential Growth}

\vskip 2mm
The solutions to the symbiotic equations (\ref{31}) grow to infinity
with increasing time $t\ra\infty$, when there are no stable fixed
points, that is, when either
\be
\label{40}
0 < b < 1 \; , \qquad g > g_c \; ,
\ee
or when
\be
\label{41}
b > 1 \; , \qquad g > 0 \; .
\ee
A similar exponential divergence at infinity occurs when the
stable fixed points exist, but the initial conditions are taken
outside of the attraction basin, so that either
\be
\label{42}
x_0 z_0 > \frac{1}{|b|} \qquad ( b < 0 , \; g > 0) \;  ,
\ee
or when
\be
\label{43}
x_0 z_0 > \frac{1}{|g|} \qquad ( b > 0 , \; g < 0) \;  ,
\ee
or when
\be
\label{44}
x_0 z_0 > {\rm max} \left \{ \frac{1}{|b|} \; , \;
\frac{1}{|g|} \right \} \qquad ( b < 0 , \; g < 0) \; .
\ee
The typical behavior of such increasing solutions is shown in Fig. 7.
When the initial conditions are outside of the attraction basin, this
means that at least one of them is sufficiently large, that is, the
initial population is so large that even the existence of parasites
cannot suppress the development of this species.

\vskip 3mm
\subsection{Finite-Time Death and Singularity}

\vskip 2mm
A specific behavior occurs under parasitic symbiotic relations, when
stable fixed points can exist, but the initial conditions are taken
outside of the attraction basin, so that
\be
\label{45}
\frac{1}{|b|} < x_0 z_0 < \frac{1}{|g|} \qquad
( b  < g < 0 )\; .
\ee
Then, at the {\it critical time} $t_c$, defined as the solution of
the equation
\be
\label{46}
x(t_c) z(t_c) = \frac{1}{|b|} \; ,
\ee
the first variable $x(t)$ sharply drops to zero, while the second
one rises to infinity:
$$
x(t) \ra 0 \; , \qquad \dot{x}(t) \ra -\infty \qquad
(t \ra t_c) \; ,
$$
\be
\label{47}
z(t) \ra +\infty \; , \qquad \dot{z}(t) \ra +\infty \qquad
(t \ra t_c) \; .
\ee
Here the overdot means time derivative.

The behavior is inverted for the case when
\be
\label{48}
\frac{1}{|g|} < x_0 z_0 < \frac{1}{|b|} \qquad
( g  < b < 0 )\; ,
\ee
for which the critical time is given as the solution of the equation
\be
\label{49}
x(t_c) z(t_c) = \frac{1}{|g|} \; .
\ee
At this time, the first solution rises to infinity, while the
second one quickly dies:
$$
x(t) \ra +\infty \; , \qquad \dot{x}(t) \ra +\infty \qquad
(t \ra t_c) \; ,
$$
\be
\label{50}
z(t) \ra 0 \; , \qquad \dot{z}(t) \ra -\infty \qquad
(t \ra t_c) \; .
\ee

The situation corresponding to the case (\ref{45}) is illustrated in
Fig. 8. The meaning of this phenomenon, when one of the species dies,
while the other rises, is easily understandable. Under the mutually
parasitic relations, each species destroys the livelihood of the
other one. That species, whose livelihood is destroyed faster, dies
out, while the other species grows at the expense of the former one.

\section{Symbiosis with asymmetric interactions}

If the symbiotic relations between species are characterized by
asymmetric interactions with the carrying capacities (\ref{25}),
then Eqs. (\ref{9}) take the form
\be
\label{51}
\frac{dx}{dt} = x \; - \; \frac{x^2}{1+ b xz} \; , \qquad
\frac{dz}{dt} = z \; - \; \frac{z^2}{1+ g x}  \; .
\ee
The following qualitatively different dynamic regimes can occur.

\vskip 3mm
\subsection{Convergence to stationary states}

\vskip 2mm
There can exist just one stable fixed point, given by
\be
\label{52}
x^* = \frac{1-b-\sqrt{(1-b)^2-4bg}}{2bg} \; , \qquad
z^* = \frac{1+b-\sqrt{(1-b)^2-4bg}}{2b} \; .
\ee
Let us introduce the critical parameter value
\be
\label{53}
g_c \equiv \frac{(1-b)^2}{4b} \;  .
\ee
The fixed point (\ref{52}) is stable and non-negative if either
\be
\label{54}
b \leq -1 \; , \qquad g \geq g_c \; ,
\ee
or if
\be
\label{55}
-1 < b \leq 0 \; , \qquad g > -1 \; ,
\ee
or if
\be
\label{56}
0 < b < 1 \; , \qquad - 1 \leq g \leq g_c \; ,
\ee
or when
\be
\label{57}
b \geq 1 \; , \qquad - 1 < g < 0 \; .
\ee
The overall region of stability is depicted in Fig. 9.

The marginal values of the fixed point, occurring at the boundary of
the stability region, are
$$
x^* = 1 \; , \qquad z^* = 1 + g \qquad ( b=0, \; g > -1) \; ,
$$
\be
\label{58}
x^* = \frac{1}{1-b} \; , \qquad
z^* = 1 \qquad ( b < 1, \; g=0) \; ,
\ee
if $g \neq g_c$, and
\be
\label{59}
x^* = \frac{2}{1-b} \; , \qquad z^* = \frac{1+b}{2b}
\qquad (g=g_c) \; ,
\ee
if $g = g_c$ and either $ b < -1$ or $0 < b < 1$.

The solutions tend to the fixed point, provided that the related
initial conditions are in the basin of attraction. When one of the
symbiotic coefficients is negative, the basin of attraction is
defined so that either
\be
\label{60}
x_0 z_0 < \frac{1}{|b|} \qquad (b <0, \; g>0)  \; ,
\ee
or if
\be
\label{61}
x_0 < \frac{1}{|g|} \qquad (b > 0, \; g < 0) \;  ,
\ee
or when
\be
\label{62}
x_0 < \frac{1}{|g|} \; , \qquad x_0 z_0 < \frac{1}{|b|}
\qquad ( b < 0, \; g < 0 ) \;  .
\ee
The convergence to the stationary state can be monotonic or
non-monotonic, as illustrated in Figs. 10 and 11 for various
initial conditions and different symbiotic coefficients. The
convergence of solutions to the stable fixed point is qualitatively
similar to that documented in the case of symmetric interactions.

\vskip 3mm
\subsection{Exponential growth at infinity}

\vskip 2mm
Another regime, which is similar to that found for symmetric
interactions, is the exponential growth at increasing time
$t\ra\infty$. This happens when there are no stable fixed
points and either
\be
0 < b < 1 \; , \qquad g > g_c \; ,
\ee
or when
\be
\label{63}
b > 1 \; , \qquad g > 0 \; ,
\ee
with the critical parameter $g_c$ given in Eq. (\ref{53}). The
same kind of exponential growth arises when there exist stable
fixed points, but the initial conditions are outside of their
basin of attraction, which occurs for
\be
\label{64}
0 < b < 1 \; , \qquad 0 < g < g_c \; ,
\ee
or when
\be
\label{65}
x_0 z_0 > \frac{1}{|b|} \qquad ( b < 0, \; g > 0) \; .
\ee
The temporal behavior is qualitatively similar to that of Fig. 7.

\vskip 3mm
\subsection{Finite-time singularity}

\vskip 2mm
A different regime occurs when the parameter $g$ is negative,
and no stable fixed point exists, which occurs when
\be
\label{66}
0 < b < \infty \; , \qquad g < -1 \; ,
\ee
or if
\be
-1 < b < 0 \; , \qquad g < -1 \; ,
\ee
or if
\be
\label{67}
b \leq -1 \; , \qquad g < g_c < -1 \; .
\ee
When the initial conditions obey the inequalities
\be
\label{68}
x_0 > \frac{1}{|g|} \; , \qquad x_0z_0 > -\; \frac{1}{b}
\qquad (b>0, \; g < 0) \; ,
\ee
or the inequalities
\be
\label{69}
x_0 > \frac{1}{|g|} \; , \qquad  x_0 z_0 > \frac{1}{|b|}
\qquad (b < 0 , \; g < 0) \; ,
\ee
then the solutions live only a finite life until the critical
time $t_c$. The value of this critical point can only be determined
numerically. As time approaches $t_c$, the first variable remains
finite, while the second one diverges:
$$
x(t) \ra x(t_c) > 0 \; , \qquad \dot{x}(t) \ra \dot{x}(t_c)
\qquad (t\ra t_c) \; ,
$$
\be
\label{70}
z(t) \ra \infty \; , \qquad \dot{z}(t) \ra \infty
\qquad (t\ra t_c)  \; .
\ee
Here $x(t_c)$ and $\dot{x}(t_c)$ imply finite values of the corresponding
functions at the critical time $t_c$.

The main difference with the case of Eq. (\ref{47}) is the
finiteness of the solution $x(t_c)$ at the critical time. The
typical behavior of the two variables in this regime is shown
in Fig. 12. The explosive divergence of one of the species
announces a change of regime: beyond $t_c$, the system of two
species will change their characteristics and, possibly, the
nature of their interactions.

The same kind of behavior also happens when the stable fixed
points exist for $b$ and negative $g$ in the region
$$
b > -1 \; , \qquad -1 < g < 0 \; ,
$$
or if
$$
b < -1 \; , \qquad g_c < g < 0 \; ,
$$
but initial conditions are taken outside of the basin of attraction,
so that $x_0,z_0$ are defined by Eqs. (\ref{68}) or (\ref{69}),
correspondingly.

\vskip 3mm
\subsection{Finite-time death}

\vskip 2mm
An interesting behavior appears for some parameters, in
particular, when the parameter $g$ is negative, meaning parasitism,
so that no stable fixed points exist or when fixed points exist but
the initial conditions are outside their domain of attraction.
For $g<0$, when the symbiosis coefficients are taken outside
of the stability region presented in Fig. 9, and initial
conditions obey the following inequalities, either
\be
\label{71}
x_0 < \frac{1}{|g|} \; , \qquad x_0 z_0 > -\; \frac{1}{b}
\qquad ( b > 0 , \; g < 0) \; ,
\ee
or if
\be
\label{72a}
x_0 < \frac{1}{|g|} \; , \qquad x_0 z_0 < \frac{1}{|b|}
\qquad ( b < 0, \; g< 0 ) \; ,
\ee
or if
\be
\label{72b}
x_0 < \frac{1}{|g|} \; , \qquad x_0 z_0 > \frac{1}{|b|}
\qquad ( b < 0, \; g< 0 ) \; ,
\ee
or when
\be
\label{72c}
x_0 > \frac{1}{|g|} \; , \qquad x_0 z_0 < \frac{1}{|b|}
\qquad ( b < 0, \; g< 0 ) \; ,
\ee
then the solutions have a finite life and one
of the species dies or exhibits a gradient
catastrophe at the {\it death time} $t_d$ (see below) found as
the solution to the equation
\be
\label{73}
|g|x(t_d) = 1 \; ,
\ee
at which either $\dot{z}(t)=-\infty$ or $\dot{x}(t)=-\infty$ for
$t\ra t_d$.

When the parameters are taken outside of the stability region
and the initial conditions obey inequalities (\ref{71}) or
(\ref{72a}), then
$$
x(t) \ra \frac{1}{|g|} \; , \qquad \dot{x}(t) \ra \dot{x}(t_d)
\qquad (t\ra t_d) \; ,
$$
\be
\label{74a}
z(t) \ra z(t_d) \; , \qquad \dot{z}(t) \ra -\infty
\qquad (t\ra t_d) \; .
\ee

When the initial conditions obey inequality (\ref{72b}), then
$$
x(t) \ra \frac{1}{|g|} \; , \qquad \dot{x}(t) \ra +\infty
\qquad ( t \ra t_d) \; ,
$$
\be
\label{74b}
z(t) \ra \frac{|g|}{|b|} \; , \qquad \dot{z}(t) \ra -\infty
\qquad ( t \ra t_d)\; ,
\ee
while, when the initial conditions satisfy inequality (\ref{72c}),
then
$$
x(t) \ra \frac{1}{|g|} \; , \qquad \dot{x}(t) \ra -\infty
\qquad ( t \ra t_d) \; ,
$$
\be
\label{74c}
z(t) \ra \frac{|g|}{|b|} \; , \qquad \dot{z}(t) \ra +\infty
\qquad ( t \ra t_d) \; .
\ee

In the cases (\ref{71}) and (\ref{72a}), the solutions (\ref{74a})
remain non-zero at $t\ra t_d$, but the time derivative of $z(t)$
tends to $-\infty$. Such a behavior corresponds to the so-called
{\it gradient catastrophe}. The value $z(t_d)$ is close to zero,
as shown in Fig. 13. This kind of behavior can be interpreted
as the extinction of the species described by the variable $z(t)$.
Therefore, after $t_d$, we can set
\be
\label{75}
z(t) \equiv 0 \qquad ( t\geq t_d) \; .
\ee
Then the continuation of the evolution of the population $x(t)$
after $t_d$ is characterized by the single equation $\dot{x}=x-x^2$.
The death of the species $z(t)$ is due to the large negative (parasitic)
symbiosis coefficient $g$, which leads the species $x$ to suppress
the population $z$ of the other species, until its complete demise
occurring at the time $t_d$.

A behavior similar to that, described by Eqs. (\ref{74b}) and
(\ref{74c}), occurs for $g<0$, when the symbiosis coefficients are
in the stability region, but the initial conditions obey inequalities
(\ref{72b}) or (\ref{72c}).

When $\{x_0,z_0\}$ obey Eq. (\ref{72b}), then again there is a
death time $t_d$, given by the same Eq. (\ref{73}). Here $z(t_d)$
is not necessarily close to zero, but it experiences the same
gradient catastrophe as in Eq. (\ref{74a}). Contrary to the case
of Eq. (\ref{74a}), the derivative $\dot{x}(t)$ tends to
$+\infty$. Strictly speaking, the coupled system of Eqs. (\ref{51})
has no solution after the time $t_d$. However, we can again interpret
the gradient catastrophe, with $\dot{z}(t)\ra-\infty$ as the death
of the species $z(t)$, extending the solution $x$ after $t_d$ using
condition (\ref{75}), as is shown in Fig. 14.

In the case when the initial conditions obey Eq. (\ref{72c}),
the evolution also ends at the death time $t_d$, defined by the same
Eq. (\ref{73}). At this point, conditions (\ref{74c}) hold true.
The difference with the case of Eqs. (\ref{74b}) is that here the
gradient catastrophe happens with $\dot{x}(t)\ra -\infty$ while
$\dot{z}(t) \ra +\infty$ for $t\ra t_d$. Hence, interpreting $t_d$
as the time of extinction of the species $x$ that are killed by
the species $z$, the subsequent evolution of the dynamics after
$t_d$ is obtained by setting
\be
\label{80}
x(t) \equiv 0 \qquad (t > t_d) \; .
\ee
The overall behavior is presented in Fig. 15. Comparing Figs. 14
and 15, we see that the dynamics essentially depends on the initial
conditions.

\section{Symbiosis without direct interactions}

When symbiosis is characterized by the influence of species on the
livelihood of each other without direct mutual interactions, the
corresponding carrying capacities are given by Eqs. (\ref{30}).
Then, the symbiotic Eqs. (\ref{9}) take the form
\be
\label{81}
\frac{dx}{dt} = x\; - \; \frac{x^2}{1+bz} \; , \qquad
\frac{dz}{dt} = z \; - \; \frac{z^2}{1+gx} \; .
\ee
These equations are symmetric with respect to the replacement
$b\ra g$ and $x \ra z$. The different possible regimes exhibited
by Eqs. (\ref{81}) are described as follows.

\vskip 3mm
\subsection{Convergence to stationary states}

\vskip 2mm
There can exist the single stable fixed point
\be
\label{82}
x^* = \frac{1+b}{1-bg} \; , \qquad
z^* = \frac{1+g}{1-bg} \; ,
\ee
which is positive and stable if either
\be
\label{83}
-1 \leq b < 0 \; , \qquad g \geq -1
\ee
or if
\be
\label{84}
b \geq 0 \; , \qquad 0 \leq g < g_c \equiv
\frac{1}{b} \; .
\ee
The corresponding stability region is shown in Fig. 16. On the
boundary of the stability region, the fixed point degenerates to
one of the values
$$
x^* = 0 \; , \qquad z^* = 1 \qquad (b=-1,\; g >-1) \; ,
$$
$$
x^* = 1 \; , \qquad z^* = 0 \qquad (b > -1,\; g=-1) \; ,
$$
\be
\label{85}
x^* = 1-z^* \; , \qquad 0 \leq z^* \leq 1 \qquad (b=g=-1) \; .
\ee
For $b>0$ and $g>0$, the basin of attraction is the whole region
in the $b-g$ plane, where inequalities (\ref{84}) are valid. When
one of the parameters $b$ or $g$ is negative, or both are negative,
then the attraction basin is defined by one of the following
conditions, either
\be
\label{86}
x_0 < \frac{1}{|g|} \qquad ( b > 0 , \; g < 0 )\; ,
\ee
or by
\be
\label{87}
z_0 < \frac{1}{|b|} \qquad ( b < 0 , \; g > 0) \;  ,
\ee
or by
\be
\label{88}
x_0 < \frac{1}{|g|} \; , \qquad z_0 < \frac{1}{|b|} \qquad
( b < 0 \, \; g < 0) \;  .
\ee
The convergence to the stationary solution can be monotonic or
non-monotonic, depending on the symbiosis parameters and initial
conditions, similarly to Figs. 2, 3, 4, and 10, 11.

\vskip 3mm
\subsection{Exponential growth at infinity}

\vskip 2mm
This behavior, which is analogous to that displayed in Fig. 7,
occurs when
\be
\label{89}
b > 0 \; , \qquad g > g_c \equiv \frac{1}{b} \; .
\ee
Then, both solutions $x(t)$ and $z(t)$ exponentially increase as
$t\ra\infty$.

\vskip 3mm
\subsection{Finite-time singularity}

\vskip 2mm
When one or both of the symbiosis parameters are negative, implying
parasitic symbiotic relations, a finite-time singularity can occur.
For the initial conditions
\be
\label{90}
x_0 > \frac{1}{|g|} \; , \qquad z_0 > -\; \frac{1}{b} \qquad
(b > 0, \; g < 0) \;  ,
\ee
there exists a critical time $t_c$, such that
$$
x(t) \ra x(t_c)\; , \qquad \dot{x}(t) \ra \dot{x}(t_c) \qquad
(t\ra t_c) \; ,
$$
\be
\label{91}
z(t) \ra +\infty\; , \qquad \dot{z}(t) \ra +\infty \qquad
(t\ra t_c) \; .
\ee
This behavior is similar to that shown in Fig. 12. Because of the
equation symmetry, there is the opposite case, occurring for
\be
\label{92}
x_0 > -\; \frac{1}{g} \; , \qquad z_0 > \frac{1}{|b|} \qquad
(b < 0, \; g > 0) \; ,
\ee
when
$$
x(t) \ra +\infty\; , \qquad \dot{x}(t) \ra +\infty \qquad
(t\ra t_c) \; ,
$$
\be
\label{93}
z(t) \ra z(t_c)\; , \qquad \dot{z}(t) \ra \dot{z}(t_c) \qquad
(t\ra t_c) \; .
\ee
If both symbiosis parameters are negative, and
\be
\label{94}
x_0 > \frac{1}{|g|} \; , \qquad z_0 > \frac{1}{|b|} \qquad
(b < 0, \; g < 0) \; ,
\ee
there exists a finite-time singularity of the same type as above,
with the particular behavior depending on the parameter values and
initial conditions.

\vskip 3mm
\subsection{Finite-time death}

\vskip 2mm
The occurrence of a gradient catastrophe, of the type described
previously in Sec. 5,  depends on the values of the symbiotic
parameters and initial conditions. If the symbiotic parameters
are outside of the stability region and the initial conditions
are such that
\be
\label{95}
x_0 < \frac{1}{|g|} \; , \qquad z_0 > -\; \frac{1}{b}
\qquad ( b > 0, \; g \leq -1) \; ,
\ee
then there exists a finite time $t_d$, determined as the solution
of the equation
\be
\label{96}
|g|x(t_d) = 1 \;  ,
\ee
where
$$
x(t) \ra \frac{1}{|g|} \; , \qquad \dot{x}(t) \ra \dot{x}(t_d)
\qquad (t \ra t_d) \; ,
$$
\be
\label{97}
z(t) \ra z(t_d) \; , \qquad \dot{z}(t) \ra -\infty \qquad
( t\ra t_d) \;  .
\ee
The gradient catastrophe occurs for the species $z$, whose
gradient tends to $-\infty$. This can be interpreted as an abrupt
collapse of $z$ to $0$ and thus as the extinction of this species.
The continuation of the solution $x$ is then obtained by setting
$z \equiv 0$ for $t > t_d$. The extinction of the species $z$ is
caused by the strong parasitic action of the species $x$ onto
species $z$, represented by the negative value of the symbiotic
parameter $g$.

Under the condition
\be
\label{98}
x_0 > -\; \frac{1}{g} \; , \qquad z_0 < \frac{1}{|b|} \qquad
( b \leq -1, \; g > 0)\; ,
\ee
the role of $x$ and $z$ is interchanged and
the finite-time extinction is observed for the species $x$. At the
finite time $t_d$, given as the solution to the equation
\be
\label{99}
|b|z(t_d) = 1  \; ,
\ee
we have
$$
x(t) \ra x(t_d) \; , \qquad \dot{x}(t) \ra -\infty \qquad
( t\ra t_d)
$$
\be
\label{100}
z(t) \ra \frac{1}{|b|} \; , \qquad \dot{z}(t) \ra \dot{z}(t_d)
\qquad (t \ra t_d)\;  .
\ee
Again, interpreting the gradient catastrophe for $x$ as the
extinction of $x$, the continuation of the solution for $z$ is obtained
by setting $x \equiv 0$ for $t > t_d$. The extinction of the
species $x$ is caused by the strong parasitic action of $z$ on
$x$, due to the negative symbiotic parameter $b$.

The gradient catastrophe can occur even when the symbiotic
parameters are inside the stability region, but with initial
conditions that are outside of the basin of attraction. Thus,
if
\be
\label{101}
x_0 < \frac{1}{|g|} \; , \qquad z_0 > \frac{1}{|b|} \qquad
(b < 0, \; g < 0 )\;  ,
\ee
then there exists a finite time $t_d$, given as the solution of
the equations
\be
\label{102}
|g|x(t_d) = |b|z(t_d) = 1 \;  ,
\ee
at which
$$
x(t) \ra \frac{1}{|g|} \; , \qquad \dot{x} \ra +\infty \qquad
(t\ra t_d) \; ,
$$
\be
\label{103}
z(t) \ra \frac{1}{|b|} \; , \qquad \dot{z} \ra -\infty \qquad
(t\ra t_d) \; .
\ee
Again, here the gradient catastrophe for $z$ can be understood as
the extinction of $z$. Conversely, when the initial conditions are
such that
\be
\label{104}
x_0 > \frac{1}{|g|} \; , \qquad z_0 < \frac{1}{|b|}
\qquad (b < 0, \; g < 0 ) \; ,
\ee
then there is the finite time $t_d$, given by the same Eqs.
(\ref{102}), when
$$
x(t) \ra \frac{1}{|g|} \; , \qquad \dot{x}(t) \ra -\infty
\qquad (t\ra t_d) \; ,
$$
\be
\label{105}
z(t) \ra \frac{1}{|b|} \; , \qquad \dot{z}(t) \ra +\infty
\qquad (t\ra t_d) \;  .
\ee

This can be interpreted as the extinction of the species $x$,
after which it can be set zero for $t > t_d$. The overall behavior is
presented in Fig. 17.

\section{Discussion}

\subsection{Summary of the main results and outlook}

\vskip 3mm

We have proposed an approach to describe the dynamics
of symbiotic relations between several species. We have argued
that the notion of symbiosis can be applied not merely to
biological systems but can also be generalized to social systems
of different nature.

The principal point of our approach is the description of
symbiosis through the influence of each species on the carrying
capacities of the others. This is in agreement with the common
understanding that symbiotic species act on the livelihood of
each other, either improving it, under beneficial relations,
or destroying it, under parasitic relations.

The general symbiotic model can be represented by several
variants characterizing different types of symbiotic relations.
We have considered three basic types of such relations, symbiosis
with mutual interactions, symbiosis with asymmetric interactions,
and symbiosis without direct interactions. In all cases, we have
provided a complete classification of all admissible dynamical
regimes.

The functional dependence of the carrying capacities on the
species variables $x$ and $z$ constitutes the principal difference
between our model and the logistic equation, where the carrying
capacity is fixed. We have chosen here the simplest functional
form of the carrying capacity characterizing symbiotic interactions.
The carrying capacity has been taken as a combination of terms not
exceeding the bilinear order in terms of the species variables. In
general, it would be possible to describe the carrying capacity as
an expansion in increasing powers of $x$ and $z$. Then, we could
use an effective summation of the expansion, obtaining a more
complicated expression, e.g., in the form of the self-similar
exponentials, as has been done for the model describing the dynamics
of a nonequilibrium financial system \cite{34}. Another modification
could be by including the effects of delays into the carrying capacity,
as has been done for the model of punctuated evolution \cite{35}.

Even in its simplest form, the proposed model of symbiosis turned
out to be sufficiently rich, exhibiting four qualitatively different
regimes: convergence to stationary states, unbounded exponential
growth, finite-time singularity, and finite-time death or extinction
of species.

The suggested model of symbiosis can be applied to a variety of
systems, biological, social, economic, financial, and so on. We
have just mentioned some of the possible applications. The main
goal of the present paper has been to advance a general parsimonious
model of symbiosis and to analyze its main dynamical regimes. Here,
we limit ourselves by the mathematical side of the problem.
Particular applications require separate investigations and will
be studied in future publications.

\vskip 5mm

\subsection{Recipe to avoid finite-time singularities}

\vskip 3mm
We have shown that singularities appear under parasitic relations,
when at least one of the symbiosis coefficients is negative.
The characteristics of real systems, of course, cannot
diverge. Therefore, the occurrence of finite-time singularities
should be understood as the manifestation of an instability
developing in the system, which signals a change of regime into
a new structural phase described by different mechanisms and thus
different equations \cite{36}. In some cases, the singularity can be
avoided by slightly modifying the model, e.g, by taking into account
higher-order powers of the variables $x$ and $z$. Such higher orders
can remove the finite-time singularity \cite{34}. In any case, when
a finite-time singularity does happen, this can be understood as the
signal that the system behavior is drastically changing at this
point, somewhat similar to the changes occurring under phase
transitions \cite{37,38}.

In order to be more specific, let us delineate how the model can be
modified in order to avoid the appearance of finite-time singularities,
leading to a more realistic behavior. Recall that finite-time
singularities appear under parasitic relations, when at least one
of the symbiosis coefficients is negative. The divergence occurs when
the effective carrying capacity either becomes zero or is negative
owing to the choice of initial conditions. Hence, mathematically,
to avoid the occurrence of such divergences, it would be sufficient
to have always positive-definite carrying capacities. This also
would be reasonable for the majority of biological systems,
though for economic and financial systems, a transient effective
negative carrying capacity can have sense, representing the leverage
of an economy over-stretching its borrowing level beyond
its capacity for reimbursement, leading to crises and bankruptcies
as illustrated by the sovereign default issues in Greece, Ireland,
and in Europe that were revealed in 2010.

With the goal of removing finite-time singularities, let us consider
the case of Sec. 4, where the finite-time singularity happens at a
critical time, as is described in subsection 4.3, when either the
carrying capacity $y_1$ or $y_2$ becomes zero. We may treat the
carrying capacities $y_1 = 1 + bxz$ and $y_2 = 1 + gxz$ as
the first terms of the general expansions
\be
\label{109}
 y_1 = \sum_{n=0}^\infty a_n(xz)^n \; , \qquad
y_2 = \sum_{n=0}^\infty b_n(xz)^n \;  ,
\ee
with $a_0 = 1$, $a_1 = b$, $b_0 = 1$, and $b_1 = g$. To find effective
limits of such expansions, we can resort to some resummation, or
renormalization, procedure. Probably, the most general procedure of
this kind is based on the self-similar approximation theory
\cite{39,40,41,42,43,44,45,46}. Invoking the variant of this theory,
employing self-similar exponential approximants \cite{47,48} yields
renormalized effective sums that are always sign defined. For instance,
the first-order exponential approximants for sums (\ref{109}) read as
$$
 y_1^* = \exp(bxz) \; , \qquad
y_2^* = \exp (g xz)  .
$$
These forms are evidently positive for any signs of $b$ and $g$, which
makes it straightforward to avoid the finite-time singularities caused
by the occurrence of zero values of $y_i$ at some finite time. The same
method of avoiding the finite-time singularities can be used for other
cases, where such divergences arise. This method, e.g., was employed
in Ref. \cite{34} for removing the finite-time singularities in the
dynamical models of financial markets. The use of the renormalized
expressions for the effective carrying capacities makes the mathematical
treatment more involved and requires separate consideration.

\vskip 5mm
\subsection{Comparison of symbiosis model with predator-prey Lotka-Volterra model}

\vskip 3mm
It is worth emphasizing that the suggested symbiosis
model is basically different from the predator-prey Lotka-Volterra model.
For two species, the general form of the latter can be written as
\be
\label{110}
 \frac{dx_1}{dt} = \gm_1 x_1 - a_1 x_1^2 + b_{12} x_1 x_2 \; , \qquad
\frac{dx_2}{dt} = \gm_2 x_2 - a_2 x_2^2 + b_{21} x_2 x_1 \; ,
\ee
where the coefficients $a_1$ and $a_2$ are positive. One gets the standard
Lotka-Volterra model \cite{13,14} for prey, with the concentration $x_1$,
and predators, with the concentration $x_2$, when $\gamma_1 > 0, a_1 = 0$,
and $b_{12} < 0$, while $\gamma_2 < 0, a_2 = 0$ and $b_{21} > 0$. In the
general case, Eqs. (\ref{110}) correspond to the competitive Lotka-Volterra
model, if the coefficients $b_{12}$ and $b_{21}$ are negative, while if
they are positive, one has the mutualistic Lotka-Volterra model \cite{49,50}.

Let us compare the mathematical structure of these equations (\ref{110}) with
the symbiosis equations studied above, say, with Eqs. (\ref{31}). It is
easy to see that their nonlinearity is of rather different forms, which
makes the behavior of their solutions principally different. One could
ask the question whether there are, maybe, some particular cases when these
equations are close to each other. For example, could it be that Eqs.
(\ref{31}) would reduce to Eqs. (\ref{110}) at small values of $xz$?
Expanding the expressions $1/y_1$ and $1/y_2$ in powers of $xz$, and
retaining the minimal terms containing the symbiosis coefficients, we get
$$
 \frac{dx}{dt} \simeq x - x^2 + bx^3 z \; , \qquad
\frac{dz}{dt} \simeq z - z^2 + g x z^3 \;  .
$$
These equations, clearly, are very different from Eqs. (\ref{110}).

But, maybe, the symbiosis and Lotka-Volterra equations could be equivalent
for the dynamics close to stationary points, when the symbiosis equations
could be reduced to polynomial forms? To check this, let us consider small
deviations from fixed points $x^*$ and $z^*$, defined as
$$
x_1 \equiv x - x^* \; , \qquad x_2 \equiv z - z^* \;  .
$$
To compare with Eqs. (\ref{110}), we need to represent Eqs. (\ref{31})
in the polynomial form of second order with respect to the small deviations
$x_1$ and $x_2$. This results in the following equations
$$
 \frac{dx_1}{dt} = A_{11} x_1 + A_{12} x_2 + B_{11} x_1^2 +B_{12} x_2^2
+ C_{12} x_1 x_2 \; ,
$$
$$
 \frac{dx_2}{dt} = A_{21} x_1 + A_{22} x_2 + B_{21} x_1^2 +B_{22} x_2^2
+ C_{21} x_2 x_1 \; .
$$
Again these equations are principally different from Eqs. (\ref{110}).

Of course, if we would limit ourselves to only small deviations of first
order in Eqs. (\ref{31}) and (\ref{110}), we would come to
linear equations. All linear equations are formally similar to each
other. However, the linearized forms obtained from Eqs. (\ref{31}) and
(\ref{110}) will have absolutely different coefficients and will
correspond to different fixed points, with different stability
properties.

We should also mention the existence of a fundamental difference
at the conceptual level. The Lotka-Volterra equations of the type
(\ref{110}) represent the dynamics of species as resulting from
pairwise interactions, associated with the quadratic and bilinear
terms. In contrast, the symbiosis equation  (\ref{31}) expresses the
interactions through the impact of species concentrations on their
carrying capacities. In principle, modeling explicitly the dynamics
of the carrying capacities as genuine coupled dynamical variables
provides a richer conceptual and technical approach to complex
ecologies. The carrying capacities act as important relevant
dynamical variables.

Therefore we have to conclude that the symbiosis and prey-predator
equations are principally different and cannot be reduced to each
other. They have different mathematical structure and different
solutions. The mathematical difference of the suggested symbiosis
model from the predator-prey models reflects the actual difference
between the processes of symbiosis and predation \cite{1,2,3,4,5,31}.

\vskip 5mm

\subsection{Examples of symbiotic types}

In conclusion, we feel it is useful to return back to concrete examples
of particular symbiotic relations treated in the paper. This would give 
the reader the feeling that the considered mathematical models are 
closely connected to real cases of symbiosis.  

Probably, one of the best known examples of symbiosis is the mutualistic 
interaction between trees and fungi. Trees ability to generate large amounts 
of biomass or store carbon is underpinned by their interactions with soil 
microbes known as mycorrhizal fungi that are a crucial part of all 
forest ecosystems. Mycorrhizas are symbiotic relationships between certain 
fungi and the roots of plants. The fine fungal threads (called hyphae) 
either ensheathe or penetrate the host plant's roots. The fungus helps the 
plant to extract nutrients and water from the soil. Fungi excel at 
procuring necessary, but scarce, nutrients such as phosphate and nitrogen. 
They also protect their hosts against harmful organisms. In return, the 
fungus receives sugars via the plant's photosynthesis. The fungus within 
the root is protected from competition with other soil microbes and gains 
preferential access to carbohydrates within the plant. As with most 
mutualistic relationships, each partner grows better in association with 
the other than it would individually. For instance, birch (Betula spp.) 
has a number of these partnerships, the most familiar being with the red 
and white fly agaric (Amanita muscaria), as well as with the chanterelle 
(Cantharellus cibarius). Scots pine has mycorrhizal associations with over 
200 species of fungi in Scotland, including another kind of chanterelle 
(Cantharellus lutescens). In fact, the majority of plants forests benefit 
from mycorrhizal relationships, and it is thought that mycorrhizas helped 
plants to colonize the land, millions of years ago \cite{Margulis}.

In the frame of our approach, such a symbiotic relation is described by the
carrying capacities (\ref{12}). There, the carrying capacity $A_1$ is the 
amount of natural resources available for trees, and the carrying capacity $A_2$,
the amount of natural resources of fungi. In their mutual interactions, trees 
and fungi produce additional resources that enlarge their carrying capacities
by providing more food for each of them. Thus, $B_1N_1N_2$ and $B_2N_2N_1$ are 
the additional amounts of food for trees and fungi, respectively, produced in
the process of their interaction. 

As is explained in the Introduction, the ideas of biological symbiosis nowadays
are widely used in economic relations, leading to the appearance of such terms
as industrial symbiosis and economic symbiosis. For example, the relations 
between employers and employees can be described in the same way as symbiotic 
relations between some biological species \cite{Weyl}. The approach, known as 
Partner Fidelity Feedback, holds that, similarly to biological species, social
symbionts have evolved to help their hosts because a healthy host automatically 
feeds back benefits to the symbionts. A cheating symbiont would seem to be 
treated like any other environmental setback, such as infertile soil, and a 
mutualistic symbiont elicits the same sort of investments that are triggered 
by the availability of new resources, like a patch of sunlight \cite{Weyl}.

In the frame of our model, the symbiotic relations between employers and 
employees should be characterized by the carrying capacities (\ref{21}). Then,
$A_1$ and $A_2$ are the natural carrying capacities available irrespectively 
of interactions. Employers provide tools to employees for producing additional
capacity $B_1N_1N_2$. As a result of the growing firm or company, the employees
also profit getting the increased capacity $B_2N_1$, say, in the form of 
increased salaries.  

Another example of industrial symbiosis between different enterprises is the 
process of sharing of services, utility, and by-product resources among diverse 
industrial actors in order to add value, reduce costs, and improve the 
environment. Industrial symbiosis is a subset of industrial ecology, with a 
particular focus on material and energy exchange. Industrial ecology is a 
relatively new field that is based on a natural paradigm, claiming that an
industrial ecosystem may behave in a similar way to the natural biological 
ecosystems \cite{Pearce}. A straightforward case is when companies collaborate 
to utilize each other's by-products and otherwise share resources, enlarging 
by this their carrying capacities.

In our approach, this situation corresponds to the carrying capacities (\ref{26}).
Then the capacities $A_1$ and $A_2$ are the resources available to each of the
industrial symbionts. Their additional resources come from the exchange of the
companies by-products, increasing their capacities by the amounts $B_1N_2$ and
$B_2N_1$, respectively. These by-products are produced by each of the companies,
without direct interactions between them. And without this exchange, the 
by-products would be lost, while due to the exchange, they increase the carrying 
capacities of the symbiotic companies.

These examples illustrate that our main idea of modeling symbiotic relations 
by taking into account the mutual influence of symbiotic species on there 
carrying capacities allows us to describe various types of symbiosis
for biological or social symbiotic systems.

\subsection{Symbiosis in microbial systems}

The symbiosis of microbial species is of special interest for several 
reasons: 
\begin{itemize}
\item it is so much widespread in nature, 
\item it can be  observed and studied in the laboratory, which is often impossible for other  wild species, 
\item it can be modeled in artificial synthetic-biology  games \cite{69,70}, and 
\item the specific features of microbial coexistence, such as resource enrichment \cite{71,72}, fluctuation enhancement \cite{73}, 
quorum sensing \cite{74,75}, group selection \cite{76,77}, development of 
cooperation \cite{77,78,79}, and evolutionary race \cite{80,81}, can be 
illuminating for understanding, characterizing, and organizing both 
technological as well as social systems \cite{82}. 
\end{itemize}

Microbial symbiosis has been a survival feature of bacteria since their 
origin. The best example of this is the presence of the energy factories 
known as mitochondria in eukaryotic cells. Mitochondria arose because of the 
symbiosis between an ancient bacterium and a eukaryote. Over evolutionary 
time, the symbiosis became permanent, and the bacterium became part of the 
host. However, even to the present day, the differences in constitution and 
arrangement of the genetic material of mitochondria and the host cell's 
nucleus attest to the symbiotic origin of mitochondria.

There are many well-known examples of bacterial mutualism. The first example 
is the presence of huge numbers of bacteria in the intestinal tract of 
warm-blooded animals such as humans \cite{Host-BacterialMutualism}. About 10 percent of the dry weight of 
a human consists of bacteria. The bacteria act to break down foodstuffs, 
and so directly participate in the digestive process. As well, some of the 
intestinal bacteria produce products that are crucial to the health of the 
host. For example, in humans, some of the gut bacteria manufacture vitamin K, 
vitamin B12, biotin, and riboflavin. These vitamins are important to the host 
but are not produced by the host. The bacteria benefit by inhabiting an 
extremely hospitable environment. The natural activities and numbers of 
the bacteria also serve to protect the host from colonization by 
disease-causing microorganisms and to educate the host immune system. 
The importance of this type of symbiosis is 
exemplified by the adverse health effects to the host that can occur when 
the symbiotic balance is disturbed by antibiotic therapy.

The skin is colonized by a number of different types of bacteria, such as 
those from the genera Staphylococcus and Streptococcus \cite{bacteria-skin}. The bacteria are 
exposed to a ready supply of nutrients, and their colonization of the skin 
helps protect that surface from colonization by less desirable microorganisms.

Bacteria themselves coexist with bacteriophages that are viruses attacking 
bacteria. Therefore, bacteria and virulent phages are often treated as prey 
and predators, respectively. Classical predator-prey systems are modeled by 
the Lotka-Volterra equations, whose solutions display oscillations in the 
populations of the 
competing species \cite{83}. However, the experimental studies of
bacteria-phage biology \cite{84,85} reveal the existence of stationary 
non-oscillating equilibrium states of populations and also the occurrence
of extinction phenomena, when one of the species becomes completely extinct. 
It has therefore been necessary to modify the classical predator-prey 
equations for accounting for the non-oscillatory stable coexistence of 
bacteria and phages \cite{86,87,88,89,90}. 

In this regard, we would like to emphasize that, in our approach, the 
coexisting equilibrium states, corresponding to stable fixed points,
appear naturally for all three models (31), (52), and (84). Under strong 
parasitic relations, dynamics with death occurring in finite time are also present,
characterizing the phenomenon of species extinction, as is shown in 
Figs. 13, 14, 15, and 17. Similar extinction of bacteria $E.coli$, attacked 
by a large population of bacteriophages $T4$, has been observed in 
experiments \cite{85}.

A detailed description of coexisting bacterial hosts and virulent phages 
requires to take into account the renewable environmental resources, 
satiation effects, the host lysis, when each infected host releases many
phages, and also spatial heterogeneity. Including these processes into 
the consideration would need to add several more equations, essentially 
complicating the dynamical system, which is not the aim of our paper.
Our main goal has been to suggest a new mechanism, not treated earlier,
that of the mutual influence of symbionts on the carrying capacities of 
each other and to demonstrate that it is not less, but, probably, even 
more important than other known factors. Taking into account only this 
mutual influence makes it possible to get a rich variety of admissible 
types of the symbiotic behavior, including the stable coexistence of 
species as well as extinction of one of them, caused by parasitic 
relations.

\vskip 1cm
{\bf Acknowledgements}

\vskip 3mm
We acknowledge financial support from the
ETH Competence Center ``Coping with Crises in Complex Socio-Economic
Systems" (CCSS) through ETH Research Grant CH1-01-08-2 and the ETH Zurich
Foundation.

\newpage

\begin{center}
{\large{\bf Figure Captions}}
\end{center}

{\bf Fig. 1}. Region of stability (shaded) in the parameter plane
$b-g$ for the fixed points in the case of symbiosis with mutual
interactions.

\vskip 1cm

{\bf Fig. 2}. Convergence to stationary states of the solutions $x(t)$
(solid line) and $z(t)$ (dashed-dotted line), as functions of time,
in the case of symmetric mutual interactions, when $x_0 > x^*$ and
$z_0 > z^*$, for different parameters $b$ and $g$ and initial
conditions $\{x_0,z_0\}$: (a) $b = -0.1$, $g = 1$, $\{0.75, 3.8\}_1$,
$\{1, 3.8\}_2$, $\{2.5, 3.8\}_3$; the fixed points being $x^*=0.730$,
$z^*=3.702$. (b) $b=0.5$, $g =0.01 < g_c = 0.086$, $\{2.5, 1.5\}_1$,
$\{4, 1.5\}_2$, $\{7, 1.5\}_3$; $x^* = 2.043$, $z^* = 1.021$. (c)
$b = -0.5$, $g = -0.1$, $\{0.8, 1\}_1$, $\{0.8, 1.5\}_2$,
$\{0.8, 2.499\}_3$, $x^* = 0.681$, $z^* = 0.936$. (d) $b = 0.5$,
$g = -0.1 < g_c = 0.086$, $\{3, 1\}_1$, $\{3, 3\}_2$, $x^* = 1.742$,
$z^* = 0.852$.

\vskip 1cm

{\bf Fig. 3}. Convergence to stationary states of the solutions $x(t)$
(solid line) and $z(t)$ (dashed-dotted line), as functions of time,
in the case of symmetric mutual interactions, when $x_0 > x^*$ but
$z_0 < z^*$, for different parameters $b$ and $g$ and initial
conditions $\{x_0,z_0\}$: (a) $b = 0.05$, $g = 0.5 < g_c = 0.603$,
$\{4, 0.05\}_1$, $\{4, 0.5\}_2$, $\{4, 2\}_3$; the fixed points being
$x^*=1.130$, $z^*=2.298$. (b) $b=0.7$, $g = 0.02$, $\{6, 0.05\}_1$,
$\{6, 0.5\}_2$, $\{6, 0.9\}_3$; $x^* = 4.258$, $z^* = 1.093$. (c)
$b = -1$, $g=0.1$, $\{0.45, 0.01\}_1$, $\{1, 0.01\}_2$,
$\{1.5, 0.01\}_3$, $x^*=0.438$, $z^* = 1.281$. (d) $b = -0.01$,
$g=1$, $\{3, 0.05\}_1$, $\{3, 2\}_2$, $\{3, 3.33\}_3$, $x^*=0.730$,
$z^* = 3.702$.

\vskip 1cm

{\bf Fig. 4}. Convergence to stationary states of the solutions $x(t)$
(solid line) and $z(t)$ (dashed-dotted line), as functions of time,
in the case of symmetric mutual interactions, when $x_0 < x^*$ and
$z_0 < z^*$, for different parameters $b$ and $g$ and initial
conditions $\{x_0,z_0\}$: (a) $b = 0.1$, $g = 0.45 < g_c = 0.468$,
$\{0.05, 2\}_1$, $\{0.5, 2\}_2$, $\{1.1, 2\}_3$; the fixed points
being $x^*=1.333$, $z^*=2.5$. (b) $b = 0.7$, $g=0.025< g_c=0.027$,
$\{0.1, 0.01\}_1$, $\{1.4, 0.01\}_2$, $\{4.9, 0.01\}_3$; $x^* = 5$,
$z^* = 1.143$. (c) $b = -1$, $g = 0.1$, $\{0.01, 0.1\}_1$,
$\{0.2, 0.1\}_2$, $\{0.487, 0.1\}_3$, $x^* = 0.4874$, $z^* = 1.051$.
(d) $b = -0.1$, $g = -1$, $\{0.01, 0.001\}_1$, $\{0.01, 0.3\}_2$,
$\{0.01, 0.51\}_3$, $x^* = 0.951$, $z^* = 0.513$.

\vskip 1cm

{\bf Fig. 5}. Comparison of the symbiotic solutions $x(t)$ (solid
line) and $z(t)$ (dashed-dotted line) with the solutions
$x(t) = z(t)$ (dashed line) of the decoupled equations (\ref{thyhbwrq})
for the same initial conditions
$x_0=z_0=0.1 < 1$, but for different symbiotic parameters $b$ and
$g$: (a) $b = 0.25$, $g = 0.1 < g_c = 0.25$; the stationary points
of the symbiotic equations being $x^* = 1.411$, $z^* = 1.164$. (b)
$b = 2$, $g = -0.5$, the fixed points of the symbiotic equations
being $x^*=3.562$, $z^*=0.360$. (c) $b = -1$, $g = 2$, the symbiotic
fixed points being $x^* = 0.293$, $z^* = 2.414$. (d) $b = -1$,
$g = -2$, the symbiotic fixed points being $x^* = 0.707$,
$z^* = 0.414$.

\vskip 1cm

{\bf Fig. 6}. Comparison of the symbiotic solutions $x(t)$ (solid
line) and $z(t)$ (dashed-dotted line) with the solutions
$x(t) = z(t)$ (dashed line) of the decoupled equations (\ref{thyhbwrq})
for the same initial conditions
$x_0=z_0=1.8 > 1$, but for different symbiotic parameters $b$ and
$g$: (a) $b =0.5, g =0.080 < g_c =0.086$; the fixed points of the
symbiotic equations being $x^*=2.823$, $z^*=1.292$. (b) $b=0.5$,
$g = -0.1$, the fixed points of the symbiotic equations being
$x^* = 1.742$, $z^* = 0.852$. (c) $b = -0.3$, $g = 1$, the
symbiotic fixed points being $x^* = 0.582$, $z^* = 2.393$. (d)
$b =-0.1$, $g=-0.3$, the symbiotic fixed points being $x^*=0.927$,
$z^* = 0.782$.

\vskip 1cm

{\bf Fig. 7}. Logarithmic behavior of the exponentially growing
solutions $x(t)$ (solid line) and $z(t)$ (dashed-dotted line) for
different symbiotic parameters and initial conditions: (a) $b = 1$,
$g = 0.1$, $x_0 = 148$, $z_0 = 0.135$. (b) $b=0.5$, $g=1>g_c=0.086$,
$x_0 = 0.368$, $z_0 = 0.05$. (c) $b = 1$, $g = 0.1 > g_c = 0$,
$x_0 = 0.135$, $z_0 = 2.72$. (d) $b = -1$, $g = 0.1$, $x_0 = 148$,
$z_0 = 0.05$.

\vskip 1cm

{\bf Fig. 8}. Logarithmic behavior of the solutions in the case of
finite-time death and singularity, $x(t)$ (solid line) and $z(t)$
(dashed-dotted line), for the parasitic relations with the symbiotic
coefficients $b = -1$, $g = -2$, under the initial conditions
$x_0 = 1.8$, $z_0 = 0.5$. For these parameters, the critical time
is $t_c = 0.87$.

\vskip 1cm

{\bf Fig. 9}. Region of stability (shaded) in the parameter plane
$b-g$ for the fixed points in the case of symbiosis with asymmetric
interactions.

\vskip 1cm

{\bf Fig. 10}. Convergence to stationary states of the solutions
$x(t)$ (solid line) and $z(t)$ (dashed-dotted line), as functions
of time, in the case of asymmetric interactions, for the same
initial conditions, such that $x_0>z_0$, with $x_0=1.8$ and
$z_0=0.1$, for different parameters $b$ and $g$: (a) $b=0.5$,
$g=0.1< g_c =0.125$; the fixed points being $x^*= 2.764$,
$z^* = 1.276$. (b) $b = 2$, $g = -0.5$; the fixed points being
$x^* = 1.618$, $z^* = 0.191$. (c) $b = -0.25$, $g = 2$; with
the fixed points $x^* = 0.638$, $z^* = 2.275$. (d) $b = -0.3$,
$g = -0.4$; the fixed points being $x^* = 0.833$, $z^* = 0.667$.

\vskip 1cm

{\bf Fig. 11}. Convergence to stationary states of the solutions
$x(t)$ (solid line) and $z(t)$ (dashed-dotted line), as functions
of time, in the case of asymmetric interactions, for the same
initial conditions, such that $x_0<z_0$, with $x_0=0.1$ and
$z_0=1.8$, for different parameters $b$ and $g$: (a) $b=0.5$,
$g=0.1<g_c=0.125$; the fixed points being $x^* = 2.764$,
$z^* = 1.276$. (b) $b = 2$, $g = -0.5$; the fixed points being
$x^* = 1.618$, $z^* = 0.191$. (c) $b = -0.25$, $g = 2$; with
the fixed points $x^* = 0.638$, $z^* = 2.275$. (d) $b = -0.3$,
$g = -0.4$; the fixed points being $x^* = 0.833$, $z^* = 0.667$.

\vskip 1cm

{\bf Fig. 12}. Logarithmic behavior of the solutions in the case
of asymmetric interactions, in the presence of the finite-time
singularity, $x(t)$ (solid line) and $z(t)$ (dashed-dotted line),
for the symbiotic coefficients $b = 2$, $g = -1.5$ and the initial
conditions $x_0 = 1.8$, $z_0 = 0.1$. The critical time here is
$t_c = 4.43$.

\vskip 1cm

{\bf Fig. 13}. Finite-time death in the case of asymmetric
interactions. The species $x(t)$ (solid line) kill the species
$z(t)$ (dashed-dotted line) at the death time $t_d$: (a) $b=-2$,
$g = -1.5$; the initial conditions are $x_0 = z_0 = 0.1$; the
death time is $t_d = 3.8$. (b) $b = 1$, $g = -1.5$; the initial
conditions are  $x_0 = 0.1$, $z_0 = 1.8$; the death time is
$t_d = 2.6$.

\vskip 1cm

{\bf Fig. 14}. Finite-time death in the case of asymmetric
interactions. The species $x(t)$ (solid line) kill the species
$z(t)$ (dashed-dotted line) at the death time $t_d$. The symbiotic
coefficients are $b = -2$, $g = -1.5$; the initial conditions are
$x_0 = 0.1$, $z_0 = 6$; the death time is $t_d = 0.559$.

\vskip 1cm

{\bf Fig. 15}. Finite-time death in the case of asymmetric
interactions. The species $z(t)$ (dashed-dotted line) kill the
species $x(t)$ (solid line) at the death time $t_d$. The symbiotic
coefficients are $b = -2$, $g = -1.5$; the initial conditions are
$x_0 = 1.8$, $z_0 = 0.1$; the death time is $t_d = 1.016$.

\vskip 1cm

{\bf Fig. 16}. Region of stability (shaded) in the parameter plane
$b-g$ for the fixed points in the case of symbiosis without direct
interactions.

\vskip 1cm

{\bf Fig. 17}. Finite-time death in the case of symbiosis without
direct interactions. Temporal behavior of solutions $x(t)$ (solid
line) and $z(t)$ (dashed-dotted line) for different symbiosis
parameters and initial conditions: (a) $b = -0.75$, $g = -0.5$,
$x_0 = 0.8$, $z_0 = 3$; the death time being $t_d = 0.204$.
(b) $b = -1.5$, $g = 1$, $x_0 = 1$, $z_0 = 0.1$; with the death
time $t_d = 2.412$. (c) $b=-1$, $g=-2$, $x_0=0.8$, $z_0=0.1$;
with the death time $t_d=4.279$. (d) $b=-2$, $g=-1$, $x_0=0.1$,
$z_0 = 1$; the death time being $t_d = 1.459$.

\newpage

\begin{figure}[ht]
\centerline{\includegraphics[width=10cm]{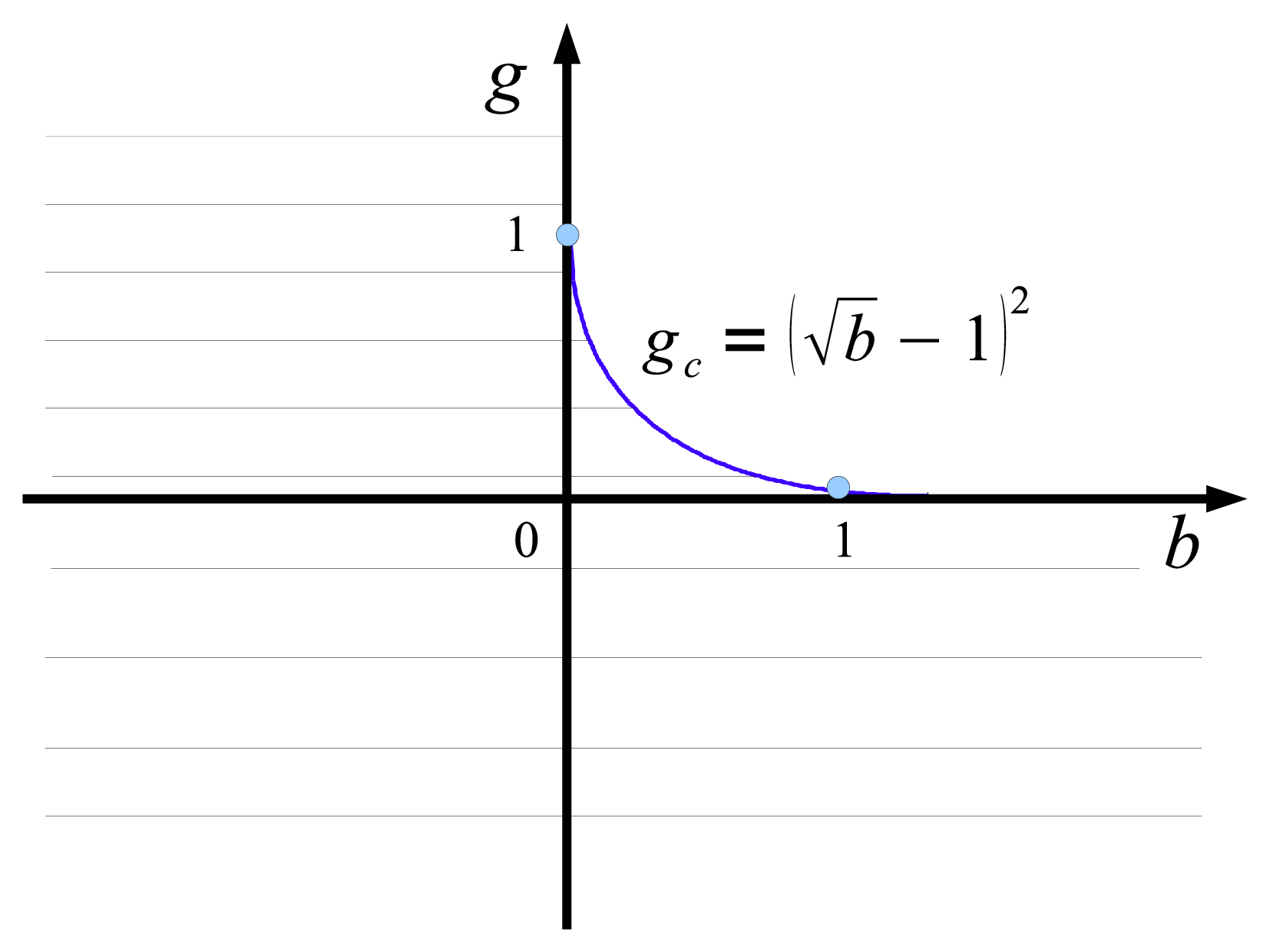}}
\caption{Region of stability (shaded) in the parameter plane
$b-g$ for the fixed points in the case of symbiosis with mutual
interactions.}
\label{fig:Fig.1}
\end{figure}

\newpage

\begin{figure}[ht]
\vspace{9pt}
\centerline{
\hbox{ \includegraphics[width=8cm]{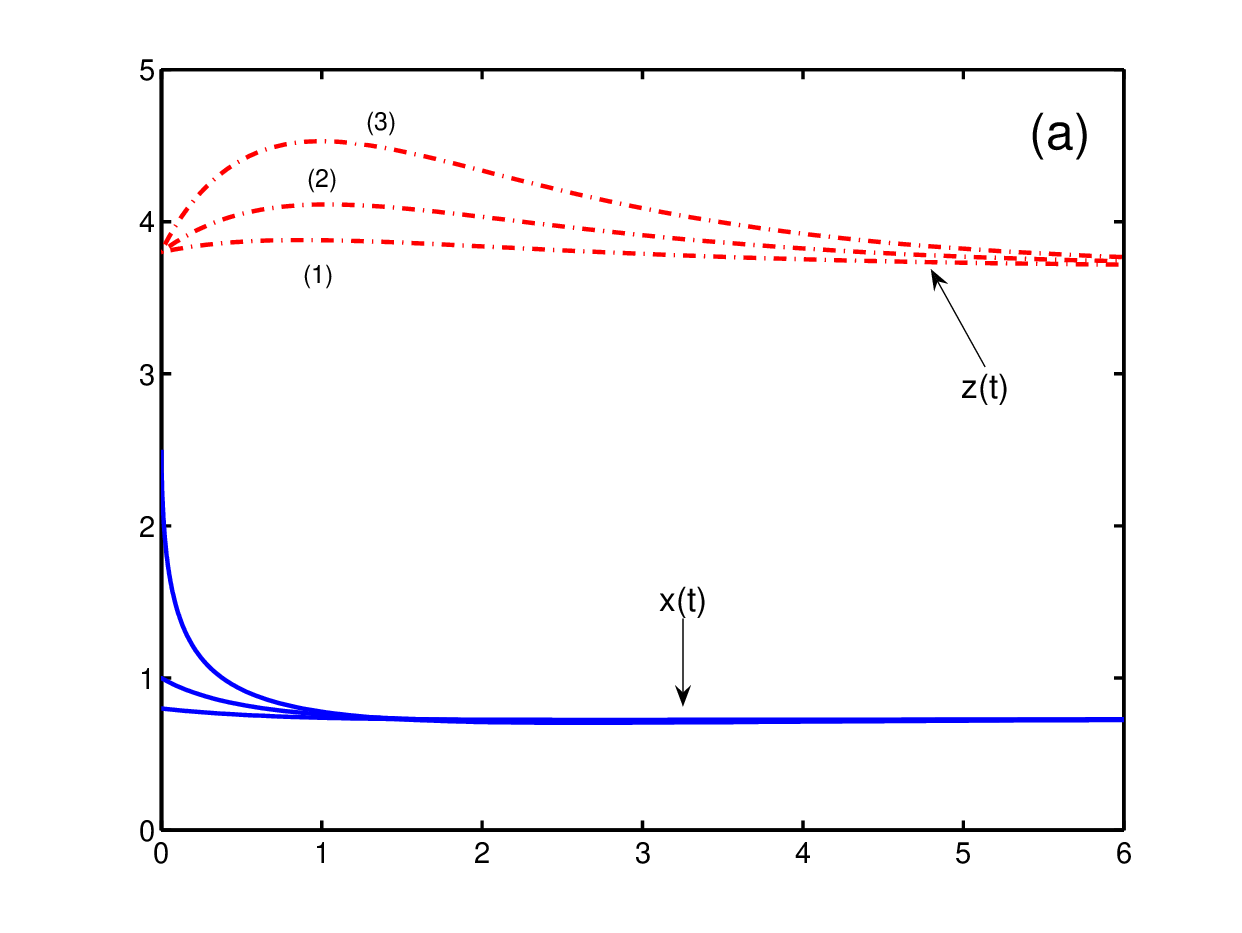} \hspace{1cm}
\includegraphics[width=8cm]{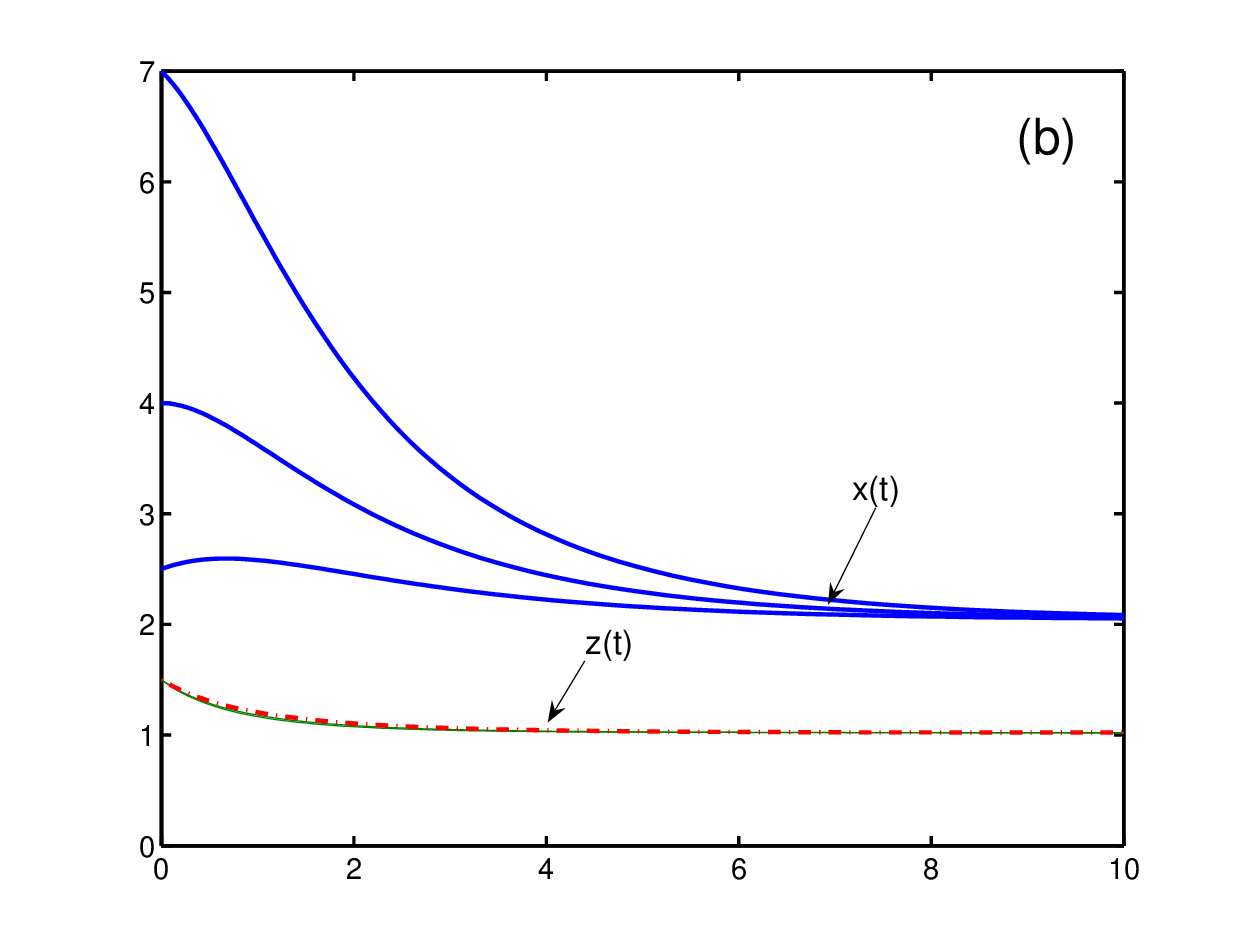} } }
\vspace{9pt}
\centerline{
\hbox{ \includegraphics[width=8cm]{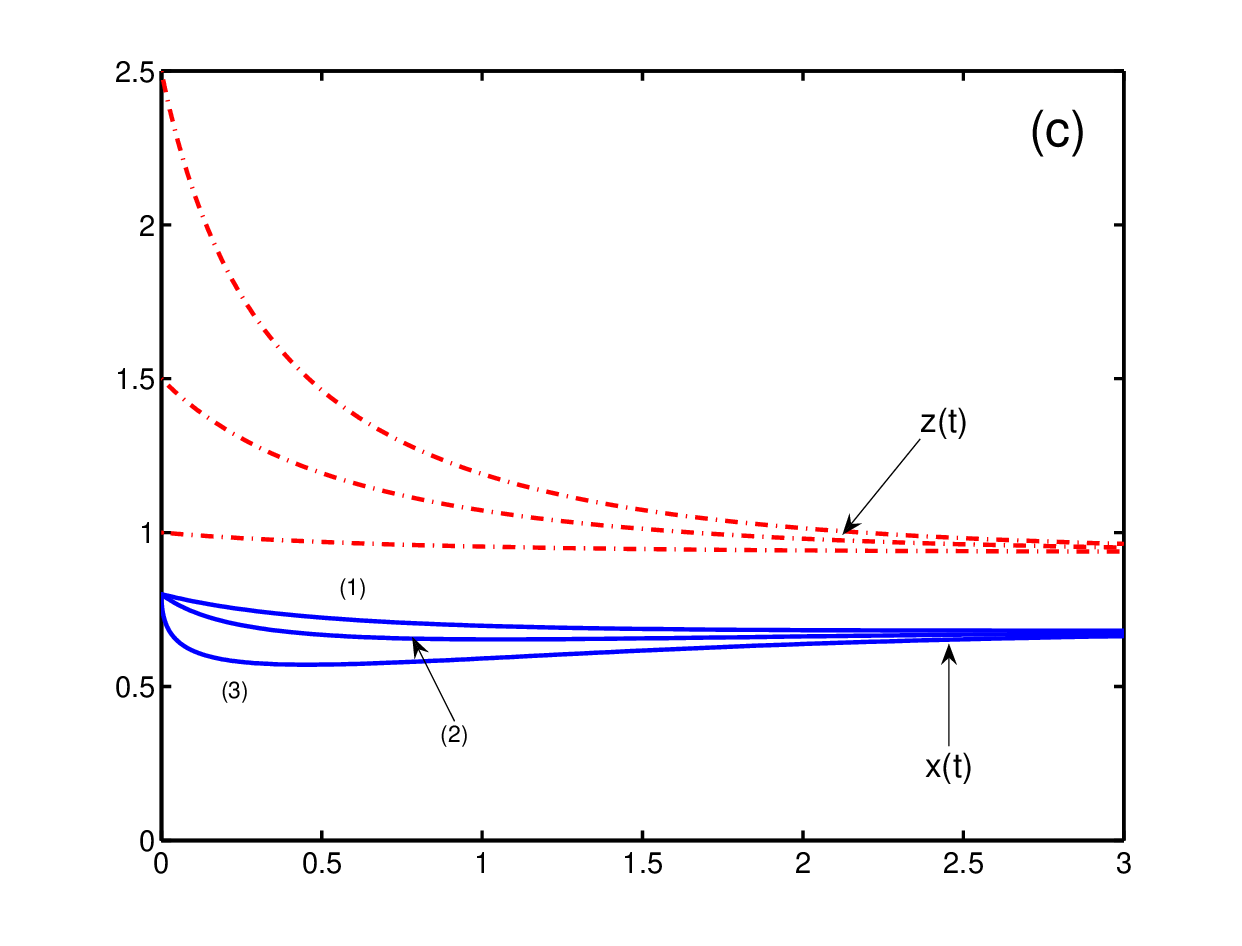} \hspace{1cm}
\includegraphics[width=8cm]{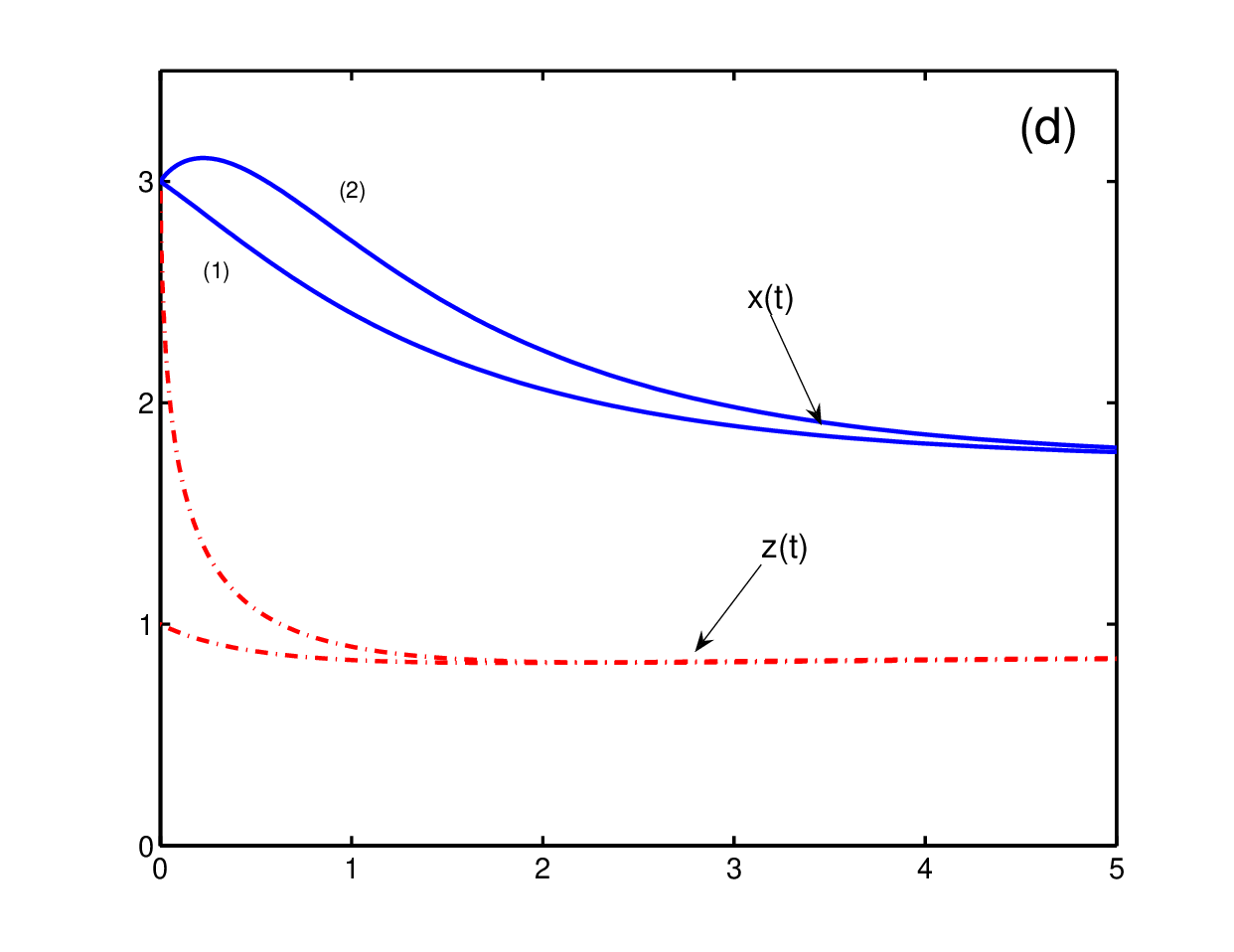} } }
\caption{Convergence to stationary states of the solutions $x(t)$
(solid line) and $z(t)$ (dashed-dotted line), as functions of time,
in the case of symmetric mutual interactions, when $x_0 > x^*$ and
$z_0 > z^*$, for different parameters $b$ and $g$ and initial
conditions $\{x_0,z_0\}$: (a) $b = -0.1$, $g = 1$, $\{0.75, 3.8\}_1$,
$\{1, 3.8\}_2$, $\{2.5, 3.8\}_3$; the fixed points being $x^*=0.730$,
$z^*=3.702$. (b) $b=0.5$, $g =0.01 < g_c = 0.086$, $\{2.5, 1.5\}_1$,
$\{4, 1.5\}_2$, $\{7, 1.5\}_3$; $x^* = 2.043$, $z^* = 1.021$. (c)
$b = -0.5$, $g = -0.1$, $\{0.8, 1\}_1$, $\{0.8, 1.5\}_2$,
$\{0.8, 2.499\}_3$, $x^* = 0.681$, $z^* = 0.936$. (d) $b = 0.5$,
$g = -0.1 < g_c = 0.086$, $\{3, 1\}_1$, $\{3, 3\}_2$, $x^* = 1.742$,
$z^* = 0.852$.}
\label{fig:Fig.2}
\end{figure}

\newpage

\begin{figure}[ht]
\vspace{9pt}
\centerline{
\hbox{ \includegraphics[width=8cm]{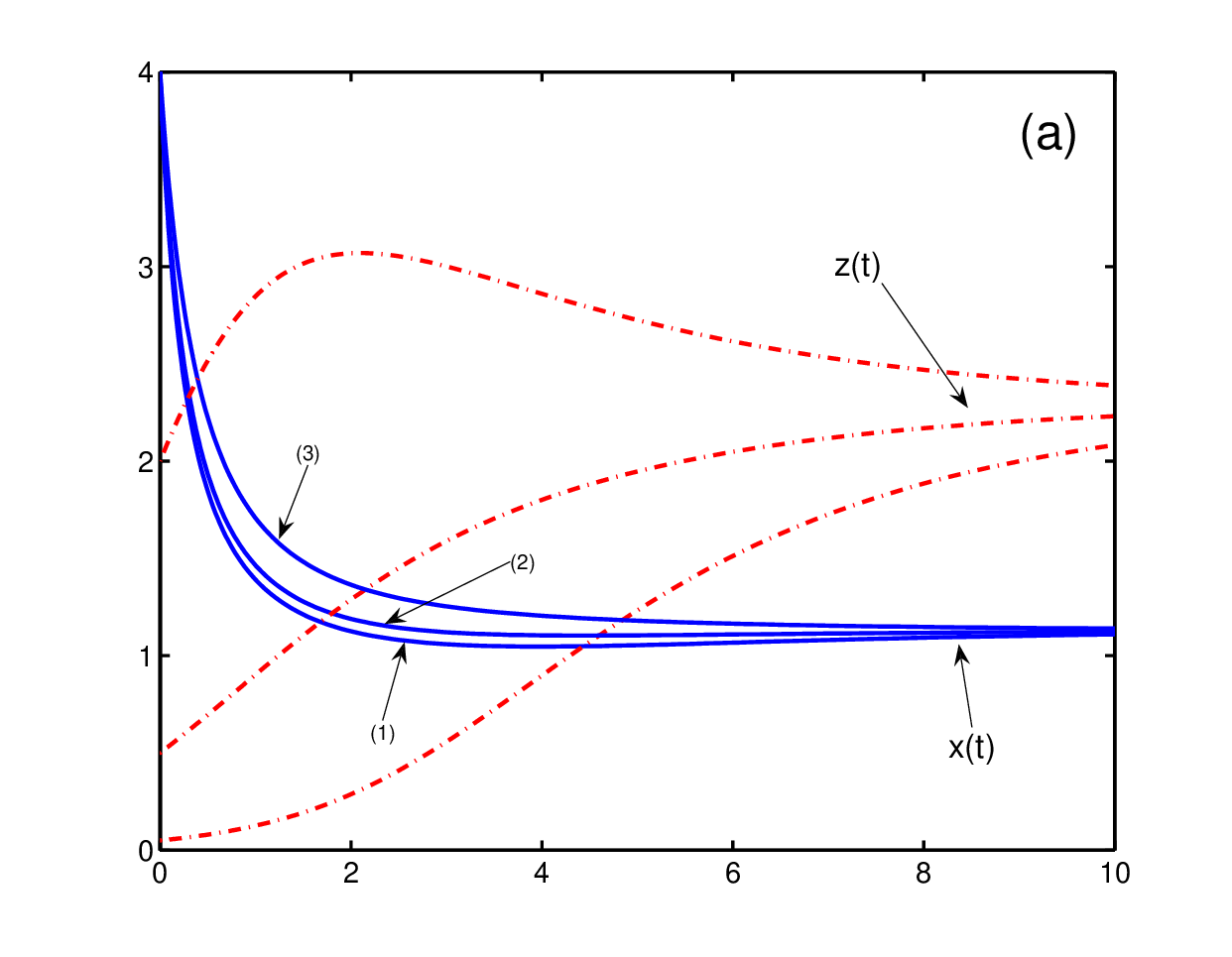} \hspace{1cm}
\includegraphics[width=8cm]{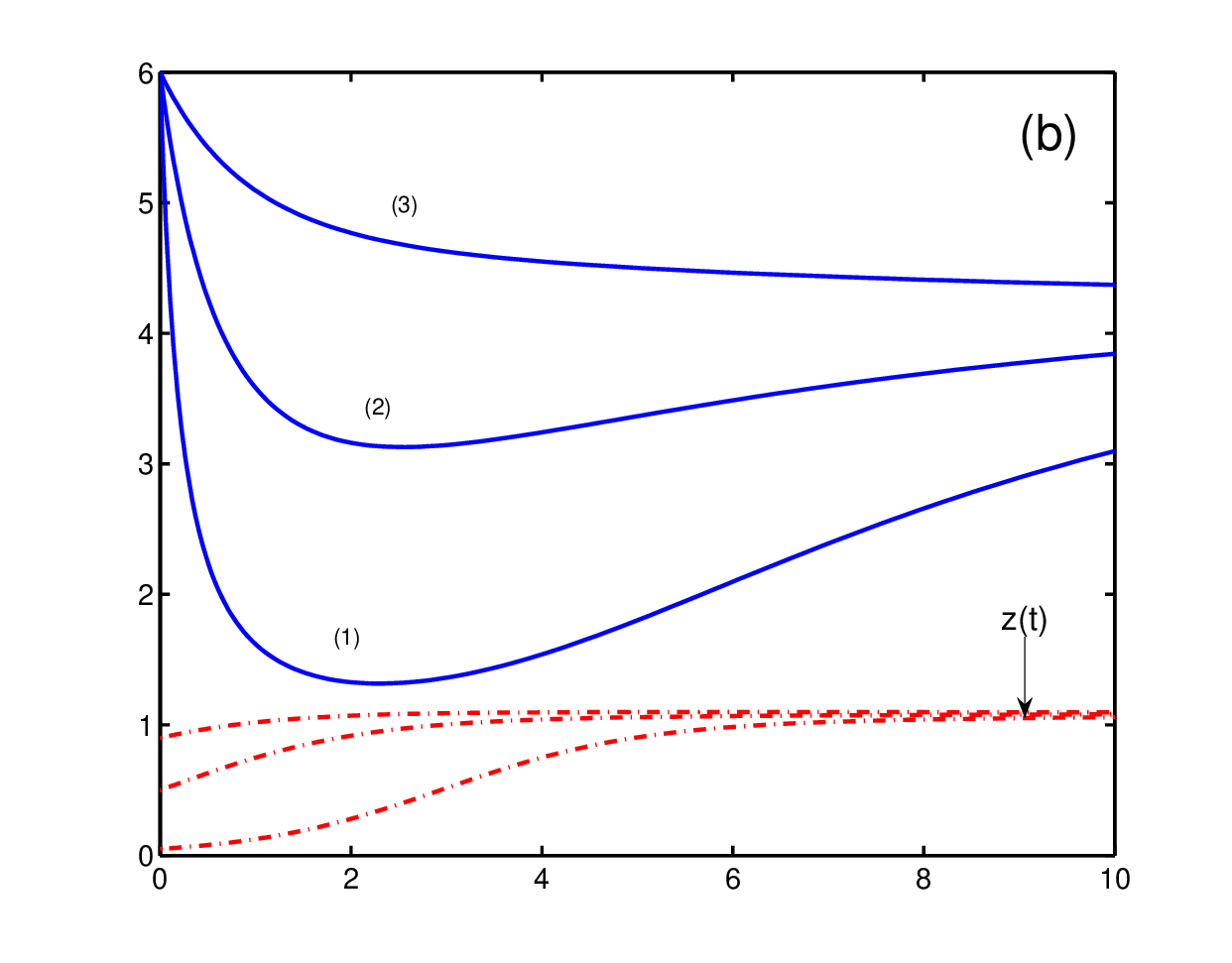} } }
\vspace{9pt}
\centerline{
\hbox{ \includegraphics[width=8cm]{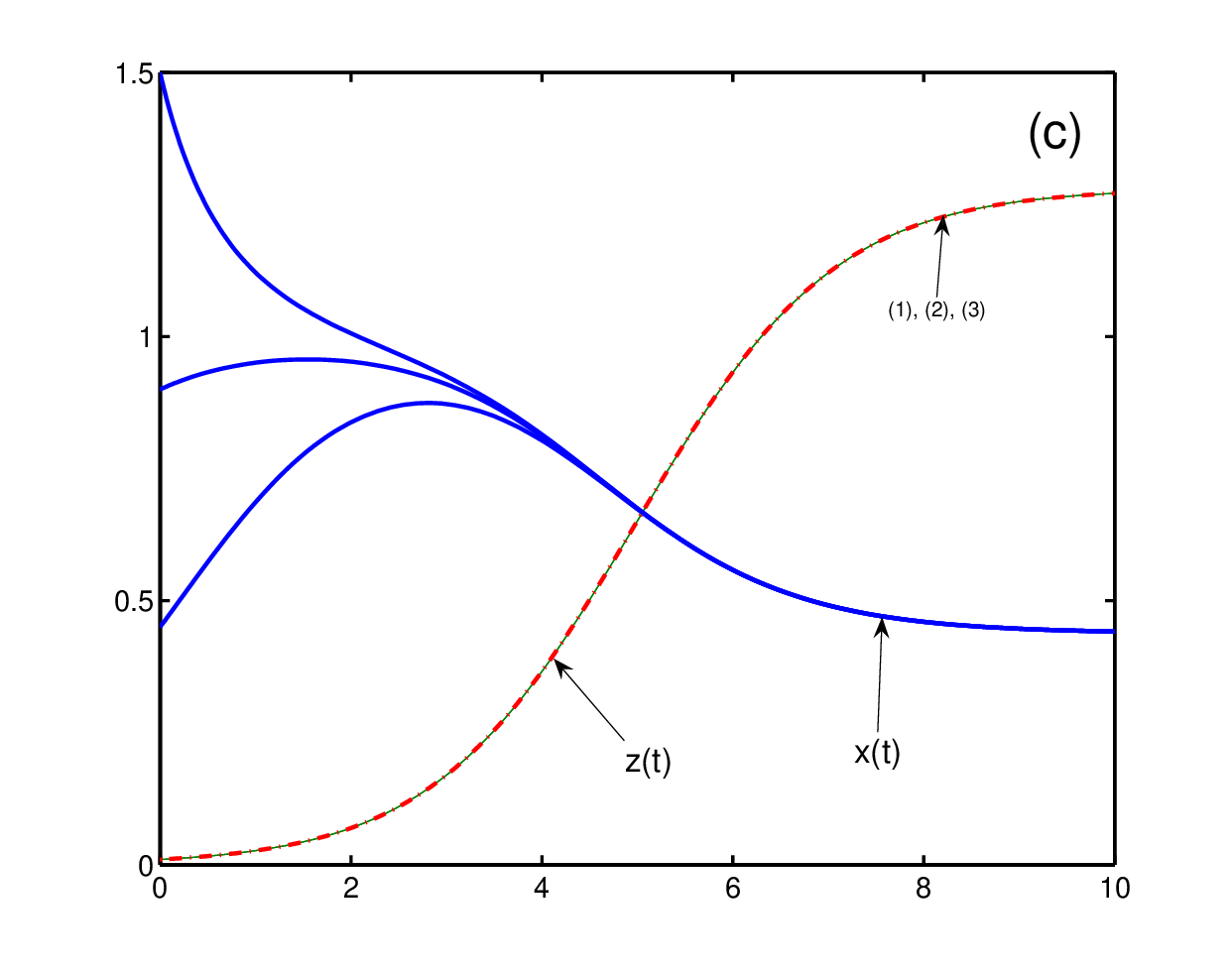} \hspace{1cm}
\includegraphics[width=8cm]{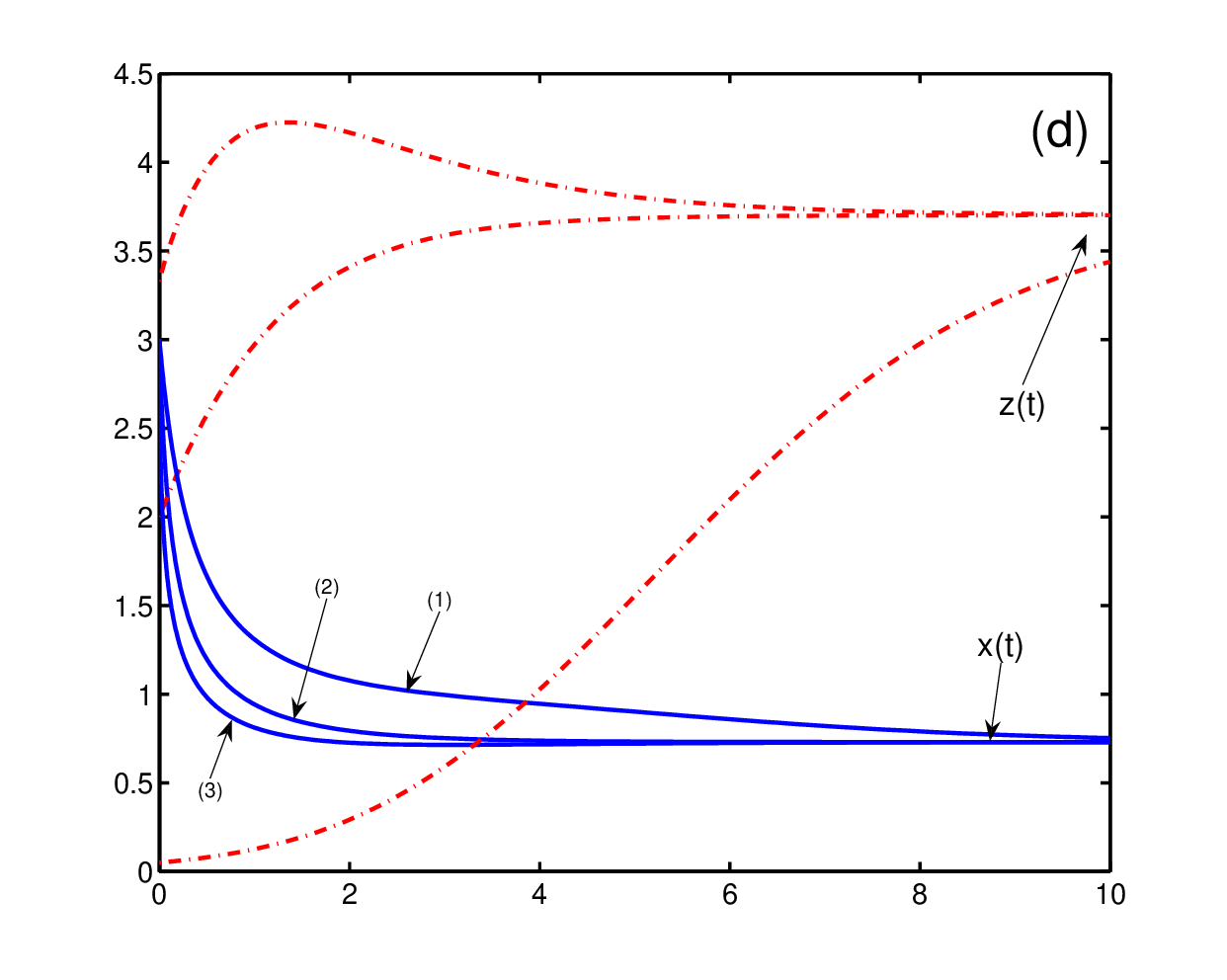} } }
\caption{Convergence to stationary states of the solutions $x(t)$
(solid line) and $z(t)$ (dashed-dotted line), as functions of time,
in the case of symmetric mutual interactions, when $x_0 > x^*$ but
$z_0 < z^*$, for different parameters $b$ and $g$ and initial
conditions $\{x_0,z_0\}$: (a) $b = 0.05$, $g = 0.5 < g_c = 0.603$,
$\{4, 0.05\}_1$, $\{4, 0.5\}_2$, $\{4, 2\}_3$; the fixed points being
$x^*=1.130$, $z^*=2.298$. (b) $b=0.7$, $g = 0.02$, $\{6, 0.05\}_1$,
$\{6, 0.5\}_2$, $\{6, 0.9\}_3$; $x^* = 4.258$, $z^* = 1.093$. (c)
$b = -1$, $g=0.1$, $\{0.45, 0.01\}_1$, $\{1, 0.01\}_2$,
$\{1.5, 0.01\}_3$, $x^*=0.438$, $z^* = 1.281$. (d) $b = -0.01$,
$g=1$, $\{3, 0.05\}_1$, $\{3, 2\}_2$, $\{3, 3.33\}_3$, $x^*=0.730$,
$z^* = 3.702$.}
\label{fig:Fig.3}
\end{figure}

\newpage

\begin{figure}[ht]
\vspace{9pt}
\centerline{
\hbox{ \includegraphics[width=8cm]{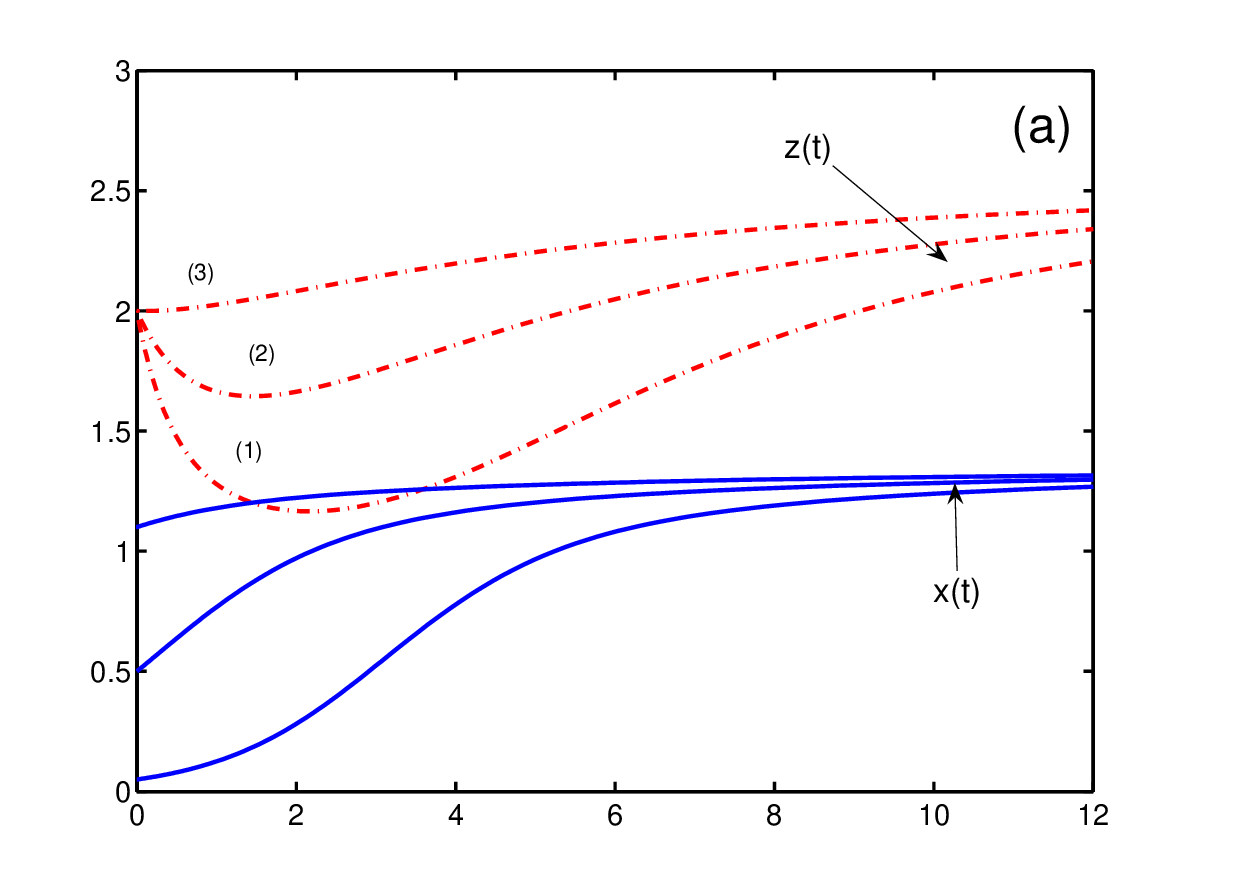} \hspace{1cm}
\includegraphics[width=8cm]{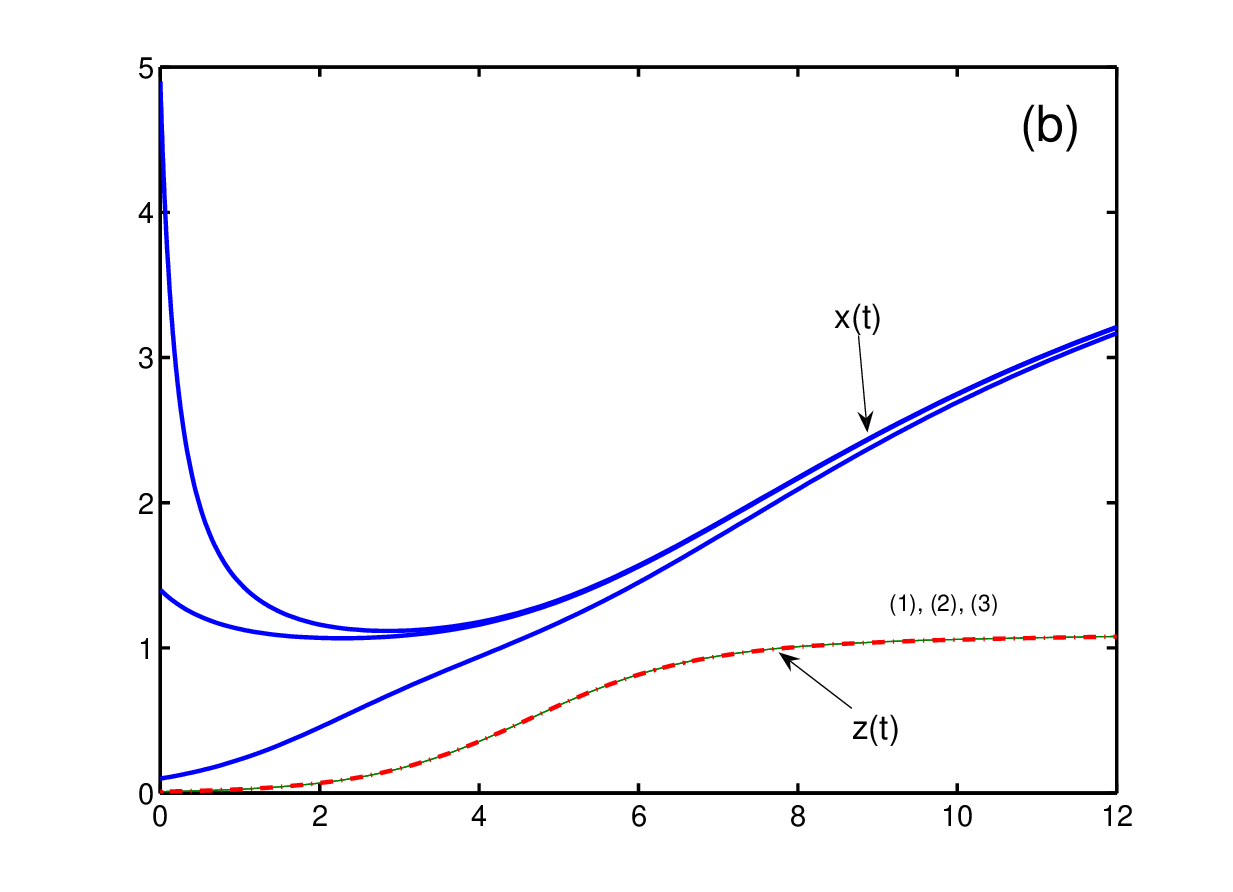} } }
\vspace{9pt}
\centerline{
\hbox{ \includegraphics[width=8cm]{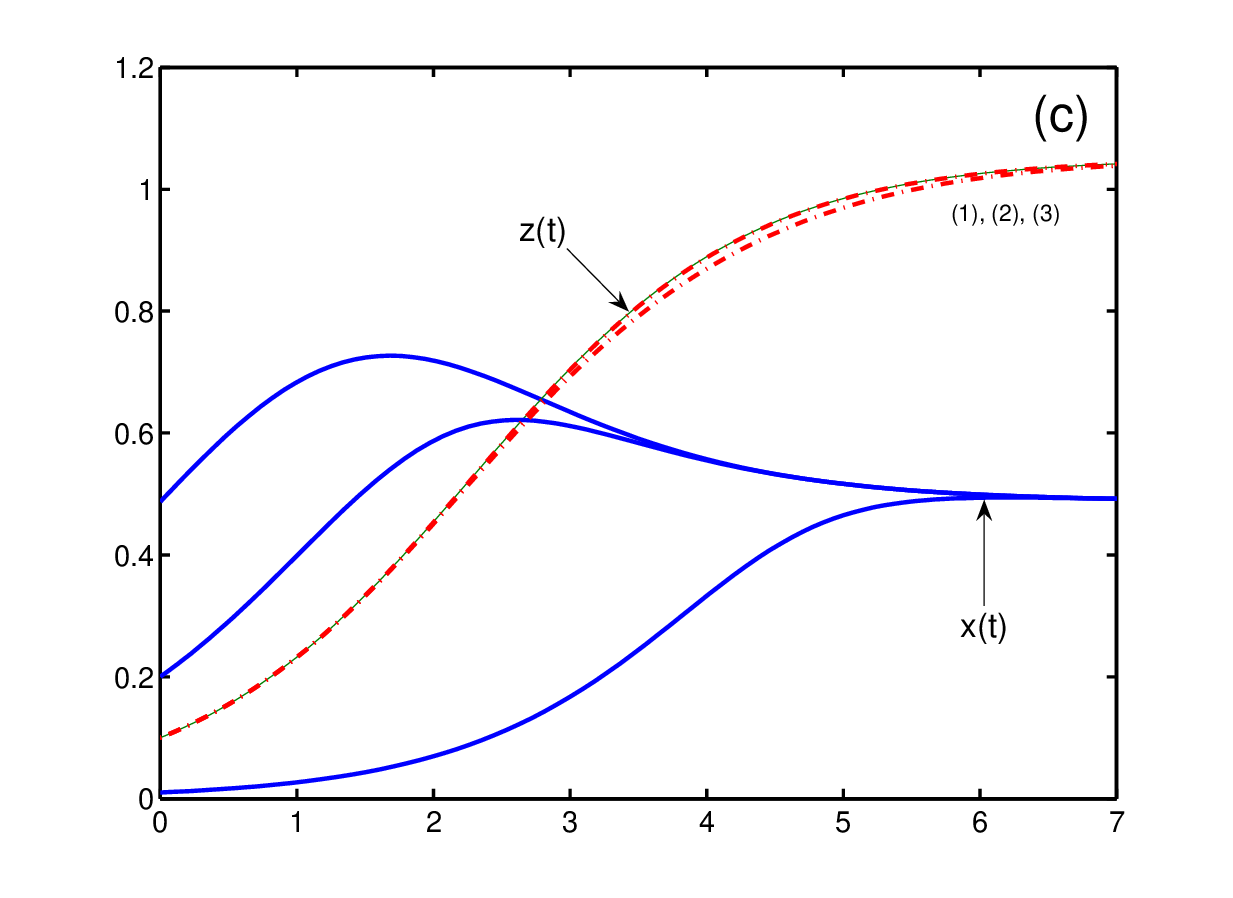} \hspace{1cm}
\includegraphics[width=8cm]{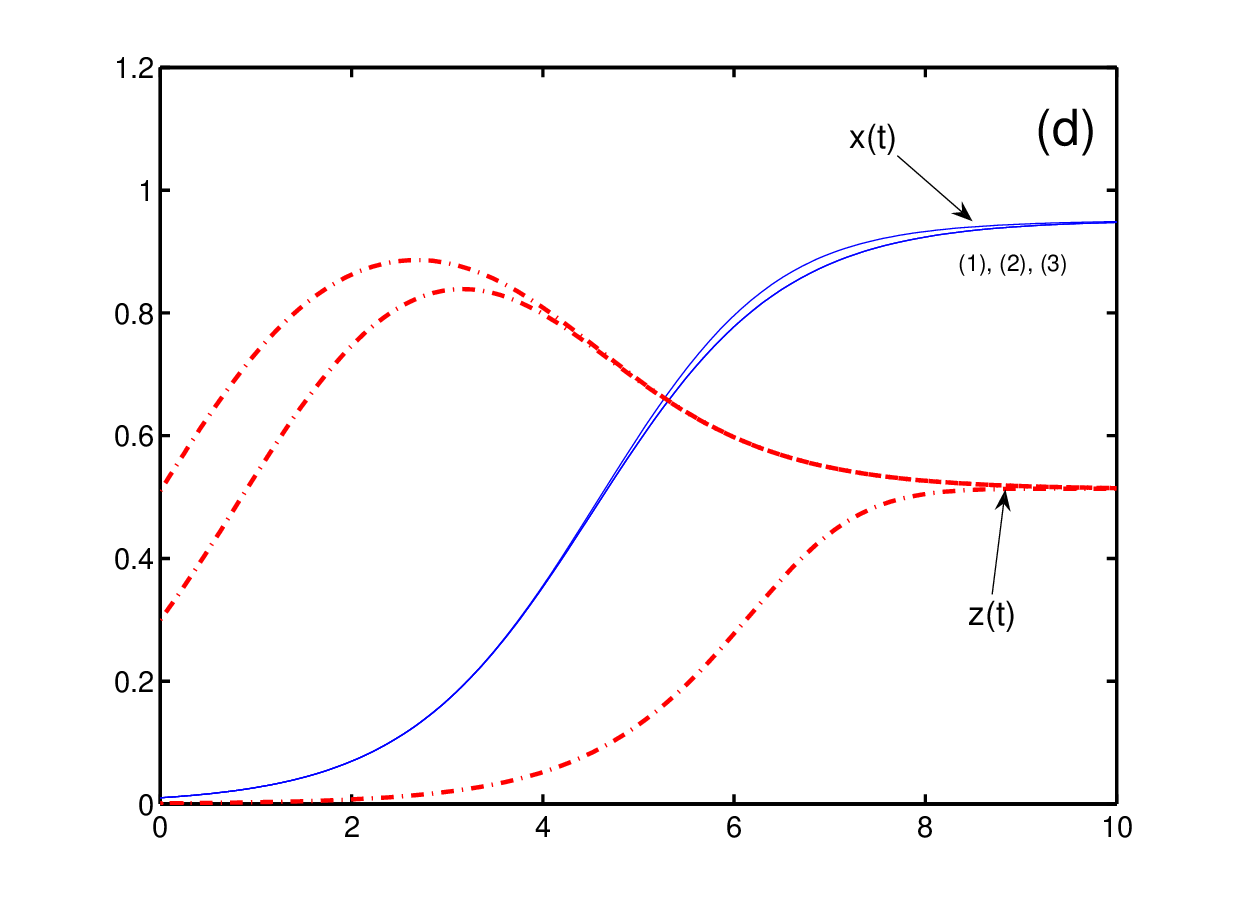} } }
\caption{Convergence to stationary states of the solutions $x(t)$
(solid line) and $z(t)$ (dashed-dotted line), as functions of time,
in the case of symmetric mutual interactions, when $x_0 < x^*$ and
$z_0 < z^*$, for different parameters $b$ and $g$ and initial
conditions $\{x_0,z_0\}$: (a) $b = 0.1$, $g = 0.45 < g_c = 0.468$,
$\{0.05, 2\}_1$, $\{0.5, 2\}_2$, $\{1.1, 2\}_3$; the fixed points
being $x^*=1.333$, $z^*=2.5$. (b) $b = 0.7$, $g=0.025< g_c=0.027$,
$\{0.1, 0.01\}_1$, $\{1.4, 0.01\}_2$, $\{4.9, 0.01\}_3$; $x^* = 5$,
$z^* = 1.143$. (c) $b = -1$, $g = 0.1$, $\{0.01, 0.1\}_1$,
$\{0.2, 0.1\}_2$, $\{0.487, 0.1\}_3$, $x^* = 0.4874$, $z^* = 1.051$.
(d) $b = -0.1$, $g = -1$, $\{0.01, 0.001\}_1$, $\{0.01, 0.3\}_2$,
$\{0.01, 0.51\}_3$, $x^* = 0.951$, $z^* = 0.513$.}
\label{fig:Fig.4}
\end{figure}

\newpage

\begin{figure}[ht]
\vspace{9pt}
\centerline{
\hbox{ \includegraphics[width=8cm]{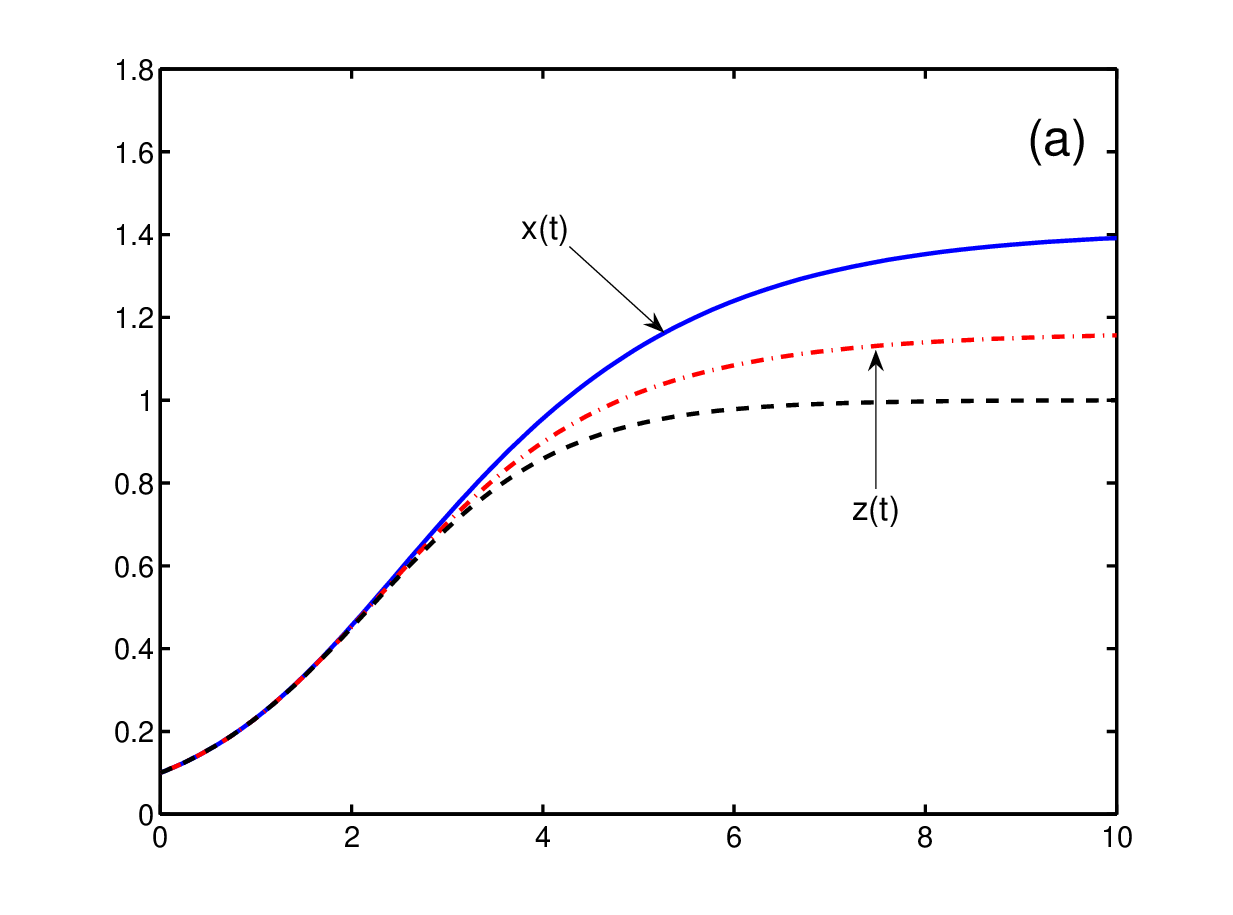} \hspace{1cm}
\includegraphics[width=8cm]{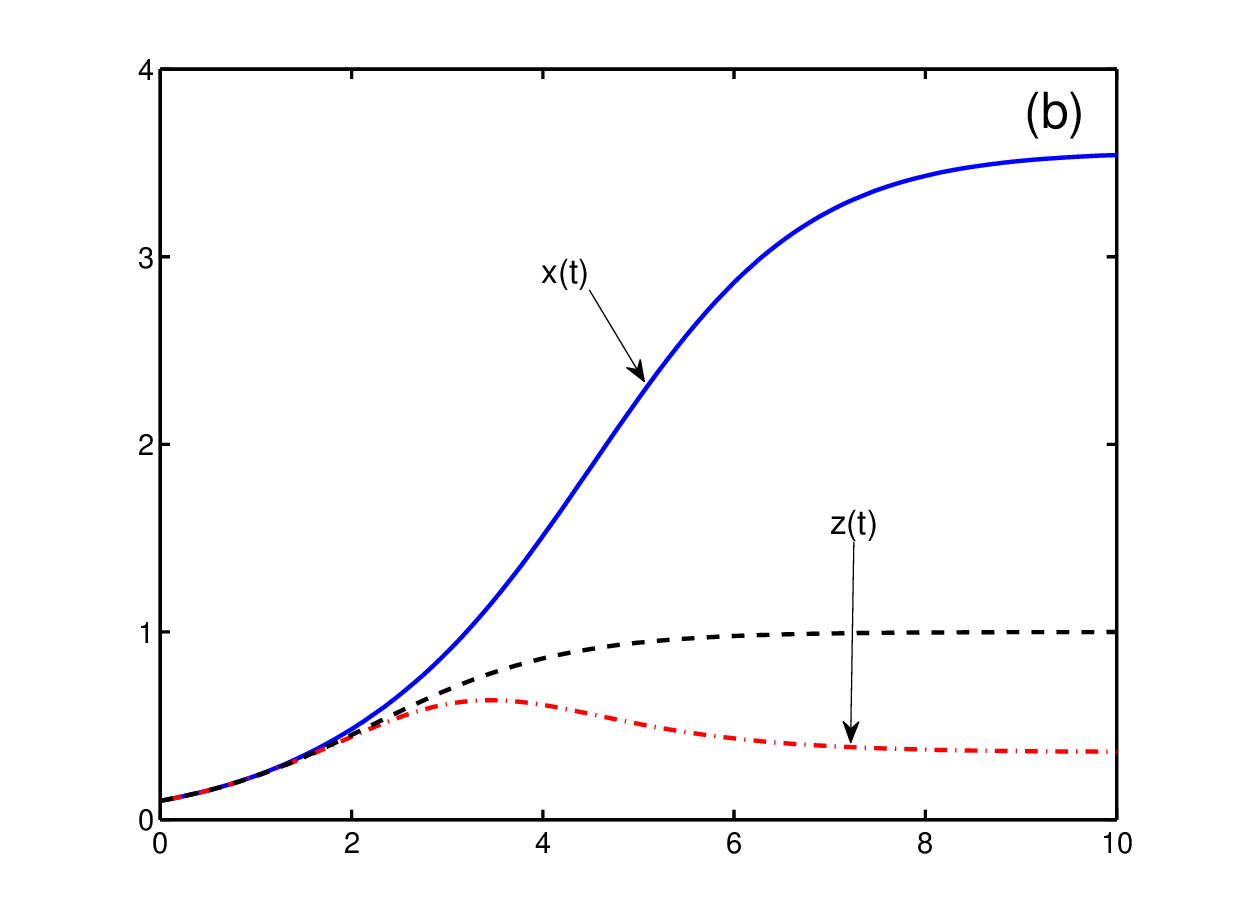} } }
\vspace{9pt}
\centerline{
\hbox{ \includegraphics[width=8cm]{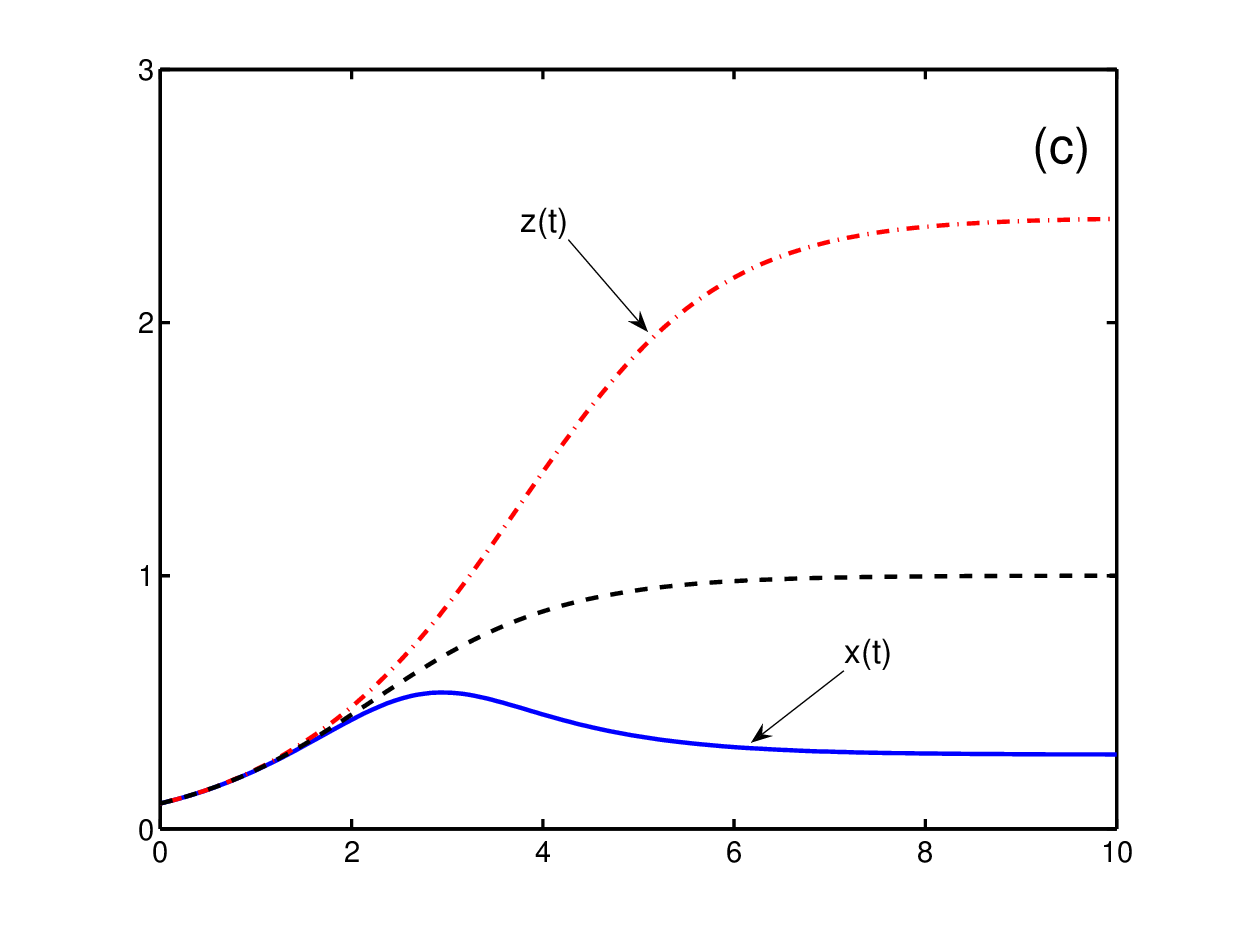} \hspace{1cm}
\includegraphics[width=8cm]{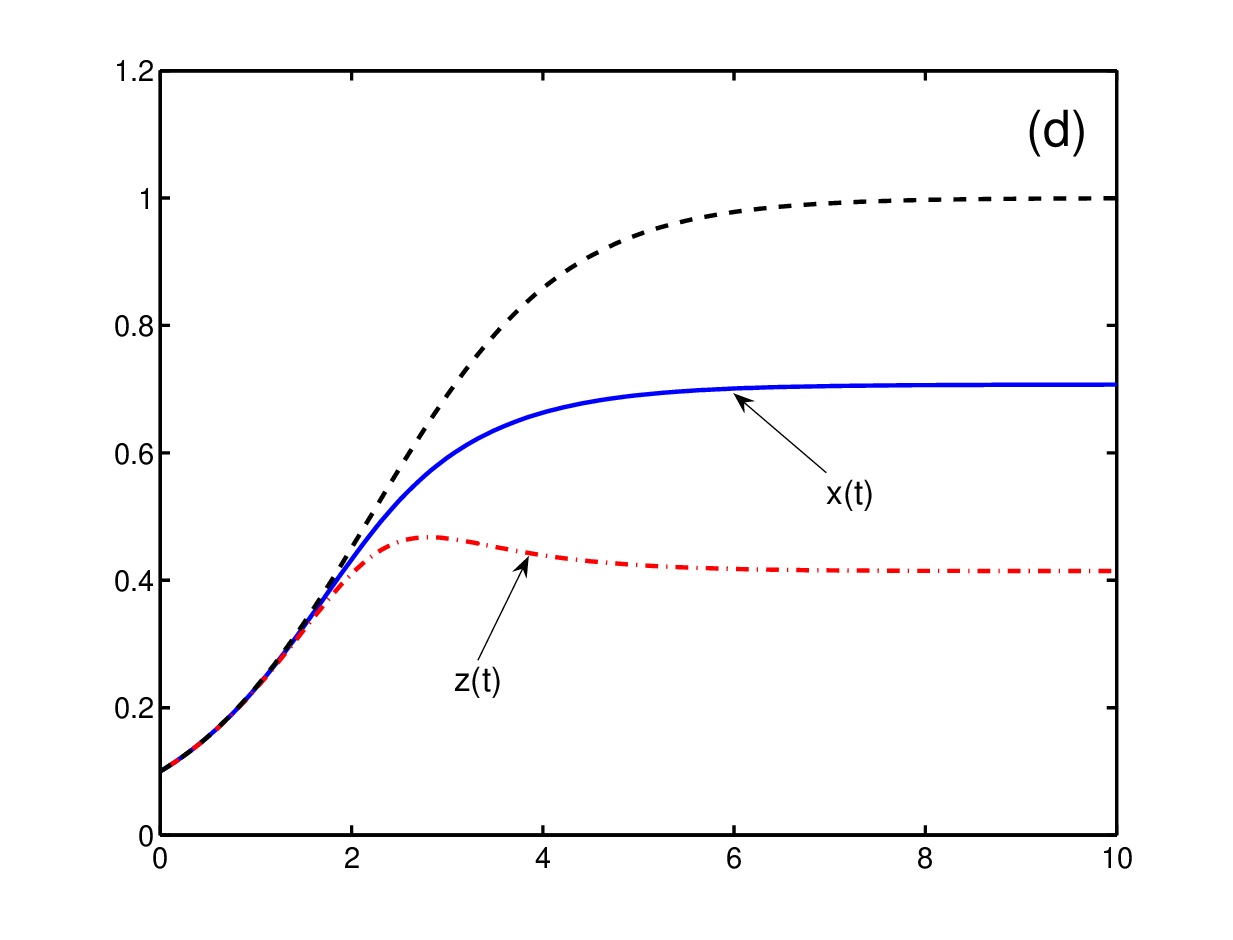} } }
\caption{Comparison of the symbiotic solutions $x(t)$ (solid
line) and $z(t)$ (dashed-dotted line) with the solutions
$x(t) = z(t)$ (dashed line) of the decoupled equations
(\ref{thyhbwrq}) for the same initial conditions
$x_0=z_0=0.1 < 1$, but for different symbiotic parameters $b$ and
$g$: (a) $b = 0.25$, $g = 0.1 < g_c = 0.25$; the stationary points
of the symbiotic equations being $x^* = 1.411$, $z^* = 1.164$. (b)
$b = 2$, $g = -0.5$, the fixed points of the symbiotic equations
being $x^*=3.562$, $z^*=0.360$. (c) $b = -1$, $g = 2$, the symbiotic
fixed points being $x^* = 0.293$, $z^* = 2.414$. (d) $b = -1$,
$g = -2$, the symbiotic fixed points being $x^* = 0.707$,
$z^* = 0.414$.}
\label{fig:Fig.5}
\end{figure}

\newpage

\begin{figure}[ht]
\vspace{9pt}
\centerline{
\hbox{ \includegraphics[width=8cm]{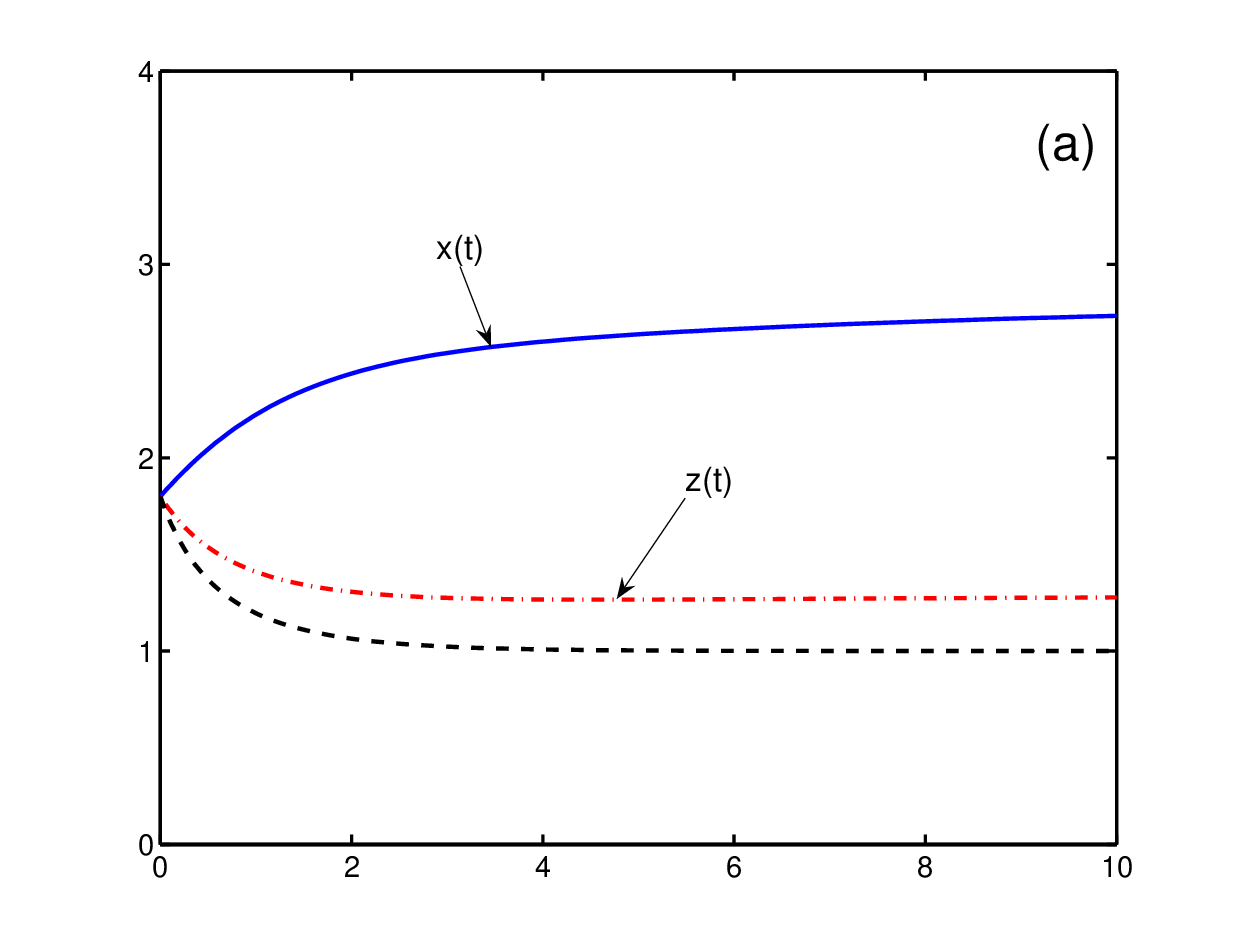} \hspace{1cm}
\includegraphics[width=8cm]{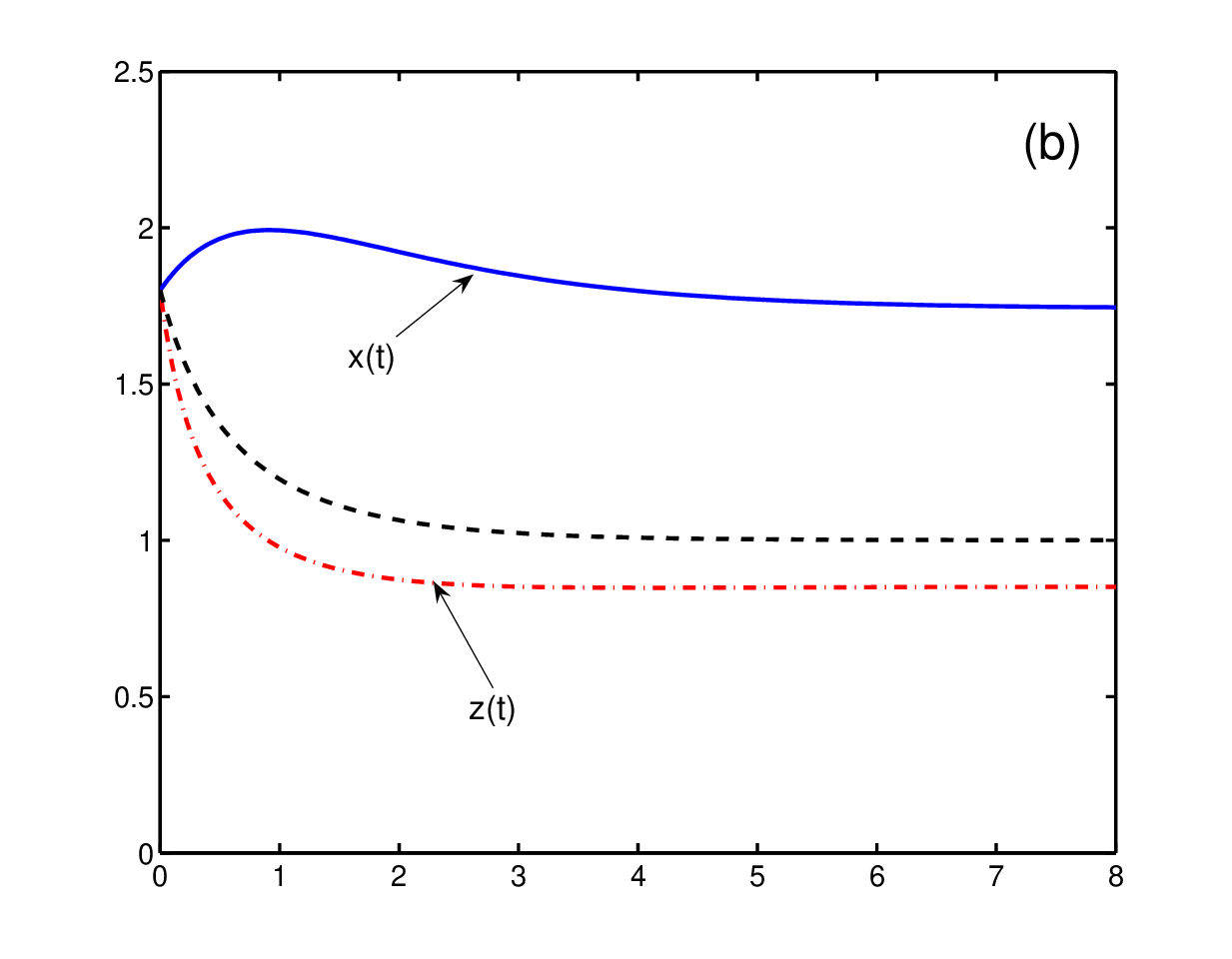} } }
\vspace{9pt}
\centerline{
\hbox{ \includegraphics[width=8cm]{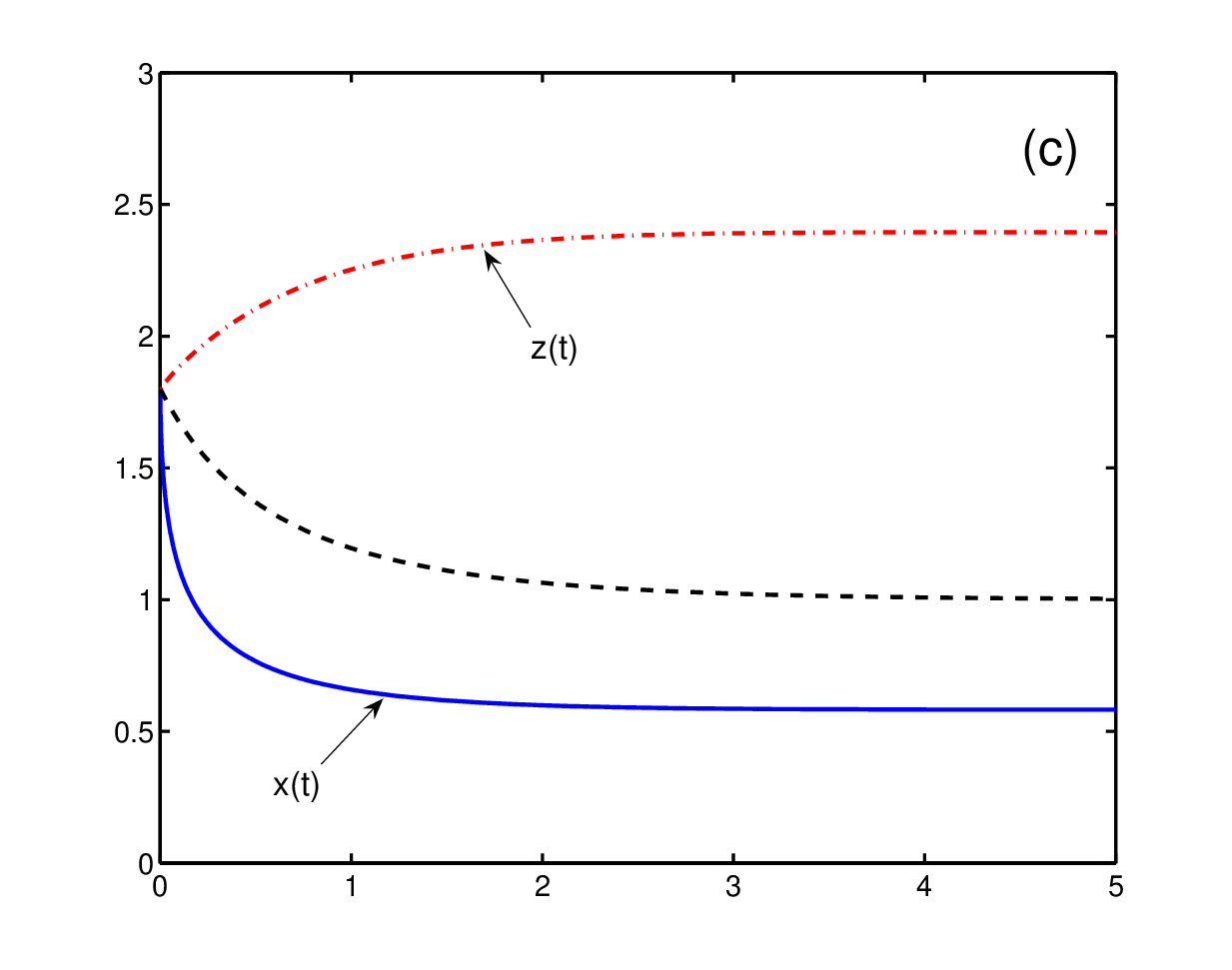} \hspace{1cm}
\includegraphics[width=8cm]{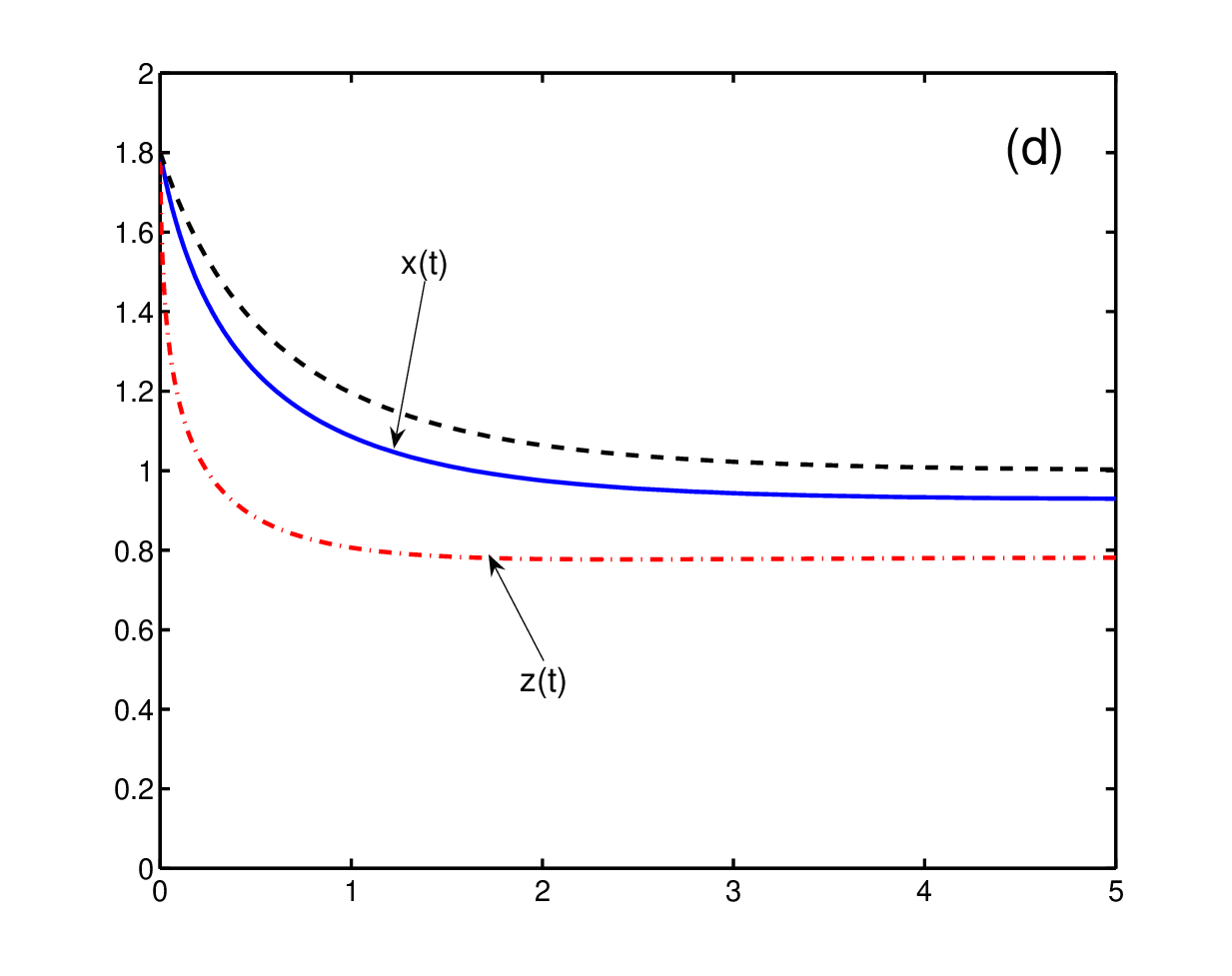} } }
\caption{Comparison of the symbiotic solutions $x(t)$ (solid
line) and $z(t)$ (dashed-dotted line) with the solutions
$x(t) = z(t)$ (dashed line) of the decoupled equations
(\ref{thyhbwrq}) for the same initial conditions
$x_0=z_0=1.8 > 1$, but for different symbiotic parameters $b$ and
$g$: (a) $b =0.5, g =0.080 < g_c =0.086$; the fixed points of the
symbiotic equations being $x^*=2.823$, $z^*=1.292$. (b) $b=0.5$,
$g = -0.1$, the fixed points of the symbiotic equations being
$x^* = 1.742$, $z^* = 0.852$. (c) $b = -0.3$, $g = 1$, the
symbiotic fixed points being $x^* = 0.582$, $z^* = 2.393$. (d)
$b =-0.1$, $g=-0.3$, the symbiotic fixed points being $x^*=0.927$,
$z^* = 0.782$.}
\label{fig:Fig.6}
\end{figure}

\newpage

\begin{figure}[ht]
\vspace{9pt}
\centerline{
\hbox{ \includegraphics[width=8cm]{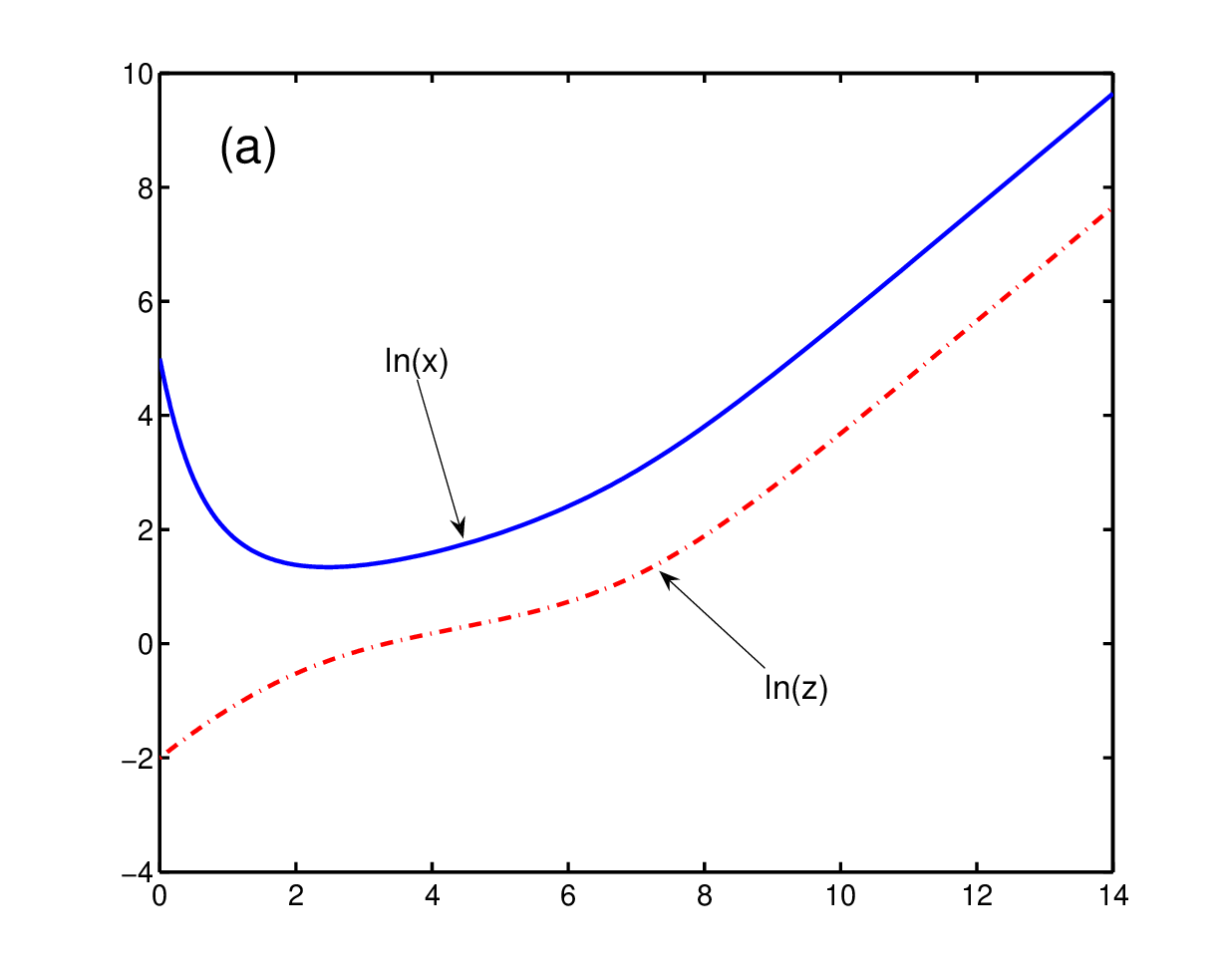} \hspace{1cm}
\includegraphics[width=8cm]{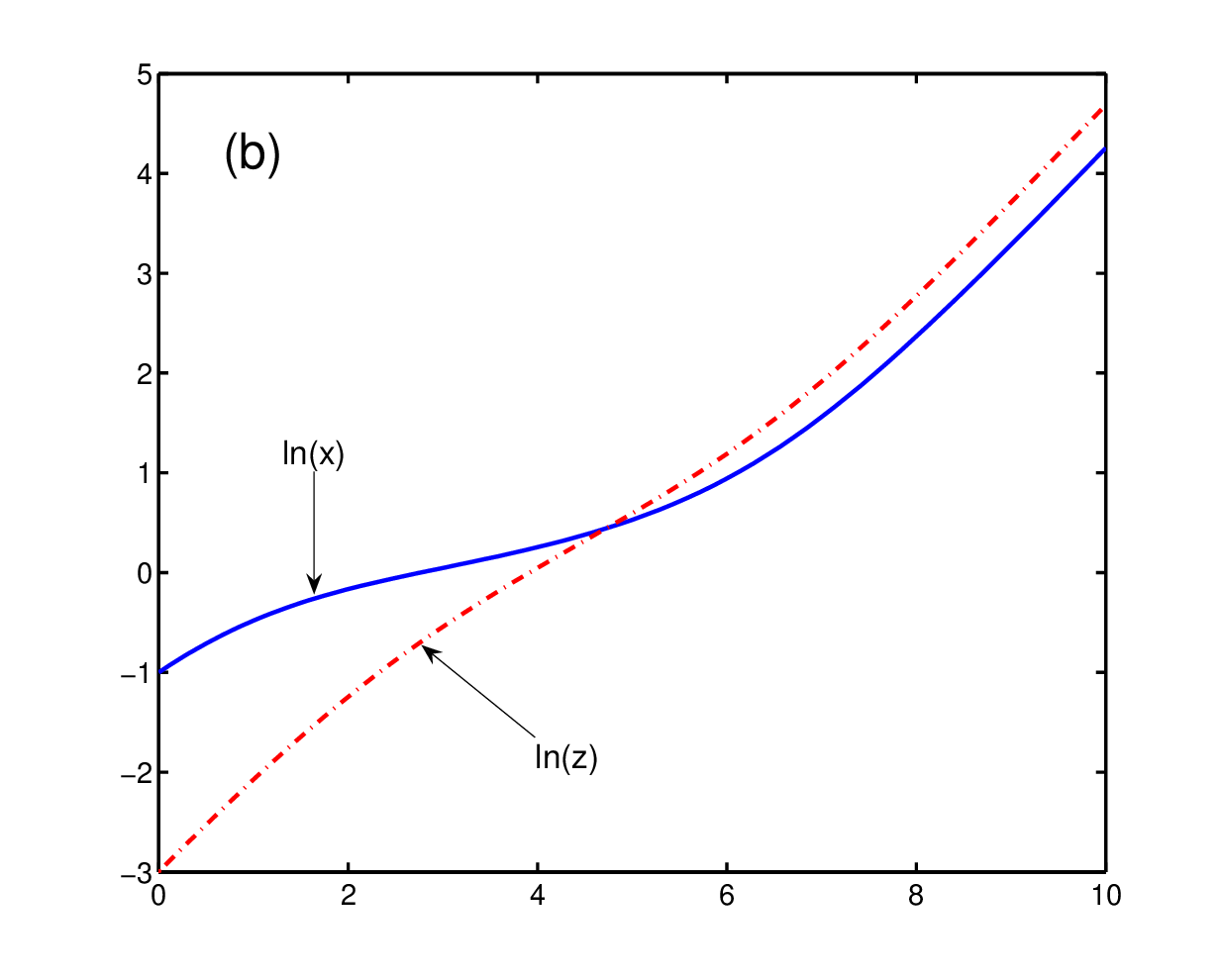} } }
\vspace{9pt}
\centerline{
\hbox{ \includegraphics[width=8cm]{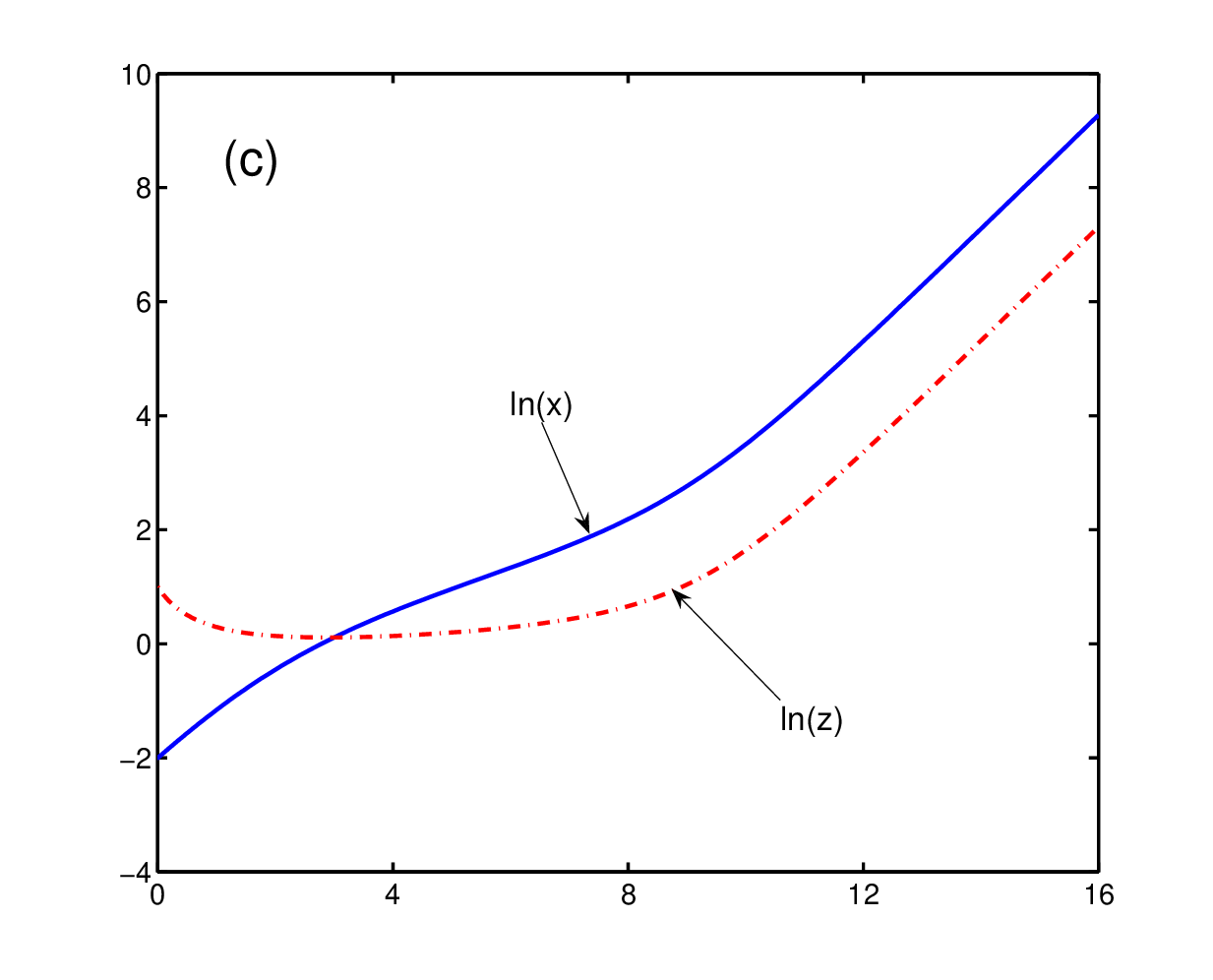} \hspace{1cm}
\includegraphics[width=8cm]{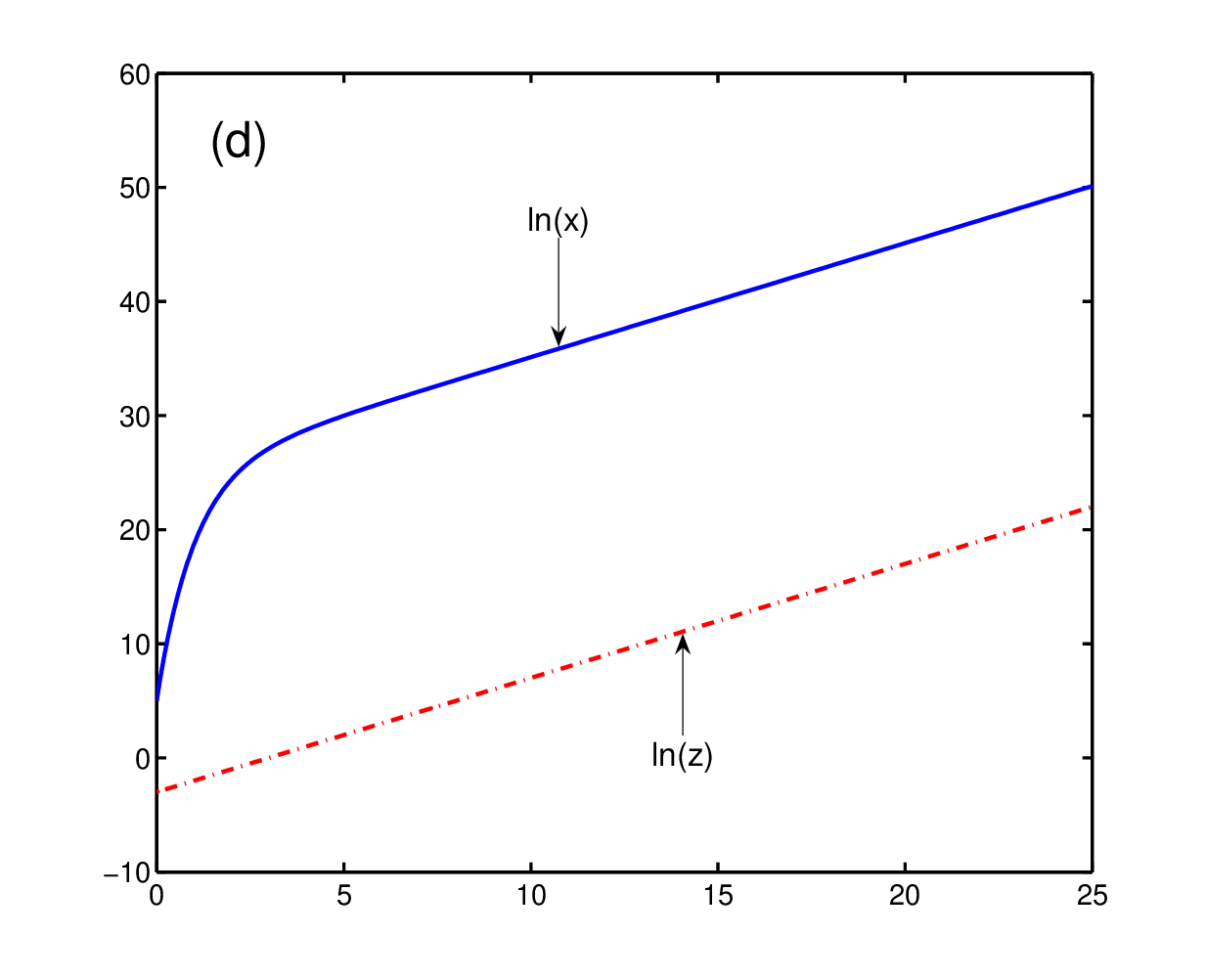} } }
\caption{Logarithmic behavior of the exponentially growing
solutions $x(t)$ (solid line) and $z(t)$ (dashed-dotted line) for
different symbiotic parameters and initial conditions: (a) $b = 1$,
$g = 0.1$, $x_0 = 148$, $z_0 = 0.135$. (b) $b=0.5$, $g=1>g_c=0.086$,
$x_0 = 0.368$, $z_0 = 0.05$. (c) $b = 1$, $g = 0.1 > g_c = 0$,
$x_0 = 0.135$, $z_0 = 2.72$. (d) $b = -1$, $g = 0.1$, $x_0 = 148$,
$z_0 = 0.05$.}
\label{fig:Fig.7}
\end{figure}

\newpage

\begin{figure}[ht]
\centerline{\includegraphics[width=12cm]{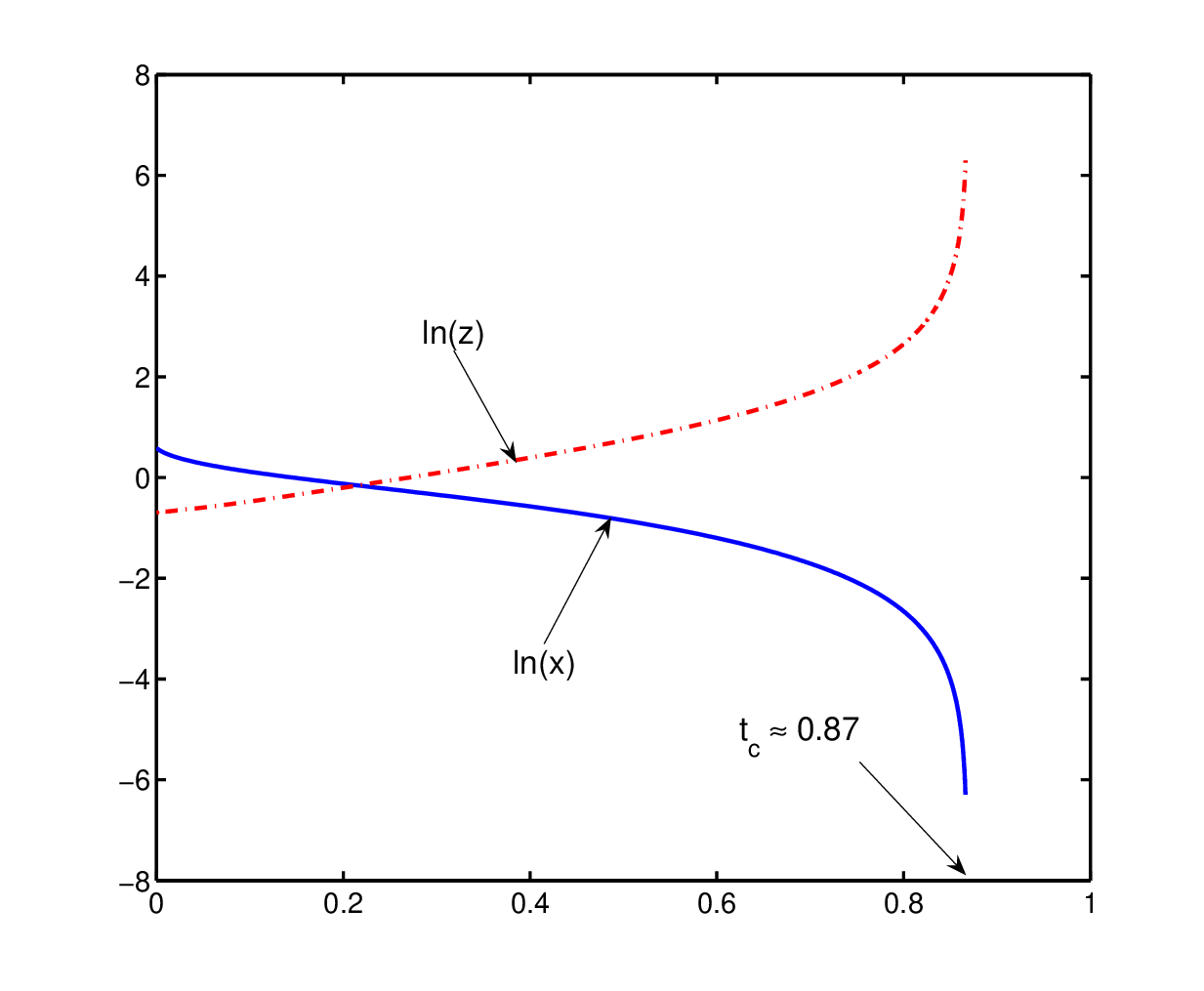}}
\caption{Logarithmic behavior of the solutions in the case of
finite-time death and singularity, $x(t)$ (solid line) and $z(t)$
(dashed-dotted line), for the parasitic relations with the symbiotic
coefficients $b = -1$, $g = -2$, under the initial conditions
$x_0 = 1.8$, $z_0 = 0.5$. For these parameters, the critical time
is $t_c = 0.87$.}
\label{fig:Fig.8}
\end{figure}

\newpage

\begin{figure}[ht]
\centerline{\includegraphics[width=12cm]{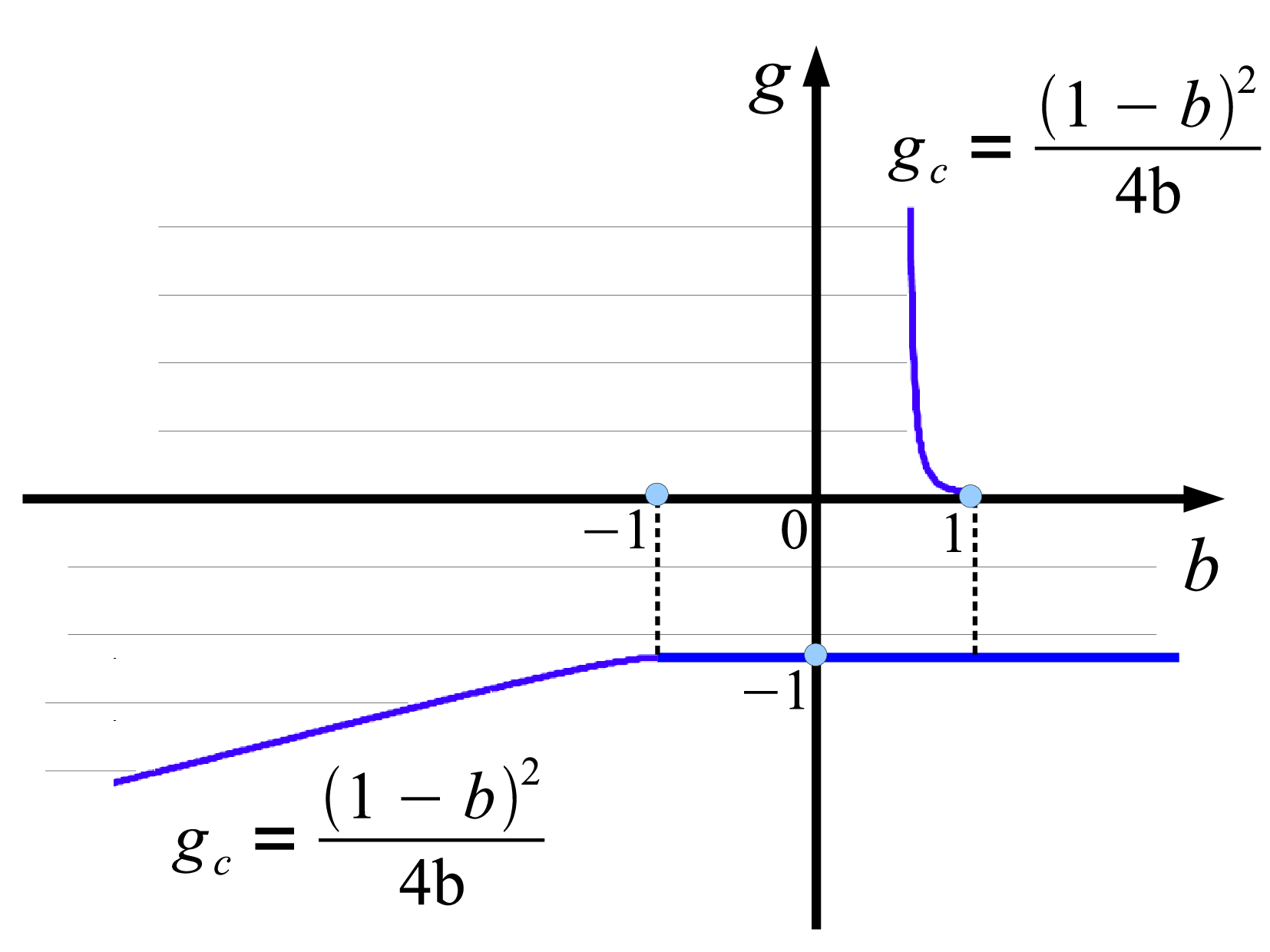}}
\caption{Region of stability (shaded) in the parameter plane
$b-g$ for the fixed points in the case of symbiosis with asymmetric
interactions.}
\label{fig:Fig.9}
\end{figure}

\newpage

\begin{figure}[ht]
\vspace{9pt}
\centerline{
\hbox{ \includegraphics[width=8cm]{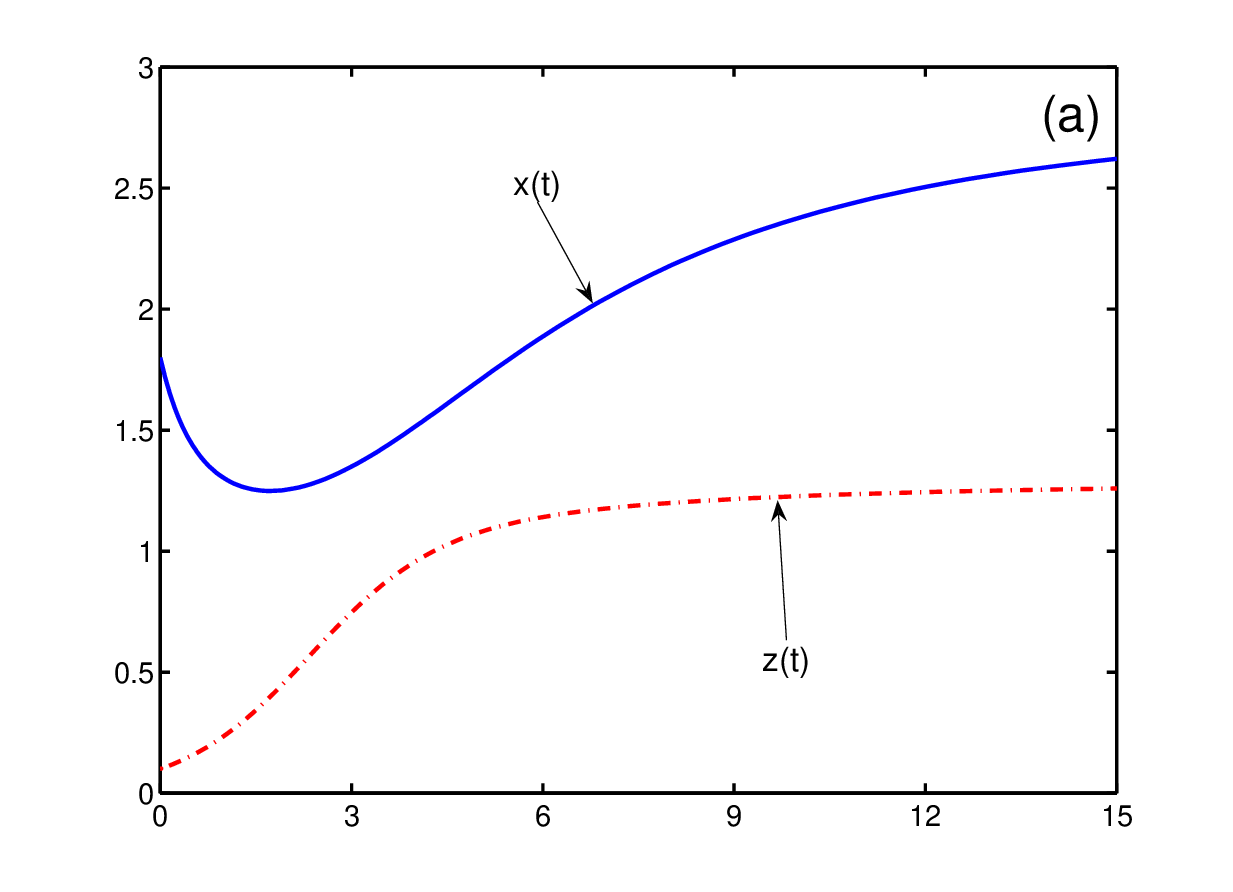} \hspace{1cm}
\includegraphics[width=8cm]{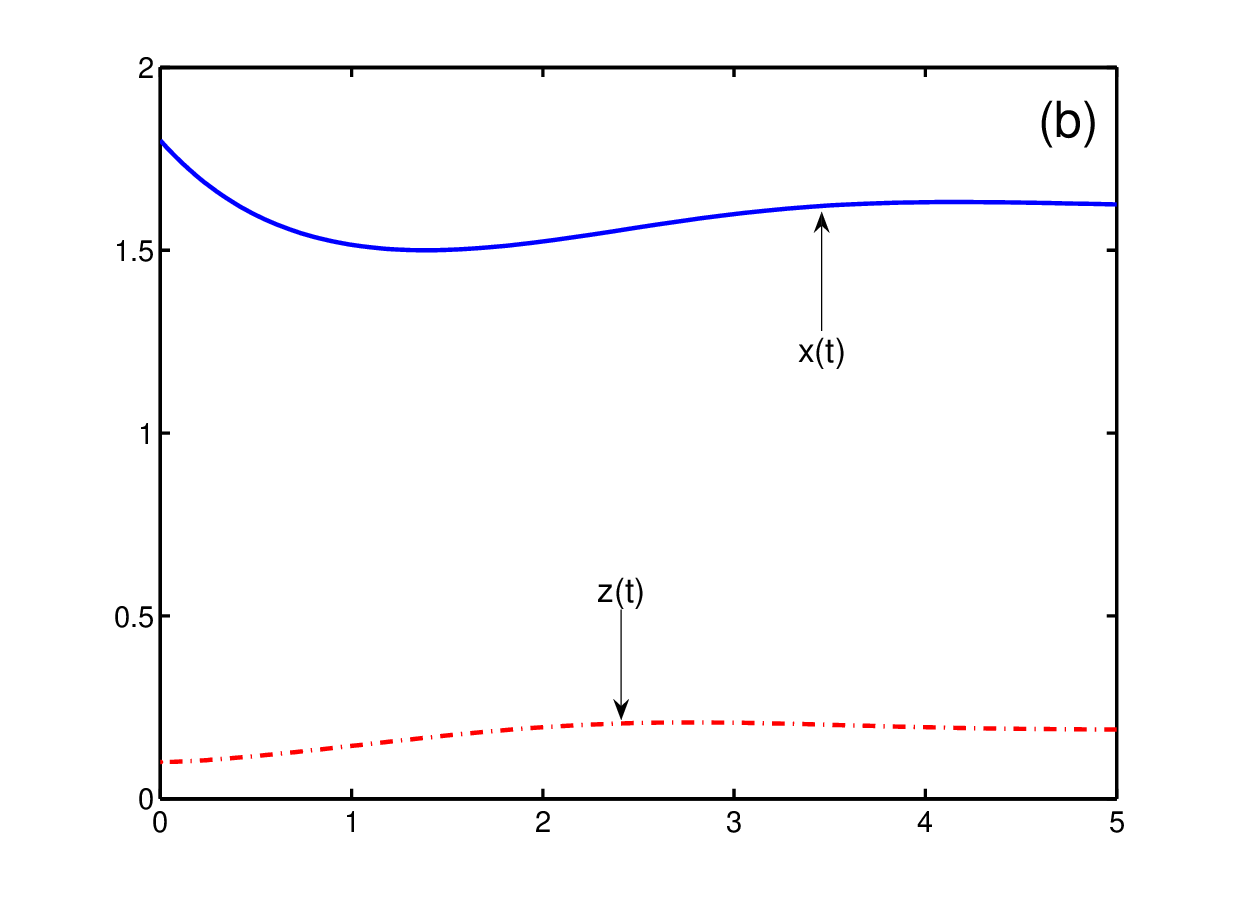} } }
\vspace{9pt}
\centerline{
\hbox{ \includegraphics[width=8cm]{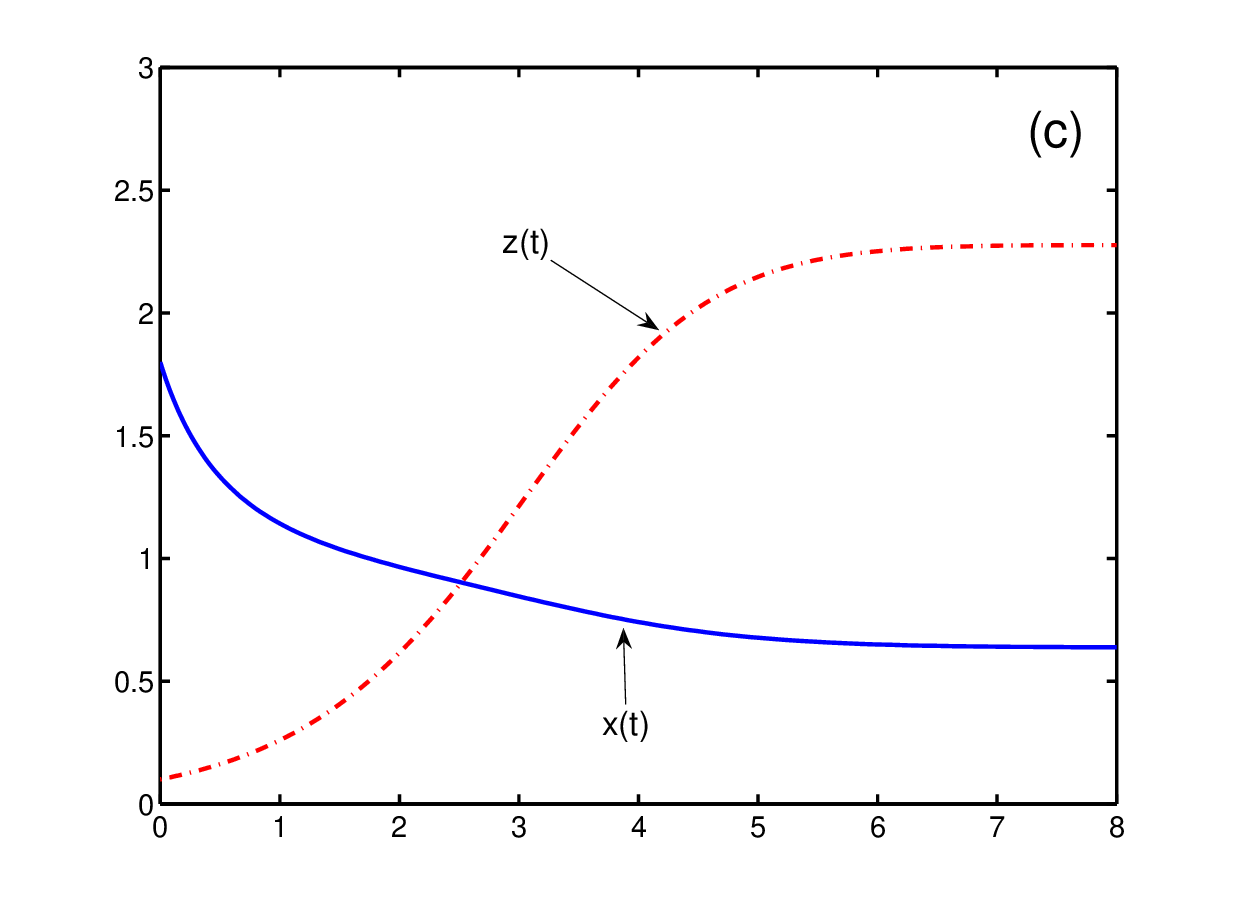} \hspace{1cm}
\includegraphics[width=8cm]{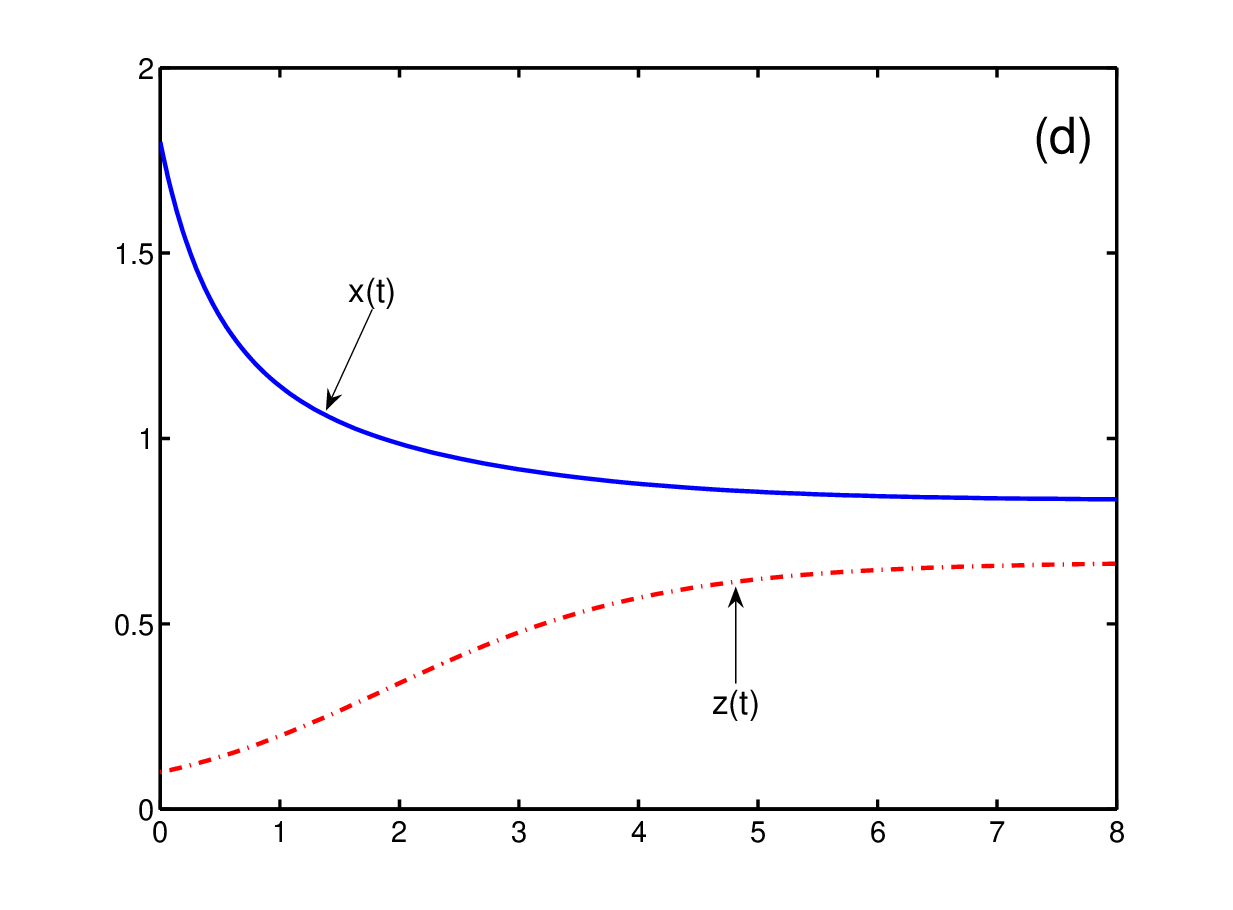} } }
\caption{Convergence to stationary states of the solutions
$x(t)$ (solid line) and $z(t)$ (dashed-dotted line), as functions
of time, in the case of asymmetric interactions, for the same
initial conditions, such that $x_0>z_0$, with $x_0=1.8$ and
$z_0=0.1$, for different parameters $b$ and $g$: (a) $b=0.5$,
$g=0.1< g_c =0.125$; the fixed points being $x^*= 2.764$,
$z^* = 1.276$. (b) $b = 2$, $g = -0.5$; the fixed points being
$x^* = 1.618$, $z^* = 0.191$. (c) $b = -0.25$, $g = 2$; with
the fixed points $x^* = 0.638$, $z^* = 2.275$. (d) $b = -0.3$,
$g = -0.4$; the fixed points being $x^* = 0.833$, $z^* = 0.667$.}
\label{fig:Fig.10}
\end{figure}

\newpage

\begin{figure}[ht]
\vspace{9pt}
\centerline{
\hbox{ \includegraphics[width=8cm]{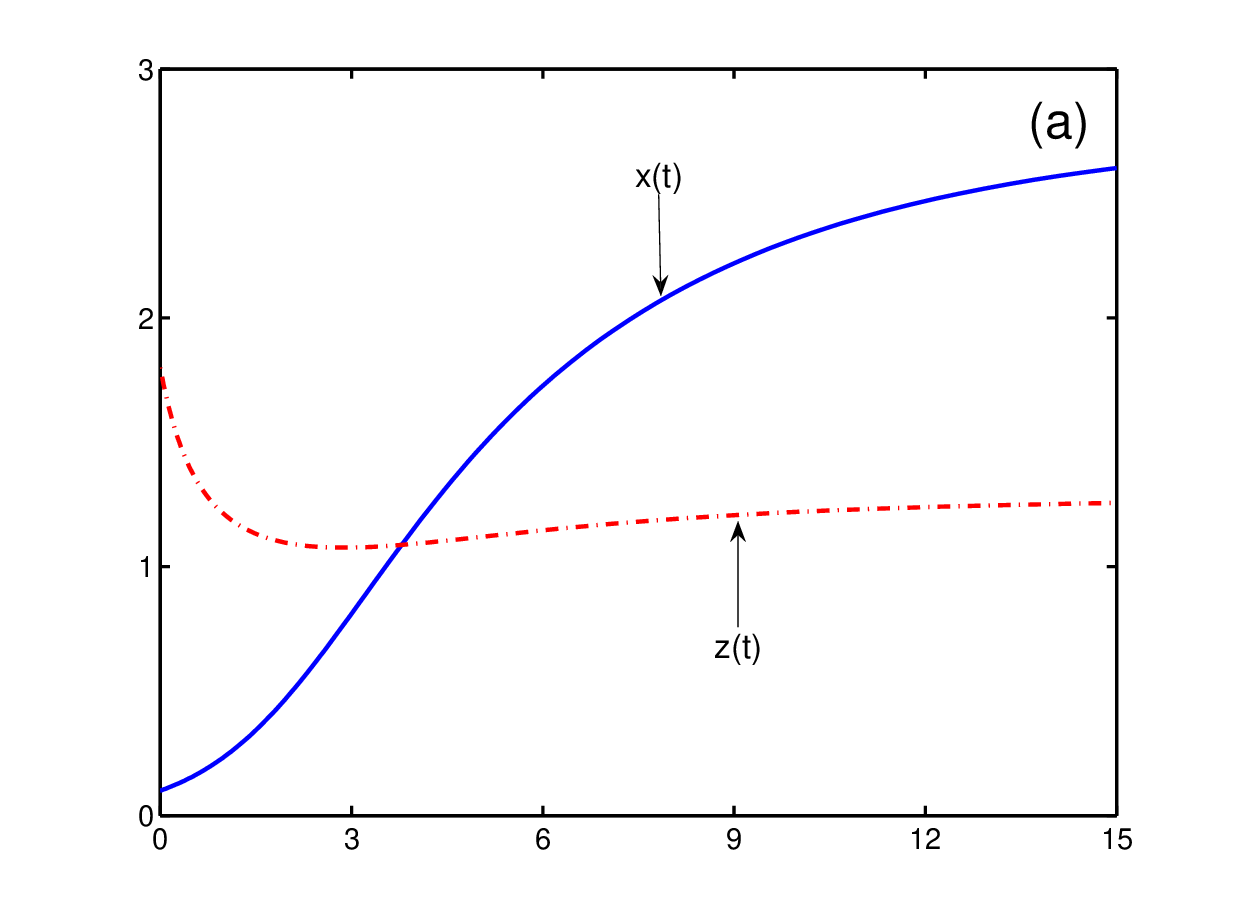} \hspace{1cm}
\includegraphics[width=8cm]{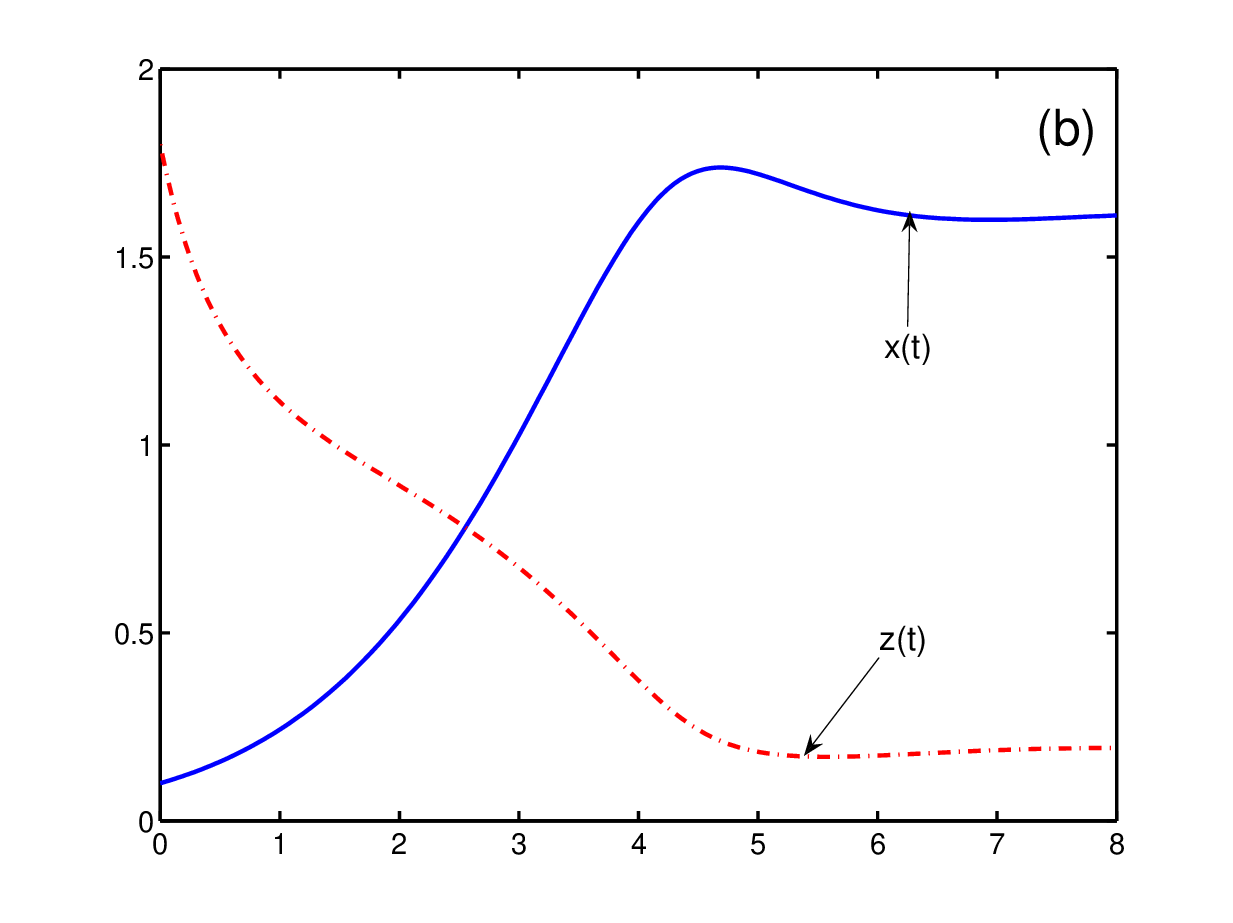} } }
\vspace{9pt}
\centerline{
\hbox{ \includegraphics[width=8cm]{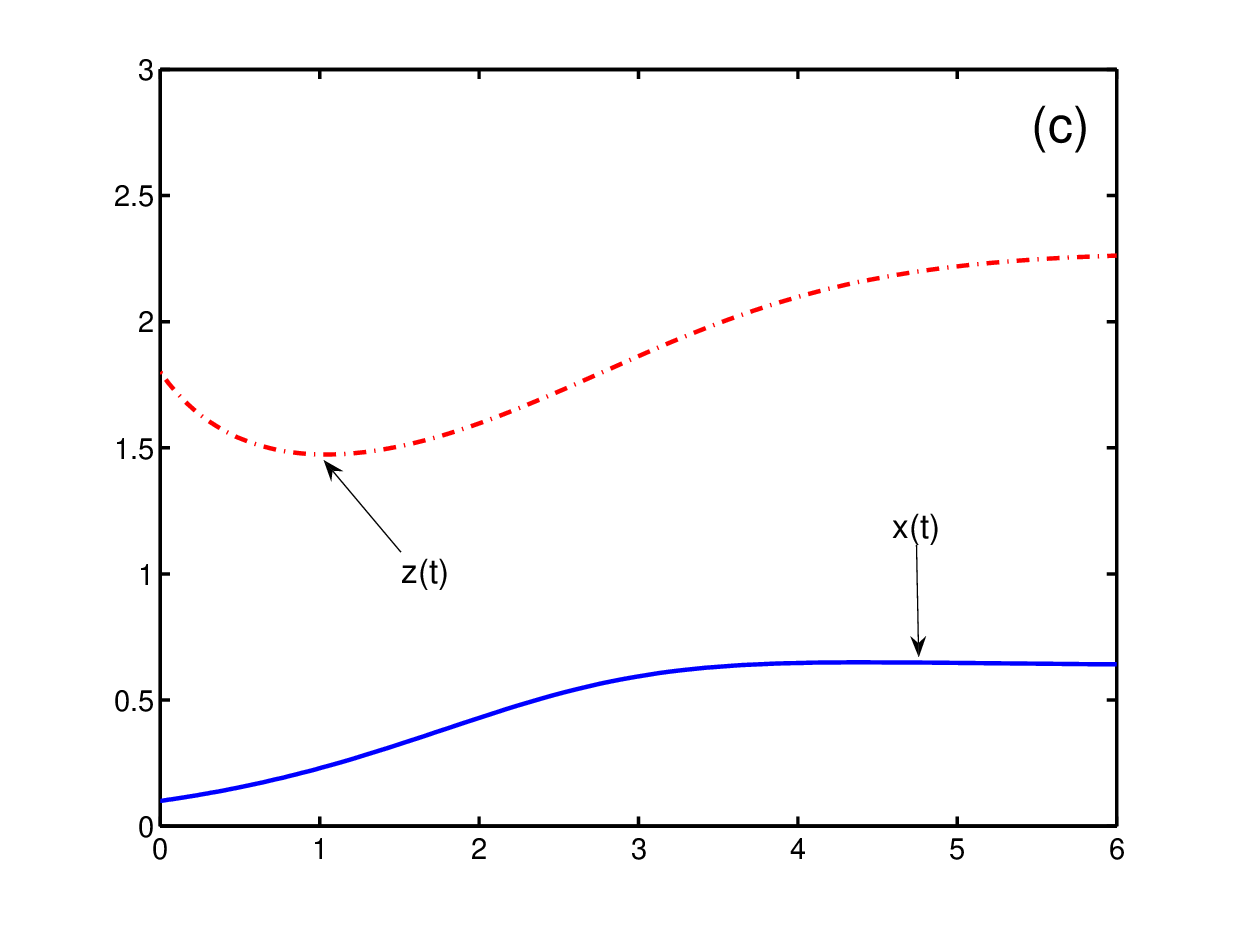} \hspace{1cm}
\includegraphics[width=8cm]{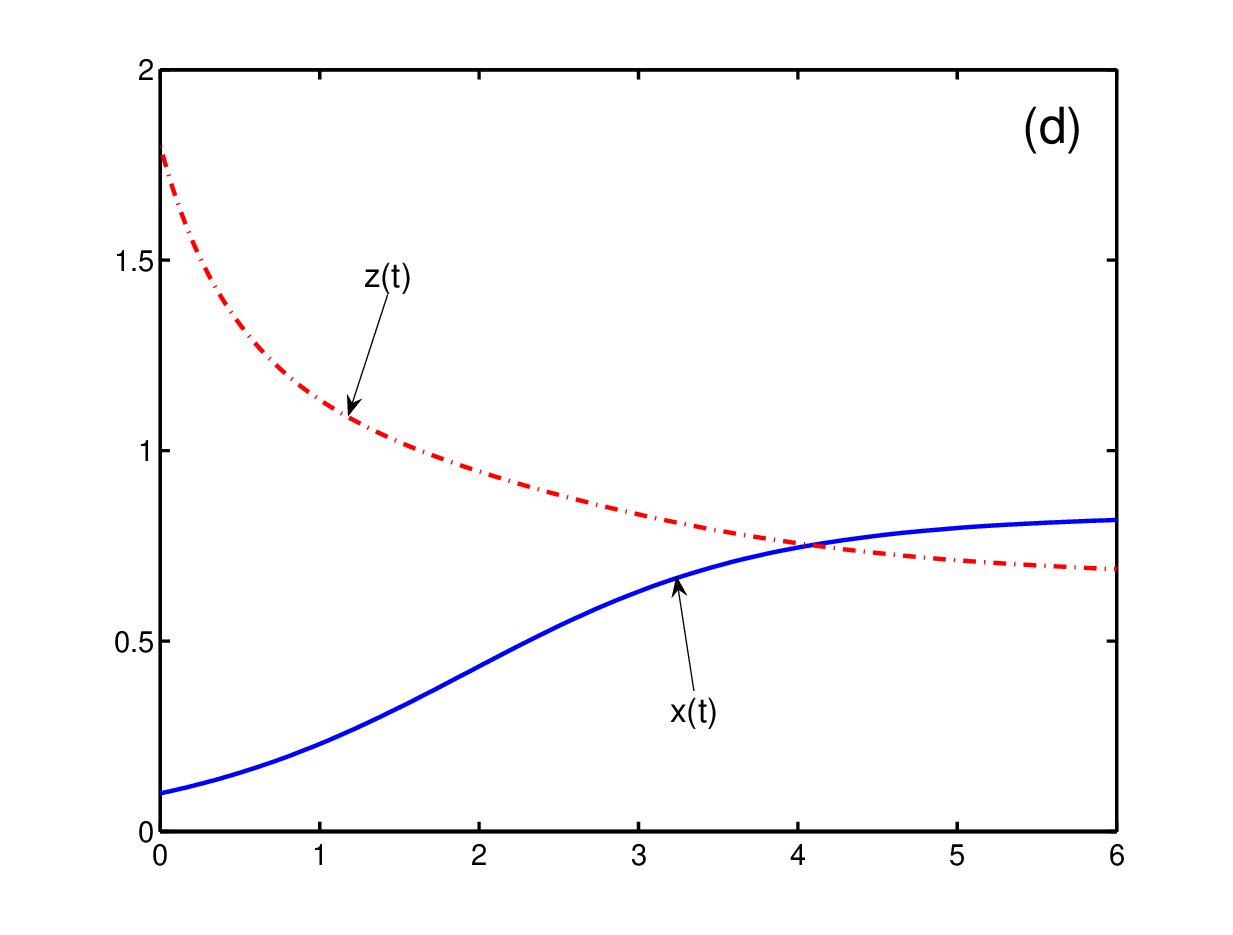} } }
\caption{Convergence to stationary states of the solutions
$x(t)$ (solid line) and $z(t)$ (dashed-dotted line), as functions
of time, in the case of asymmetric interactions, for the same
initial conditions, such that $x_0<z_0$, with $x_0=0.1$ and
$z_0=1.8$, for different parameters $b$ and $g$: (a) $b=0.5$,
$g=0.1<g_c=0.125$; the fixed points being $x^* = 2.764$,
$z^* = 1.276$. (b) $b = 2$, $g = -0.5$; the fixed points being
$x^* = 1.618$, $z^* = 0.191$. (c) $b = -0.25$, $g = 2$; with
the fixed points $x^* = 0.638$, $z^* = 2.275$. (d) $b = -0.3$,
$g = -0.4$; the fixed points being $x^* = 0.833$, $z^* = 0.667$.}
\label{fig:Fig.11}
\end{figure}

\newpage

\begin{figure}[ht]
\centerline{\includegraphics[width=12cm]{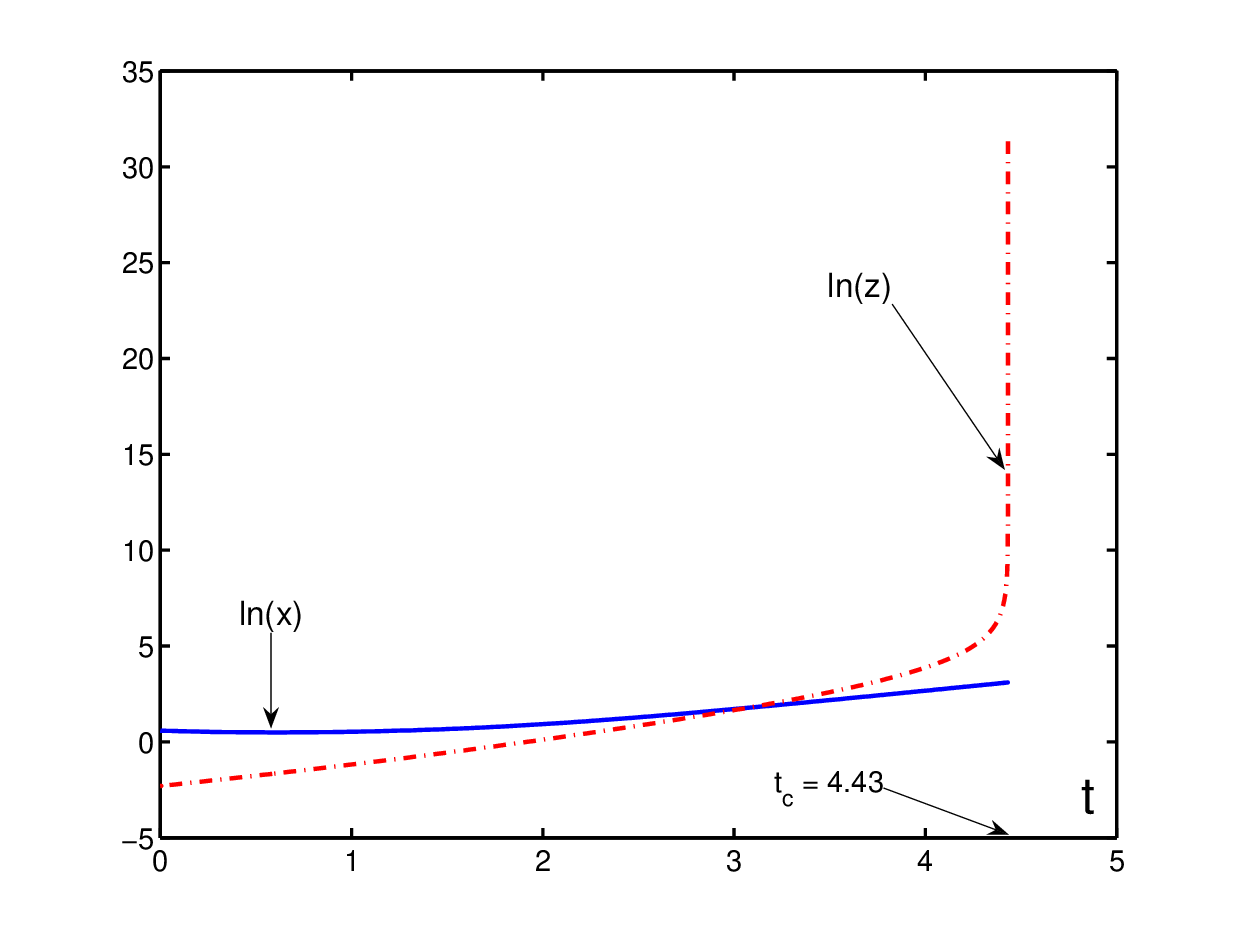}}
\caption{Logarithmic behavior of the solutions in the case
of asymmetric interactions, in the presence of the finite-time
singularity, $x(t)$ (solid line) and $z(t)$ (dashed-dotted line),
for the symbiotic coefficients $b = 2$, $g = -1.5$ and the initial
conditions $x_0 = 1.8$, $z_0 = 0.1$. The critical time here is
$t_c = 4.43$.}
\label{fig:Fig.12}
\end{figure}

\newpage

\begin{figure}[ht]
\vspace{9pt}
\centerline{
\hbox{ \includegraphics[width=8cm]{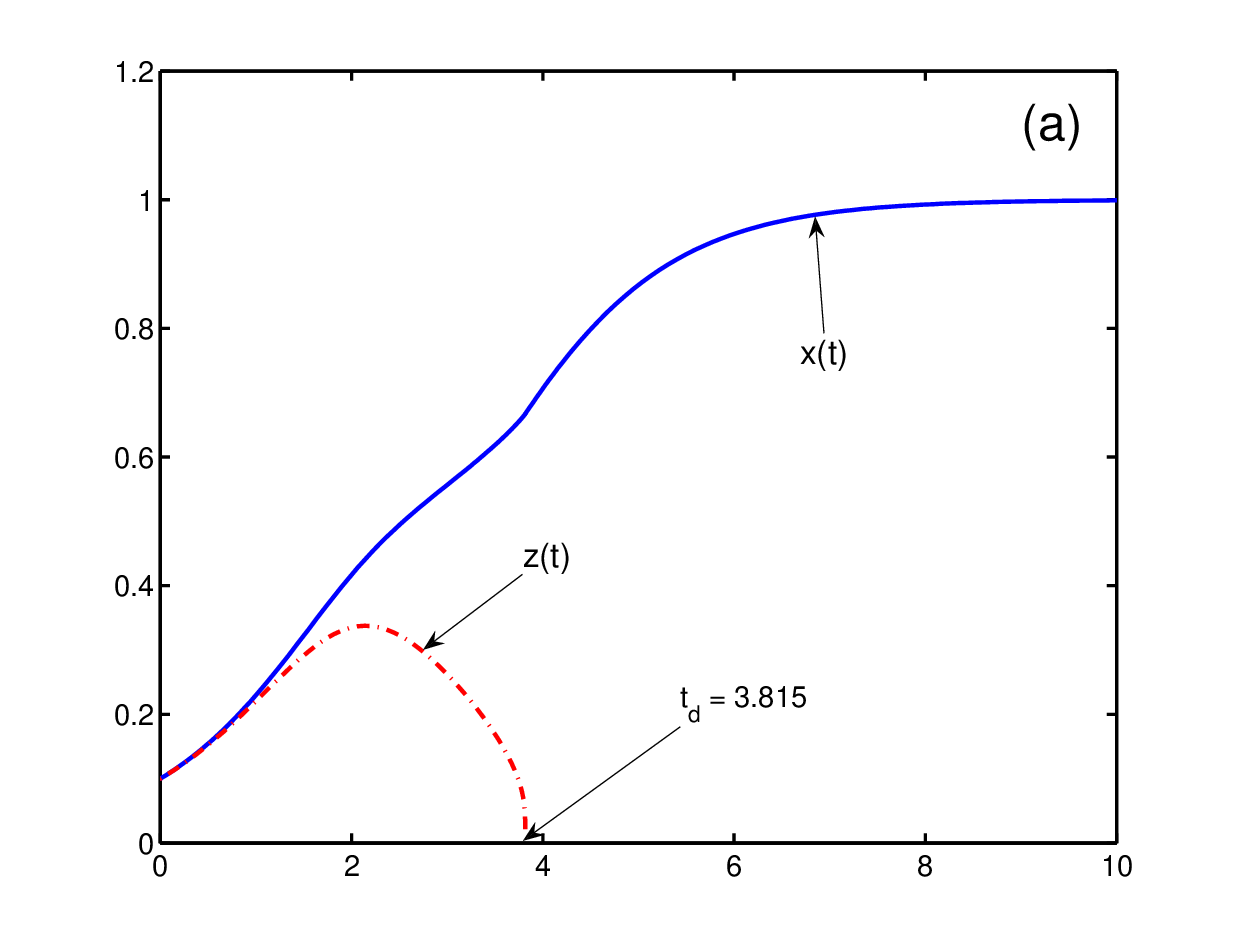} \hspace{1cm}
\includegraphics[width=8cm]{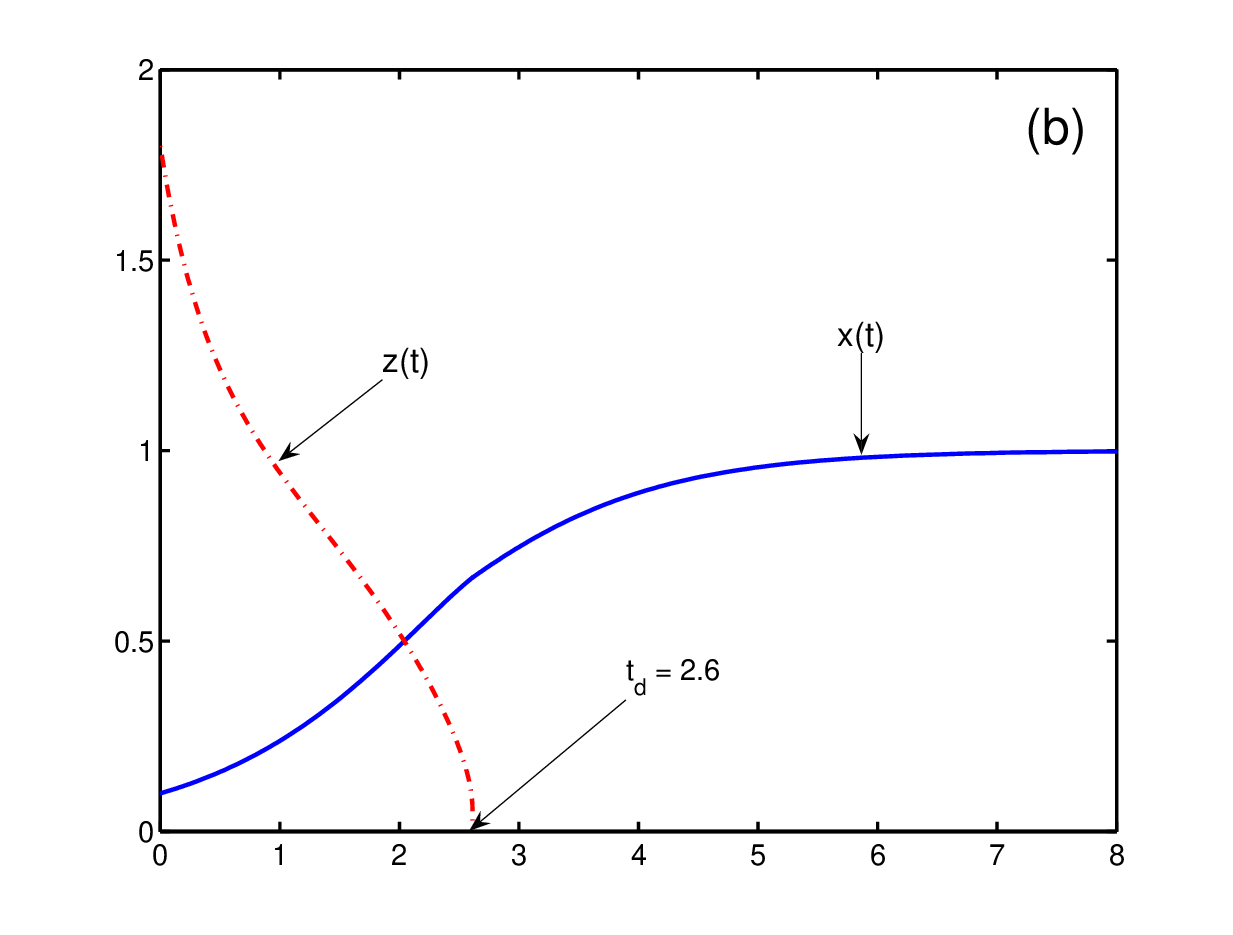} } }
\caption{Finite-time death in the case of asymmetric
interactions. The species $x(t)$ (solid line) kill the species
$z(t)$ (dashed-dotted line) at the death time $t_d$: (a) $b=-2$,
$g = -1.5$; the initial conditions are $x_0 = z_0 = 0.1$; the
death time is $t_d = 3.8$. (b) $b = 1$, $g = -1.5$; the initial
conditions are  $x_0 = 0.1$, $z_0 = 1.8$; the death time is
$t_d = 2.6$.}
\label{fig:Fig.13}
\end{figure}

\newpage

\begin{figure}[ht]
\centerline{\includegraphics[width=12cm]{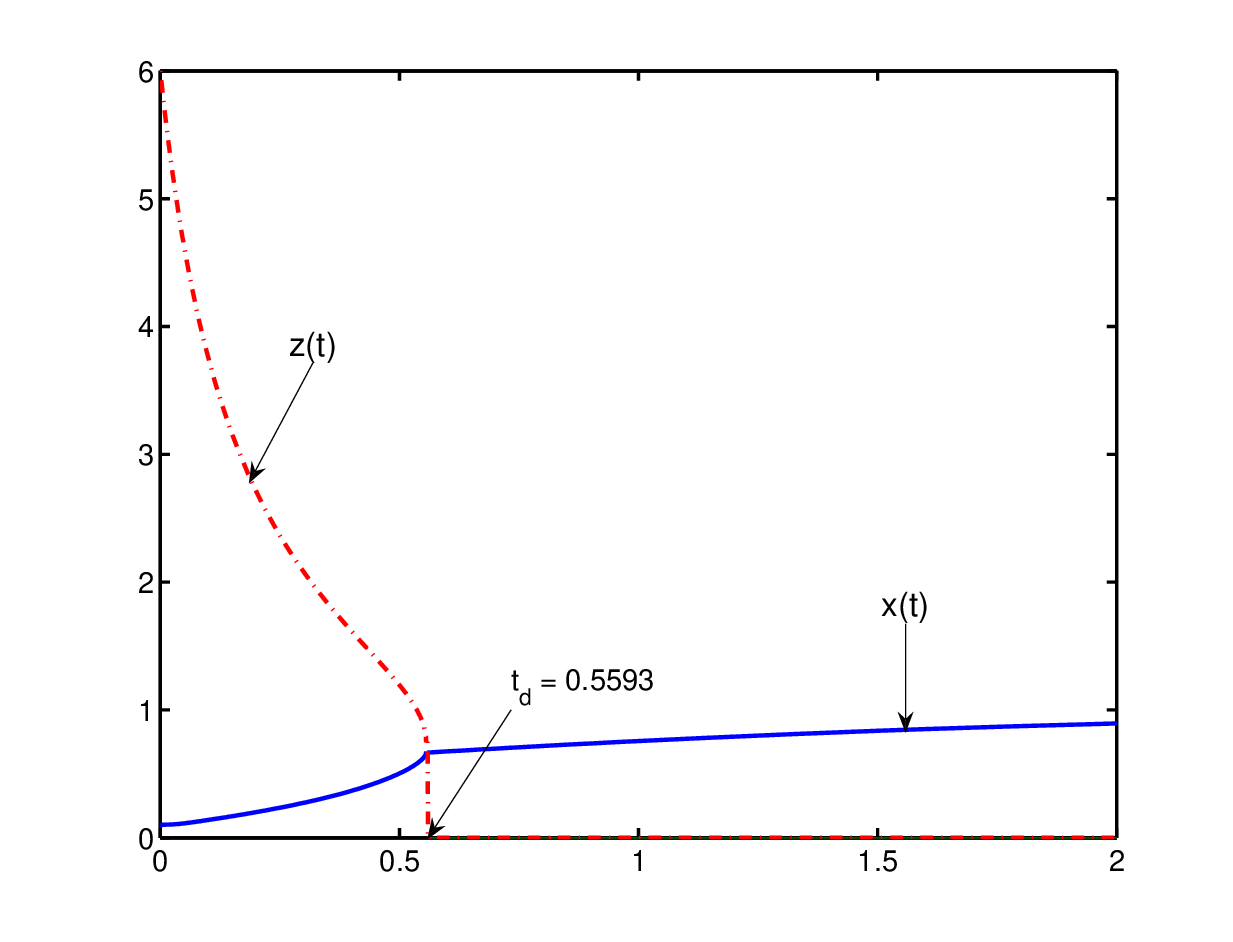}}
\caption{Finite-time death in the case of asymmetric
interactions. The species $x(t)$ (solid line) kill the species
$z(t)$ (dashed-dotted line) at the death time $t_d$. The symbiotic
coefficients are $b = -2$, $g = -1.5$; the initial conditions are
$x_0 = 0.1$, $z_0 = 6$; the death time is $t_d = 0.559$.}
\label{fig:Fig.14}
\end{figure}

\newpage

\begin{figure}[ht]
\centerline{\includegraphics[width=12cm]{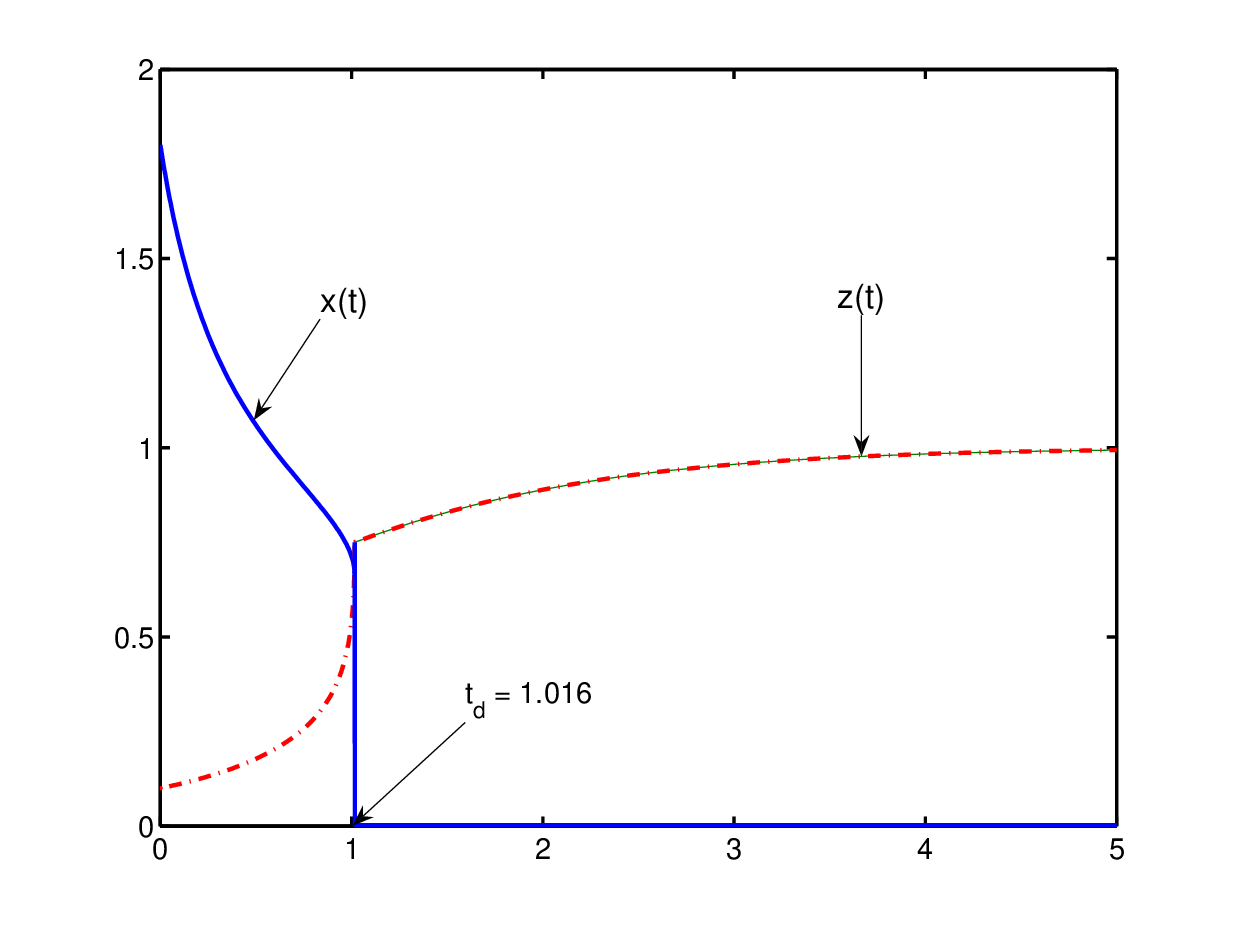}}
\caption{Finite-time death in the case of asymmetric
interactions. The species $z(t)$ (dashed-dotted line) kill the
species $x(t)$ (solid line) at the death time $t_d$. The symbiotic
coefficients are $b = -2$, $g = -1.5$; the initial conditions are
$x_0 = 1.8$, $z_0 = 0.1$; the death time is $t_d = 1.016$.}
\label{fig:Fig.15}
\end{figure}

\newpage

\begin{figure}[ht]
\centerline{\includegraphics[width=12cm]{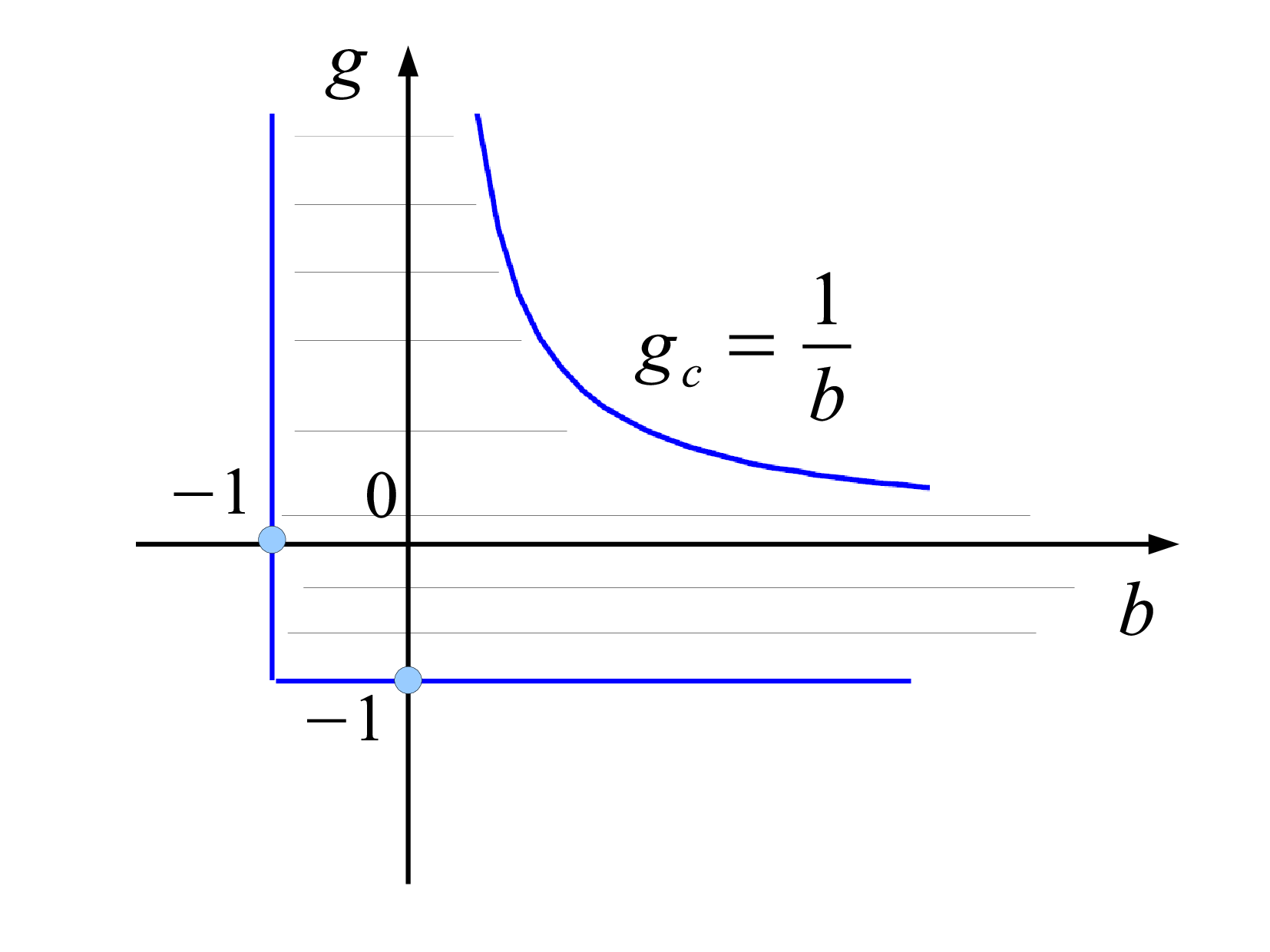}}
\caption{Region of stability (shaded) in the parameter plane
$b-g$ for the fixed points in the case of symbiosis without direct
interactions.}
\label{fig:Fig.16}
\end{figure}

\newpage

\begin{figure}[ht]
\vspace{9pt}
\centerline{
\hbox{ \includegraphics[width=8cm]{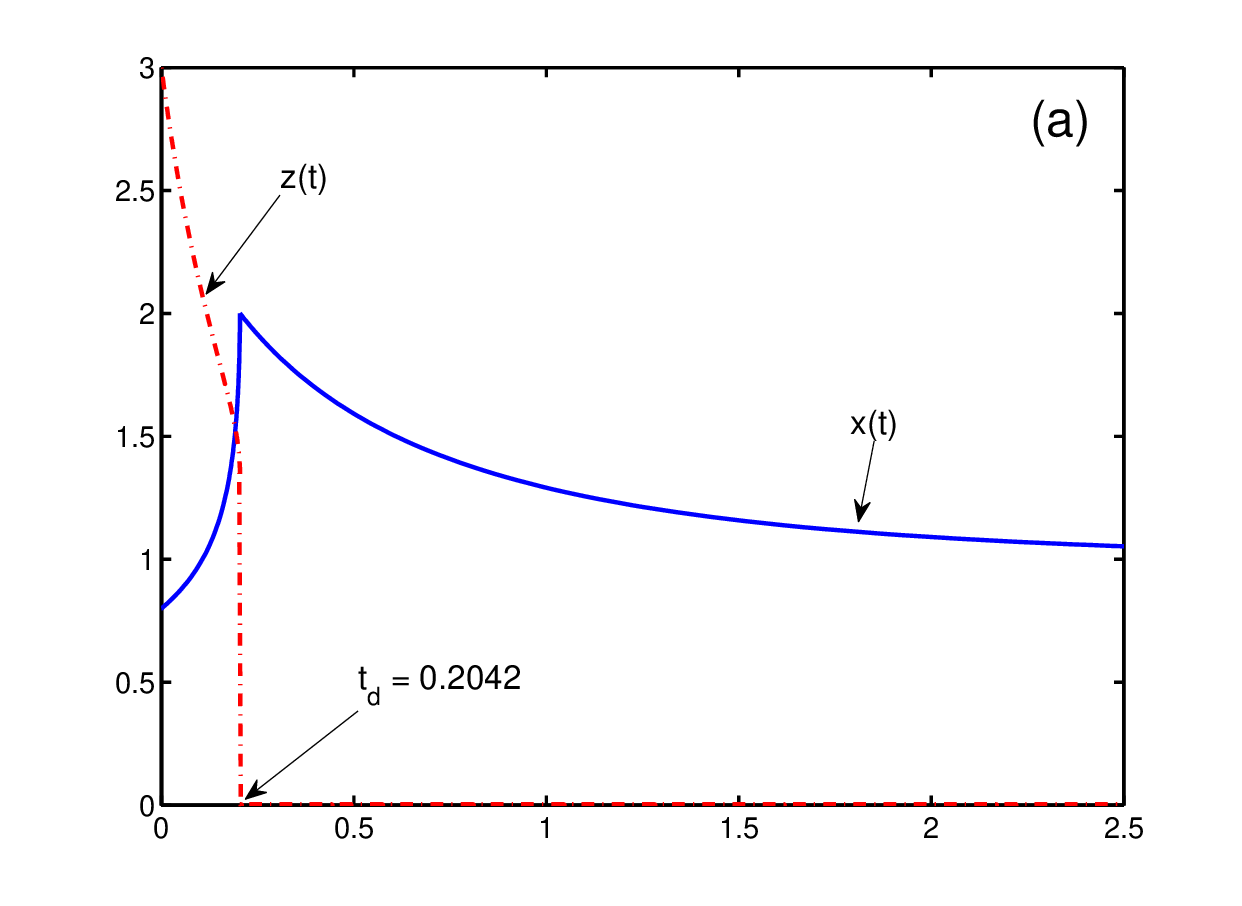} \hspace{1cm}
\includegraphics[width=8cm]{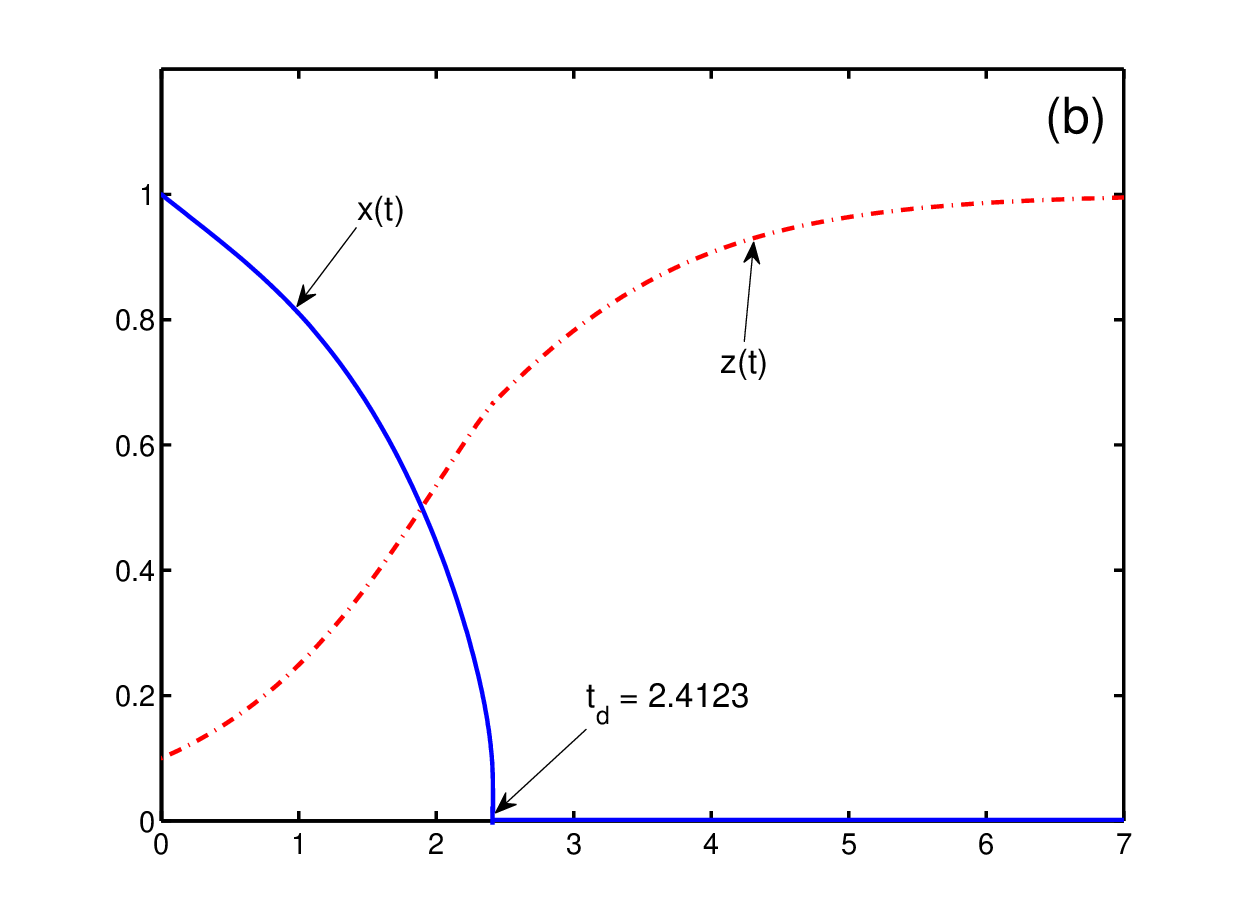} } }
\vspace{9pt}
\centerline{
\hbox{ \includegraphics[width=8cm]{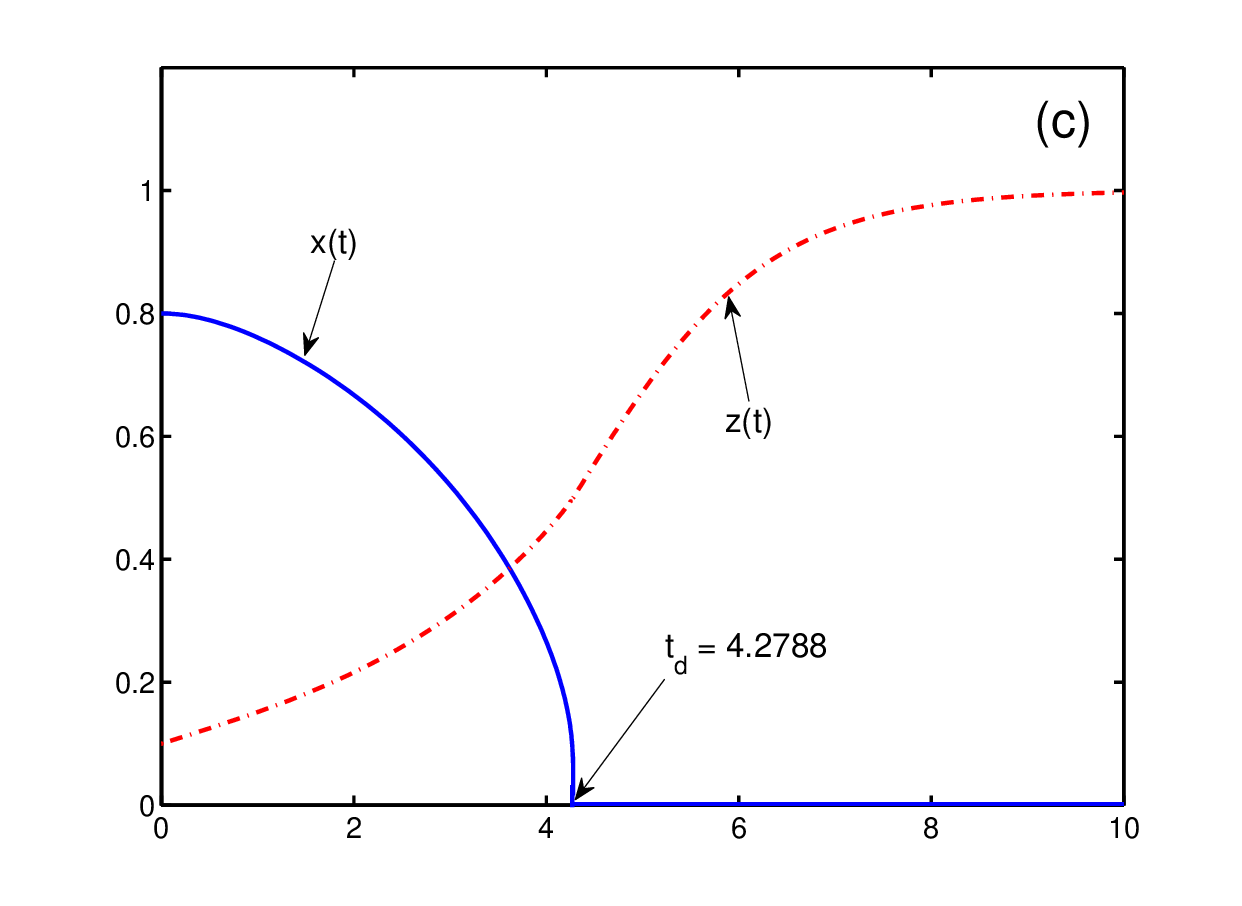} \hspace{1cm}
\includegraphics[width=8cm]{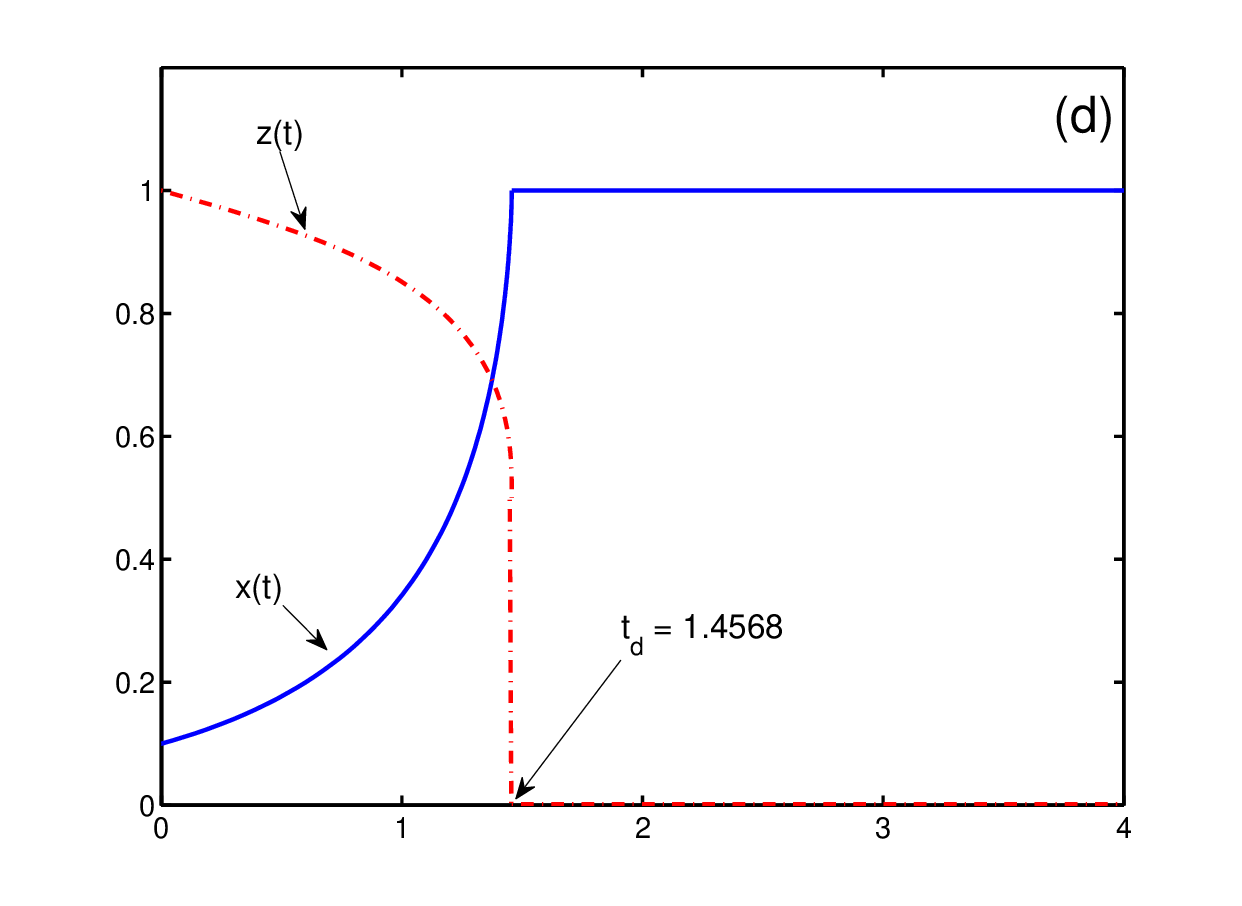} } }
\caption{Finite-time death in the case of symbiosis without
direct interactions. Temporal behavior of solutions $x(t)$ (solid
line) and $z(t)$ (dashed-dotted line) for different symbiosis
parameters and initial conditions: (a) $b = -0.75$, $g = -0.5$,
$x_0 = 0.8$, $z_0 = 3$; the death time being $t_d = 0.204$.
(b) $b = -1.5$, $g = 1$, $x_0 = 1$, $z_0 = 0.1$; with the death
time $t_d = 2.412$. (c) $b=-1$, $g=-2$, $x_0=0.8$, $z_0=0.1$;
with the death time $t_d=4.279$. (d) $b=-2$, $g=-1$, $x_0=0.1$,
$z_0 = 1$; the death time being $t_d = 1.459$.}
\label{fig:Fig.17}
\end{figure}

\end{document}